# RegionTrack: A Trace-based Sound and Complete Checker to Debug Transactional Atomicity Violations and Non-Serializable Traces

XIAOXUE MA†, City University of Hong Kong
SHANGRU WU†, City University of Hong Kong
ERNEST POBEE, City University of Hong Kong
XIUPEI MEI, City University of Hong Kong
HAO ZHANG, City University of Hong Kong
BO JIANG, Beihang University
WING-KWONG CHAN, City University of Hong Kong

Atomicity is a correctness criterion to reason about isolated code regions in a multithreaded program when they are executed concurrently. However, dynamic instances of these code regions, called transactions, may fail to behave atomically, resulting in transactional atomicity violations. Existing dynamic online atomicity checkers incur either false positives or false negatives in detecting transactions experiencing transactional atomicity violations. This paper proposes RegionTrack. RegionTrack tracks cross-thread dependences at the event, dynamic subregion, and transaction levels. It maintains both dynamic subregions within selected transactions and transactional happens-before relations through its novel timestamp propagation approach. We prove that RegionTrack is sound and complete in detecting both transactional atomicity violations and non-serializable traces. To the best of our knowledge, it is the first online technique that precisely captures the transitively closed set of happens-before relations over all conflicting events with respect to every running transaction for the above two kinds of issues. We have evaluated RegionTrack on 19 subjects of the DaCapo and the Java Grande Forum benchmarks. The empirical results confirm that RegionTrack precisely detected all those transactions which experienced transactional atomicity violations and identified all non-serializable traces. The overall results also show that RegionTrack incurred 1.10x and 1.08x lower memory and runtime overheads than Velodrome and 2.10x and 1.21x lower than Aerodrome, respectively. Moreover, it incurred 2.89x lower memory overhead than DoubleChecker. On average, Velodrome detected about 55% fewer violations than RegionTrack, which in turn reported about 3%-70% fewer violations than DoubleChecker.

† Co-first author.

## 1 INTRODUCTION

*Atomicity*, also known as conflict-serializability [5][16], is a non-interference property. An earlier survey [23] reports that atomicity violation bugs account for 70% of the examined non-deadlock concurrency bugs in multithreaded programs. Detecting and localizing these concurrency bugs are needed for program debugging. In addition, these concurrency bugs only exist at limited places of a multithreaded program. Identifying these erroneous parts, for instance, helps the diagnose step in the test-diagnose-fix cycle [3][55].

We firstly introduce some terminologies to ease us to describe our work. In a multithreaded program, multiple threads may execute their code regions concurrently. An *atomicity specification* [5][16] is a set of code regions, called *atomic regions*, each expecting to exhibit the atomicity property. An atomic region exhibits the *atomicity property* if it satisfies the following condition. The condition is that for any execution trace α of the program, there is an equivalent trace α' of α such that in α', each instance *tx* of the atomic region, called *transaction*, can execute without interleaving with events of other code regions. In other words, *tx* executes serially in α' [14][15][16]. If *tx* in α cannot be reordered to become a serial execution in some equivalent trace of α, a ***transactional atomicity violation*** (or *atomicity violation* for short) [5][16] is said to occur on *tx*. If all transactions in α execute serially, then α is a *serial trace* [16]. Any trace equivalent to a serial trace is *serializable*, otherwise, ***non-serializable*.**

For brevity, we classify existing techniques to detect these two issues from multithreaded programs into two board categories: graph-based techniques and vector-clock-based techniques.

Exemplified techniques of the first category are Velodrome [16] and DoubleChecker [5]. They build a *transactional* happens-before (HB) graph along the trace α. Such a graph tracks the program orders, lock release-acquisition synchronization orders, and cross-thread memory access dependences over the transactions in α. Whenever adding a new cross-thread HB edge between two transactions to the graph, they aim to identify any cycle (i.e., strongly connected components) formed in the graph. The presence of such a cycle in the graph indicates that the trace is non-serializable. Each transaction is modeled as a node in the graph. In general, these techniques select a transaction in the located cycle to blame for the occurrence of the transactional atomicity violation. Nonetheless, they only capture either a *superset* of the happens-before relations at the event level (e.g., DoubleChecker) or a *subset* of them (e.g., Velodrome) into the graph. As such, a transaction without transactional atomicity violation may be reported by the former as a blamed transaction (i.e., a false positive), and a transaction to be blamed may be missed by the latter (i.e., a false negative). Due to the needs of cycle localization, these techniques must keep those graph nodes reachable from not-yet-finished transactions, increasing their memory footprints. For instance, in the experiment reported in [5], both DoubleChecker and Velodrome ran out of memory on a few benchmarks. The researchers [5] altered the test harness or configured the benchmarks to reduce the memory footprints of their techniques so that the traces could be analyzed.

Exemplified techniques for the second category are the work presented in Chapter 3 of Wu's thesis [53] and Aerodrome [52]. They detect either transactional atomicity violation [53] or non-serializable traces [52], but not both. The work of Wu [53] (Algorithms 1 and 2 in this paper) is a prototype of the technique we propose in this paper, which we will introduce when we present RegionTrack in Section 4. It captures the transitive closure of the happens-before relations at the event level for each transaction into its set of transactional happens-before relations. It recovers the loss in precision in the partial order among transactions (due to abstraction of the event-level happens-before relations to the transaction level) by dividing each ambiguous transaction into subregions, each with its own vector clock. Since transactional happens-before relations are captured, the underlying events can be discarded once their belonging transaction or subregion is identified. However, this technique is inapplicable to identify non-serializable traces precisely, which we will present in Section 4. Aerodrome [52] generates an abstraction of transactional happens-before relations from the event-level happens-before relations. Each event has its vector clock to be tracked at the event level. When a transaction ends, Aerodrome locates each vector clock of the related memory access events, threads, and lock release events. It then checks whether the vector clock forms a partial order with the beginning event of the transaction to track transactional happens-before relations. If this is the case, updates of the vector clocks of the related events and threads are performed to capture the partial order found. This strategy suffices to identify non-serializable traces precisely. However, it does not keep the ordering of events. So, this idea cannot detect transactional atomicity violations efficiently.

In this paper, we present a novel vector-clock-based technique called RegionTrack. RegionTrack precisely detects both transactions with transactional atomicity violation occurrences and non-serializable traces. We observe that in the online processing of events, a transaction that starts and completes a cycle in a transactional HB graph must be a *currently running transaction* of a thread — any finished transaction involved in that cycle can only be an intermediate transaction involved in the cycle. As such, it should be possible to discard the analysis program state for any transaction once that transaction has completed, *providing that* the partial orders among the currently running transactions can be kept intact (irrespective to whether a cycle has been formed).

RegionTrack develops a set of novel algorithms by using the above observation as a guiding principle. To detect transactional atomicity violations, it uses a mechanism of forward propagation of event timestamps. This mechanism precisely captures the transactional happens-before relations kept at the subregion level (Algorithms 1 and 2 in this paper). To detect non-serializable traces, RegionTrack designs a novel transactional vector clock mechanism. This mechanism, on the other hand, maintains the reversal frontier of transactional happens-before relations of the currently running transaction $tx_j$ for each thread $j$ (Algorithms 3 and 4). In RegionTrack, each thread maintains a vector clock (called transactional vector clock). Suppose that a new cross-thread transactional happens-before relation $tx_i \rightsquigarrow tx_j$ based on a happens-before-relation $e_i \rightarrowtail e_j$ where $e_i \in tx_i$ and $e_j \in tx_j$ at the event level is created. Upon the creation of $e_i \rightarrowtail e_j$, RegionTrack conducts a recursive round of timestamp forward- and back-propagation. The propagation will update the timestamps in the transactional vector clocks of all these threads that can transitively see $tx_j$ by keeping $tx_j$'s timestamp in them. So, each related thread can directly see $tx_j$ through its transactional vector clock. Thereby, when a subsequent transactional happens-before relation is created, any timestamp propagations along any permutations of previously traversed propagation subpaths will be avoided, which speeds up the maintenance of the reversal frontiers of transactional happens-before relations for the transaction. We also prove by theorems that RegionTrack is sound and complete in detecting both transactional atomicity violations and non-serializable traces.

We have implemented RegionTrack in the Jikes RVM [2][4] to show its feasibility. We have also evaluated it on 19 subjects from the DaCapo [7] and Java Grande Forum [48] benchmark suites. The experiment results showed that RegionTrack was precise and reported more transactional atomicity violations than Velodrome. We have also inspected all the transactional atomicity violations reported by RegionTrack being true positives. Besides, RegionTrack reported all non-serializable traces. RegionTrack incurred a slightly lower time overhead than Velodrome on 15 out of 19 benchmarks and lower memory overhead on 8 out of 19 benchmarks. On average, RegionTrack incurred 1.10x and 1.08x lower memory and runtime overheads than Velodrome and 2.10x and 1.21x lower than Aerodrome, respectively. Moreover, it incurred 2.89x lower memory overhead than DoubleChecker and ran 2.95x slower, albeit that DoubleChecker ran on a more efficient infrastructure [47]. On average, Velodrome detected about 55% fewer violations than RegionTrack, which in turn reported about 3%-70% fewer violations than DoubleChecker. We also observed that owing to a much higher memory footprint requirement, DoubleChecker frequently ran out of memory on some subjects in the experiment.

The main contribution of this work is threefold: (1) This work presents RegionTrack. To our best knowledge, RegionTrack is the *first* technique that is both sound and complete in locating transactions with transactional atomicity violation occurrences and identifying non-serializable traces. (2) RegionTrack is the *first* dynamic online technique that locates transactional atomicity violations without the need for explicit enumeration of the transaction order. (3) This work reports a comprehensive experiment to show the feasibility of RegionTrack and to validate its effectiveness and efficiency.

The rest of the paper is organized as follows. Sections 2 to 6 present the preliminaries, a motivating example, RegionTrack and its evaluation. Section 6 reviews the related work. Section 7 concludes this work.

## 2 PRELIMINARIES

In this section, we review the terminologies and preliminaries necessary to introduce our work.

### 2.1 Events, Transaction and Serial Trace

A trace α is a sequence of *n* events: $\alpha = \langle e_1, e_2, \ldots, e_i, \ldots, e_n \rangle$. Each event $e_i$ represents an operation below:

- $r(t, x)$ and $w(t, x)$: thread *t* reads a value from variable *x* and writes a value to variable *x*, respectively.
- $acq(t, m)$ and $rel(t, m)$: thread *t* acquires and releases a lock *m*, respectively.
- $begin(t, l)$ and $end(t, l)$: thread *t* marks the beginning and the ending of an atomic region *l*.

We denote the thread performing the event $e_i$ by $T(e_i)$.

Following [16], we denote the sequence of events executed by a thread *t* in between a matching pair of events $begin(t, l)$ and $end(t, l)$ as a (regular) transaction $tx = \langle begin(t, l), \ldots, e_x, \ldots, end(t, l) \rangle$, where $T(e_x) = t$. For an event $e_x$ in *tx*, we denote it as $e_x \in tx$. We denote the beginning event and the ending event of the transaction *tx*, by *tx.begin* and *tx.end*, respectively. We also denote the thread preforming the transaction *tx* by $T(tx)$.

We further denote the transaction containing the event $e_x$ by $Trans(e_x)$. Moreover, if an event *e* does not belong to any atomic region, then *e* by itself forms a (unary) transaction. Besides, we denote the sequence of events executed by a thread *t* starting from $begin(t, l)$ but not executing the event $end(t, l)$ as a *currently running transaction* of *t*, referred to as $tx = \langle begin(t, l), \ldots, e_x, \ldots \rangle$, where $T(e_x) = t$.

If all the events of a transaction *tx* are executed consecutively without being interleaved by events of other threads, then *tx* is said to execute *serially*. If all transactions in α execute serially, α is a *serial trace* [16].

### 2.2 Transactional Happens-Before Relation, Transactional Atomicity Violations and Non-Serializable Traces

Two events in a trace α *conflict* with each other if any one of the following three conditions is satisfied [5][16]: (1) They both access (i.e., read or write) the same variable, and at least one of the operations is a write. (2) They acquire or release the same lock. (3) They are both executed by the same thread.

Suppose that in a trace α, an event $e_i$ appears before another event $e_j$. If $e_i$ conflicts with $e_j$, then $e_i$ *happens before* $e_j$, denoted as $e_i \rightarrowtail e_j$. Happens-before (HB) relation is transitive over the set of events in trace α.

With respect to a cross-thread HB relation $e_i \rightarrowtail e_j$, where $e_i \in tx_i$ and $e_j \in tx_j$, the HB relation $e_i \rightarrowtail e_j$ is called as an ***outgoing*** and ***incoming*** *HB relation* of $tx_i$ and $tx_j$, respectively.

For two transactions or subregions $tx_i$ and $tx_j$ in a trace α, if there are two events $e_i \in tx_i$ and $e_j \in tx_j$ forming $e_i \rightarrowtail e_j$, then $tx_i$ ***transactional happens before*** $tx_j$, denoted by $tx_i \rightsquigarrow tx_j$. Transactional happens-before (THB) relation is transitive over the whole set of transactions or subregions in trace α.

We also call the THB relation $tx_i \rightsquigarrow tx_j$ as ***outgoing*** and ***incoming*** THB relations of $tx_i$ and $tx_j$, respectively.

Suppose that there is a sequence of THB relations $tx_i \rightsquigarrow tx_{i+1}$ for $1 \le i \le n$ such that each relation $tx_i \rightsquigarrow tx_{i+1}$ is formed by their underlying HB relations $e_i \rightarrowtail e_{i+1}$ (where $e_i \in tx_i$ and $e_{i+1} \in tx_{i+1}$). If the event (denoted as *x*) in the HB relation for $tx_{i-1} \rightsquigarrow tx_i$ and the event (denoted as *y* ) in the HB relation for $tx_i \rightsquigarrow tx_{i+1}$ follow the program order of $tx_i$, (i.e., $x \rightarrowtail y$) for $1 < i < n$, then the sequence is ***increasing***, otherwise, ***non-increasing***.

In Figure 1(a), we have $tx_1 \rightsquigarrow tx_2$ (due to $e_1 \rightarrowtail e_3$) and $tx_2 \rightsquigarrow tx_3$ (due to $e_2 \rightarrowtail e_4$), $tx_1 \rightsquigarrow tx_3$ is non-increasing. In Figure 1(b), we have $tx_2 \rightsquigarrow tx_3$ (due to $e_2 \rightarrowtail e_4$) and $tx_3 \rightsquigarrow tx_1$ (due to $e_5 \rightarrowtail e_6$ and $e_6 \rightarrowtail e_7$), $tx_2 \rightsquigarrow tx_1$ is increasing.

If neither $e_i \rightarrowtail e_j$ nor $e_j \rightarrowtail e_i$, then $e_i$ and $e_j$ commute [14][16]. If a trace α' is obtained by repeatedly swapping adjacent commuting events of α, then α' and α are equivalent in behavior and called *equivalent* [15][16].

A trace is *serializable* if it is equivalent to some serial trace; otherwise, the trace is *non-serializable*.

A transaction *tx* is *serializable* in a trace α if *tx* can be executed serially in some equivalent trace α' of α [15][16], otherwise *non-serializable*. A transactional atomicity violation [5][16] occurring on the transaction *tx* means that the transaction *tx* cannot be reordered to become a serial execution in some equivalent trace α' of α.

## 2.3 Vector Clock Representation and Tracking Partial Order Relations

Vector clocks (VCs) [25][56] are a test-of-time approach for tracking happens-before relations [21][25]. It is a partial order relation. A VC records a timestamp for each thread in a trace. A typical VC-based technique maintains a vector clock $V(t)$ for each thread $t$. Each index position $V(t)[t]$ keeps the current timestamp of thread $t$. For instance, $V(t)[u]$ keeps the timestamp for the last event of thread $u$ that happens before the current event of $t$. The significance of VCs is that it can precisely capture the transitive closure of happens-before relations in a concurrent system. It has been widely and successfully used in numerous static and dynamic concurrency analysis techniques [9][17][57][58].

VCs have two kinds of operations: increment operation and join operation. An increment operation (denoted by $\boldsymbol{inc(V(t))}$) increments the timestamp kept at position $V(t)[t]$ by 1. A join operation (denoted by ⊔) is defined as $V_1 \sqcup V_2 = \lambda t.\ max(V_1[t], V_2[t])$. We denote the vector clock of event $e$ as $\boldsymbol{V(e)}$.

The basic model [25][56] of VC is to have one logical clock for each concurrent entity (thread in our case), and a vector clock is a vector of these clocks. In tracking events, each thread triggering an event will increase its clock in its own VC by one. An example of such an event in RegionTrack is a *begin(t, l)* event. When a transaction $tx_i$ of a thread $t$ generates, say, a write event, which is later read by a transaction $tx_j$ of another thread $u$, *t's* VC at the moment of issuing the read event will be available to $u$. Hence, $u$ performs a join operation between *t's* VC and its own VC to capture *the* last timestamps of all threads visible to these two transactions. Thus, the transactions that happens-before $tx_i$ become visible to $tx_j$. If the timestamps of any transactions that happens-before $tx_i$ are not in partial order with the corresponding timestamps in the vector clock of $tx_j$, a causal violation is detected by the VC approach.

When using VCs to capture the happens-before relations along a trace α, traditional representation techniques [21][25] increment $V(t)$ whenever thread $t$ executes an event $e$. Then, if $e$ conflicts with a past event $e'$ (before the appearance of $e$ in α) of another thread, a join operation between their VCs is performed to track the cross-thread happens-before relation $e' \rightarrowtail e$. After that, $e$ is associated with a shadow state (i.e., a copy) of $V(t)$ to indicate the vector clock of $e$. This guarantees that for any two events $e$ and $e'$ in α, $e' \rightarrowtail e$ if and only if $V(e')[t] \leq V(e)[t]$ for each thread $t$ (denoted as $V(e') \sqsubseteq V(e)$) [21][25].

## 2.4 Transactional Happens-Before Graph and Cycle

A transactional happens-before graph $G$ intends to capture the happens-before relations between transactions along a trace α. Each node in $G$ represents a transaction. Suppose that there are two transactions $tx_i$ and $tx_j$ in α. If there are two events $e_i \in tx_i$ and $e_j \in tx_j$ such that $e_i \rightarrowtail e_j$, then a THB edge $tx_i \rightsquigarrow tx_j$ is added to $G$. To identify non-serializable traces, Velodrome [16] provides a sufficient and necessary condition.

**Theorem 1** [16]. A trace α is non-serializable if and only if there are two transactions, $tx_i$ and $tx_j$, in trace α such that (1) $T(tx_i) \neq T(tx_j)$, (2) $tx_i \rightsquigarrow tx_j$, and (3) $tx_j \rightsquigarrow tx_i$.

Theorem 1 implies that $tx_i$ and $tx_j$ forms a cycle in the THB graph for α. For Velodrome [16] and DoubleChecker [5], identifying non-serializable traces equals to detecting cycles in the THB graph.

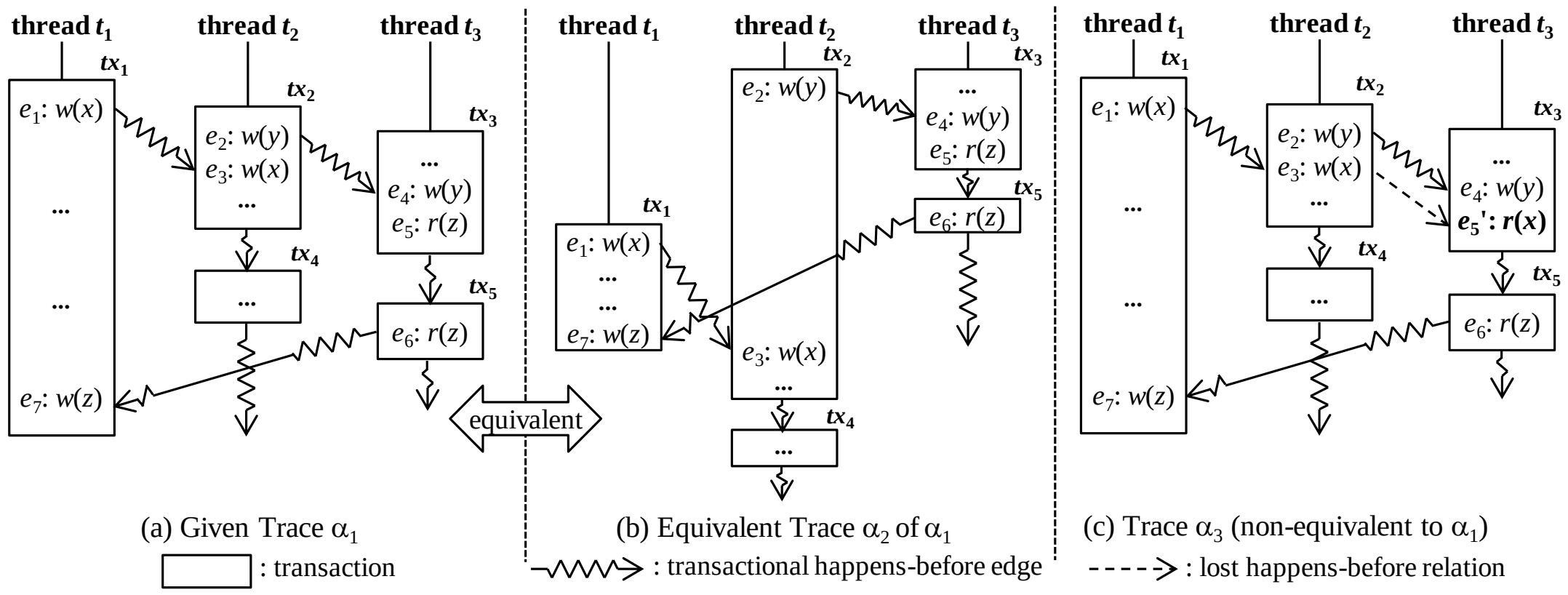

**Figure 1. Three examples of execution trace diagram**

If the transactional happens-before relation $tx_i \rightsquigarrow tx_i$ that inferred from the cyclic sequence $\langle tx_i \rightsquigarrow tx_j, tx_j \rightsquigarrow tx_i \rangle$ is increasing, then the **cyclic sequence** (or the **cycle** in terms of THB graph) is increasing, otherwise, non-increasing. An increasing cyclic sequence implies a non-serializable trace ***and*** a transactional atomicity violation, whereas, a non-increasing cyclic sequence implies a non-serializable trace only [16].

## 3 MOTIVATING EXAMPLE

Figure 1(a) shows a trace $\alpha_1 = \langle e_1, e_2, e_3, e_4, e_5, e_6, e_7 \rangle$ having seven events ($e_1$ to $e_7$), three threads ($t_1$ to $t_3$) and five transactions ($tx_1$ to $tx_5$). Thread $t_1$, $t_2$ and $t_3$ executes $tx_1$ only, $tx_2$ and $tx_4$, and $tx_3$ and $tx_5$, respectively. We denote a write event and a read event on a variable $x$ by $w(x)$ and $r(x)$, respectively. In $\alpha_1$, $e_1$, $e_2$ and $e_6$ conflict with $e_3$, $e_4$, and $e_7$, respectively, resulting in the following HB relations: $e_1 \rightarrowtail e_3$, $e_2 \rightarrowtail e_4$, and $e_6 \rightarrowtail e_7$.

We first present how Velodrome [16] and DoubleChecker [5] handle $\alpha_1$. Each technique builds a transactional HB graph (denoted as $G_1$). On processing $e_1$ and $e_2$, no edge is added to $G_1$ because $e_1$ and $e_2$ do not conflict with each other. On processing $e_3$, $e_1 \rightarrowtail e_3$ is formed. The edge $tx_1 \rightsquigarrow tx_2$ is added to $G_1$. Similarly, on processing $e_4$, $e_2 \rightarrowtail e_4$ is formed, and the edge $tx_2 \rightsquigarrow tx_3$ is added to $G_1$. There is no edge added to $G_1$ when processing $e_5$. Next, $t_2$ ends $tx_2$ and starts $tx_4$, an (intra-thread) edge $tx_2 \rightsquigarrow tx_4$ is added to $G_1$ for maintaining the program order. Similarly, the edge $tx_3 \rightsquigarrow tx_5$ is added to $G_1$. On processing $e_6$, no edge is added. On processing $e_7$, $e_6 \rightarrowtail e_7$ is formed, and the edge $tx_5 \rightsquigarrow tx_1$ is added to $G_1$. As such, a ***cycle*** (denoted as $\boldsymbol{c_1}$) is formed in $G_1$: $tx_1 \rightsquigarrow tx_2 \rightsquigarrow tx_3 \rightsquigarrow tx_5 \rightsquigarrow tx_1$.

Note that prior to the execution of $e_7$, the trace is still serializable.

**DoubleChecker**: When $e_7$ is executed, DoubleChecker detects $c_1$. It reports $\alpha_1$ as non-serializable. It also blames $tx_1$ as a transactional atomicity violation according to the convention that the transaction $tx_1$ containing $e_7$ completes the cycle $c_1$.

However, although $\alpha_1$ is a non-serializable trace, every transaction in $\alpha_1$, including $tx_1$, is serializable. The trace $\alpha_2$ shown in Figure 1(b) is equivalent to $\alpha_1$ by the following reasoning: In trace $\alpha_1$, event $e_1$ conflicts with neither $e_2$ nor events $e_4$ to $e_6$, and event $e_3$ also does not conflict with events $e_4$ to $e_7$. Thus, $e_1$ commutes with $e_2$ and $e_4$–$e_6$, and $e_3$ commutes with events $e_4$ to $e_7$. As a result, $\alpha_2$ is an equivalent trace of $\alpha_1$. Since $tx_1$ can be executed serially in $\alpha_2$ (i.e., an equivalent trace of $\alpha_1$), $tx_1$ in $\alpha_1$ is serializable. Thus, DoubleChecker reports a **false positive** in transactional atomicity violation, although it correctly reports $\alpha_1$ as non-serializable.

**Velodrome**: To avoid reporting false positives, Velodrome records the timestamps (i.e., event orders) of the events at the head and tail positions of each edge in its transactional HB graph. It only blames [16] (i.e., identifies) a transaction when the cycle is "*increasing*" according to the timestamps. For instance, Velodrome records ($tx_1$, 1) ⇝ ($tx_2$, 3) for edge $tx_1$ ⇝ $tx_2$ in $\alpha_1$, where integers 1 and 3 are the timestamps of events $e_1$ and $e_3$, respectively. Similarly, edges ($tx_2$, 2) ⇝ ($tx_3$, 4), ($tx_3$, 5) ⇝ ($tx_5$, 6), and ($tx_5$, 6) ⇝ ($tx_1$, 7) are recorded. Consider the subsequence $tx_1$ ⇝ $tx_2$ ⇝ $tx_3$ in the cycle $c_1$. They are generated by the edges ($tx_1$, 1) ⇝ (**$tx_2$, 3**) and (**$tx_2$, 2**) ⇝ ($tx_3$, 4). The timestamp on the incoming edge to $tx_2$ (i.e., 3) is greater than the timestamp on the outgoing edge from $tx_2$ (i.e., 2), indicating that the cycle $c_1$ is non-increasing. The transactional HB edges in $c_1$ do not reflect the underlying happens-before relation on events. Therefore, Velodrome reports a non-serializable trace on $\alpha_1$ but does not identify any transaction, say $tx_1$, to be responsible for $c_1$. This situation makes the debugging difficult because Velodrome does not provide precise information about which transaction developers should examine further.

Velodrome keeps at most one edge from each pair of transaction nodes in a transactional HB graph to manage its runtime overhead. In this way, it avoids traversing the same set of nodes multiple times to locate cycles. Consider another trace $\alpha_3$ (which is not equivalent to $\alpha_1$) in Figure 1(c). When executing event $e_5'$, it forms the HB edge $e_3 \rightarrowtail e_5'$, and $tx_1$ in $\alpha_3$ is not serializable. Observe that edges $e_2 \rightarrowtail e_4$ and $tx_2$ ⇝ $tx_3$ have already been recorded. As such, Velodrome does not update the timestamps of $tx_2$ ⇝ $tx_3$. Thus, the detected cycle is non-increasing. So, Velodrome does not report any transactional atomicity violation on $tx_1$. That is, without keeping the relation $e_3 \rightarrowtail e_5'$, Velodrome **misses reporting** $tx_1$ as non-serializable in $\alpha_3$.

In the algorithmic designs of Velodrome and DoubleChecker, a transaction node corresponding to a finished transaction should still be kept in the transactional HB graph for subsequent cycle localization. The node cannot be deleted until the node is no longer reachable from any other node corresponding to a currently running transaction of any thread.

To perform cycle localization, both Velodrome and DoubleChecker have to traverse their transactional HB graphs. They search for paths along the edges to look for cycles. In $\alpha_1$, when executing $e_7$, the cross-thread transactional HB edge $tx_5$ ⇝ $tx_1$ is added, and a path traversal is performed, which attempts to reach $tx_5$ from $tx_1$. In the course of traversal, if these techniques step into $tx_4$ (i.e., $tx_1$ ⇝ $tx_2$ ⇝ $tx_4$), then no cycle can be detected along this path, which unavoidably spends time on this unsuccessful attempt. In general, there are far more acyclic paths than cyclic paths in a graph. Also, an edge in the graph may be searched multiple times for multiple invocations of cycle localizations. To the best of our knowledge, many existing atomicity checkers normally [13][33][35] use the above graph search strategy for cycle localization for their focal kinds of concurrency bugs.

We next briefly review how AeroDrome handles $\alpha_3$. Section 4.4 discusses AeroDrome in greater detail.

**AeroDrome**: When executing $e_3$, the THB relation $tx_1$ ⇝ $tx_2$ (due to $e_1$ ⇝ $e_3$) is transitive to the vector clock of thread $t_2$. Similarly, on processing $e_4$, the THB relation $tx_2$ ⇝ $tx_3$ (due to $e_2$ ⇝ $e_4$) is transitive to the vector clock of thread $t_3$. On processing $e_5$, the inferred THB relation $tx_1$ ⇝ $tx_3$ is transitive to the vector clock of thread $t_3$. Thus, on processing $e_7$, AeroDrome finds that $V(tx_1.begin) \sqsubseteq V(e_6)$ holds, and reports $\alpha_1$ as a non-serializable trace. However, AeroDrome does not include any procedure to detect transaction atomicity violations. Hence, AeroDrome is unable to report any transactional atomicity violations on $tx_1$.

A typical trace contains numerous transactions. Only a limited of them are erroneous. In typical software development, the test-diagnose-fix cycle is often used [3][55]. One purpose of identifying erroneous transactions in a trace is to support the diagnosis step in such a cycle. The above three techniques either miss reporting true positives (if ever reporting) or report false negatives on transactional atomicity violations. They are imprecise in localizing transactional atomicity violations, compromising their potentials in program debugging.

In the next section, we present RegionTrack to address the limitations illustrated in this motivating example.

## 4 RegionTrack

In this section, we present RegionTrack.

## 4.1 Overview

RegionTrack precisely detects transactional atomicity violations (i.e., increasing cyclic sequence of transactional happens-before relations) without explicitly searching for cyclic sequences. It also precisely determines whether the trace is non-serializable (i.e., both increasing and non-increasing cyclic sequences of transactional happens-before relations). It leverages the forward propagation mechanism of traditional vector clocks to locate transactional atomicity violations. It also designs a new mechanism of transactional vector clock to propagate timestamps of transactions to precisely capture transactional happens-before relations for identifying non-increasing cyclic sequences of transactional happens-before relations. RegionTrack is the first work that is both sound and complete in detecting transactional atomicity violations and identifying non-serializable traces. It is also very efficient.

RegionTrack consists of Algorithms 1, 3, and 4. For ease of presentation, we firstly present Algorithms 1 and 2, which suffice to detect precisely transactional atomicity violations. We then present how to extend Algorithm 2 to become Algorithms 3 and 4 to include the detection of non-serializable traces.

For ease of readers' reference, the following notations are used in our algorithms:

- $C(t)$: the current transaction node for thread $t$.
- $V(t)$: the current vector clock of thread $t$.
- $V(e)$: the vector clock of event $e$, which refers to a shadow state of $V(T(e))$ when executing $e$.
- $W(x)$: the vector clock of the last write event to the variable $x$.
- $R(t, x)$: the vector clock of the last read event of the variable $x$ performed by thread $t$.
- $L(m)$: the vector clock of the last release event of lock $m$.
- $TV(t)$: the transactional vector clock of the thread $t$.
- $Tid$: the set of currently running thread indexes.

Note that, $W(x)$, $R(t, x)$, and $L(m)$ are special notations of $V(e)$ for *last* write events, *last* read events, and *last* release events, respectively.

## 4.2 Our Necessary and Sufficient Condition to Quantify Non-Serializable Transactions

Along a trace, RegionTrack directly locates transactions that experience transactional atomicity violations. This section presents the corresponding necessary and sufficient conditions that RegionTrack is built on.

**Theorem 2**. A transaction $tx$ is not serializable in a trace $\alpha$ if and only if there is an event $e_m \in tx$ and another event $e_x$ in $\alpha$ such that (1) $T(e_x) \neq T(e_m)$, (2) $e_x \rightarrowtail e_m$, and (3) $tx.begin \rightarrowtail e_x$.

**Proof**. ***Sufficiency***: Given the condition $e_m \in tx$, we must have $e_m \rightarrowtail tx.end$. Because $tx.begin \rightarrowtail e_x$ and $e_x \rightarrowtail e_m$, we should have $tx.begin \rightarrowtail e_x \rightarrowtail tx.end$. That is, event $e_x$ cannot be commuted outside $tx$. As we also know, $T(e_x) \neq T(e_m)$ and $T(e_m) = T(tx)$, therefore, $tx$ must be interleaved by the conflicting event $e_x$ executed by some other thread. By definition, $tx$ is not serializable.

***Necessity***: Suppose that a transaction $tx$ is not serializable in trace $\alpha$. In this case, by definition, $tx$ conflicts with at least one event, say $e_x$, belonging to a thread different from $T(tx)$, and this event $e_x$ cannot be commuted outside $tx$. Therefore, we must have $tx.begin \rightarrowtail e_x \rightarrowtail tx.end$ and $T(e_x) \neq T(tx)$. Because $tx.end$ cannot conflict with any event from any other thread, we can only obtain $e_x \rightarrowtail tx.end$ transitively via some intermediate event in between these two events in trace $\alpha$. As a result, $tx$ must contain one event, say $e_m$, where $e_x \rightarrowtail e_m$, and $T(e_x) \neq T(e_m)$. So, we have proven Theorem 2. (Note that $tx.begin \rightarrowtail e_x$ is also obtained transitively, but we do not need it

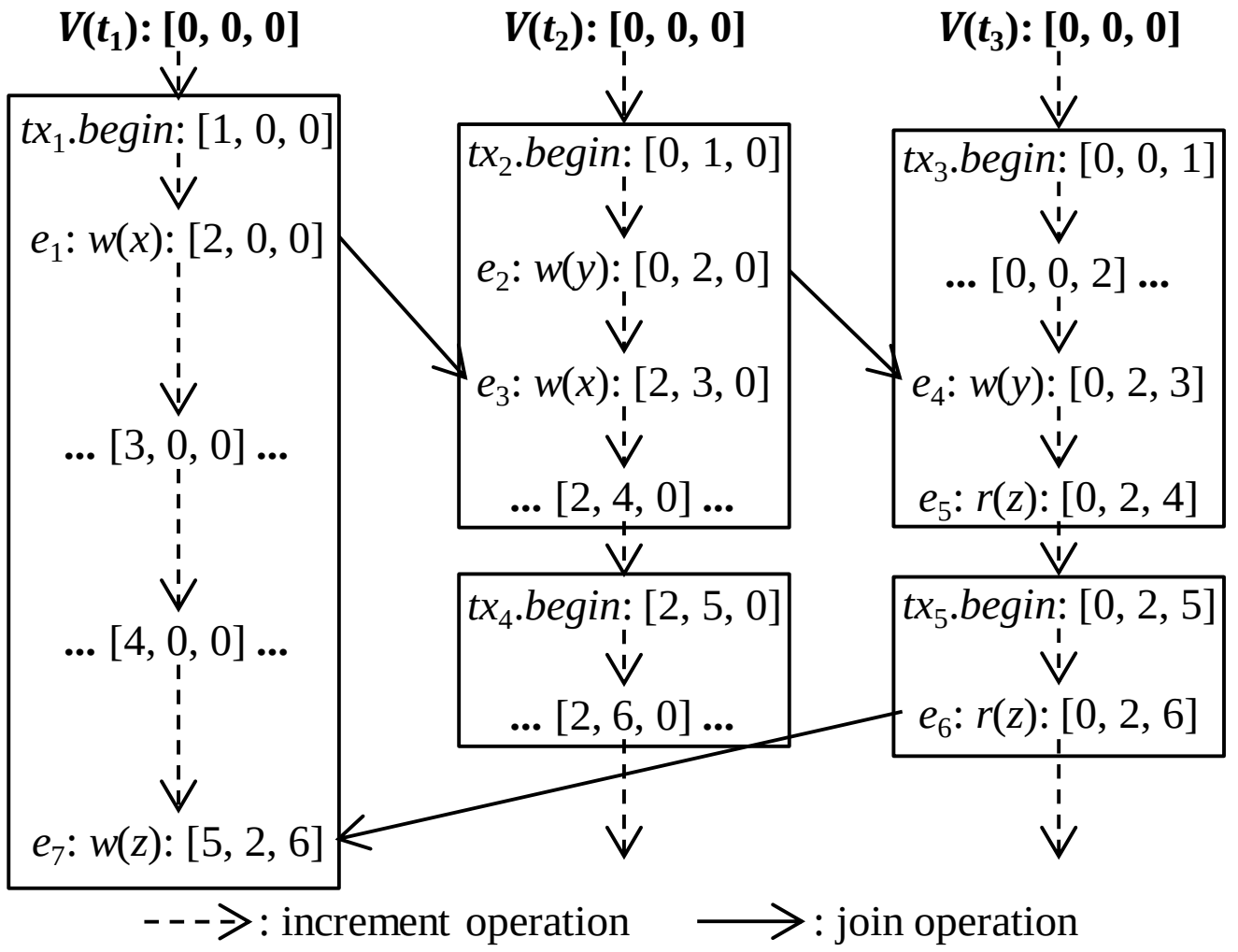

**Figure 2. Illustration of using traditional representation technique of vector clocks along the trace $\alpha_1$**

in proving Theorem 2.) □

Based on Theorem 2, we need not keep the full set of happens-before relations along a trace explicitly through THB graphs to detect transactional atomicity violations.

Theorem 2 shows that at the moment the happens-before relation $e_x \rightarrowtail e_m$ appears in the trace, where $e_m \in tx$ and $T(e_x) \neq T(e_m)$, if we can determine whether $tx.begin \rightarrowtail e_x$ holds, we have precisely determined whether $tx$ experiences a transactional atomicity violation.

Figure 2 illustrates how to use VCs to capture the happens-before relations along $\alpha_1$ and avoid the false positive case incurred by DoubleChecker shown in Figure 1(a). The technique of VCs initializes the timestamps of threads $t_1$ to $t_3$ as 0. On processing $tx_1.begin$, it increments $V(t_1)$, and copies $V(t_1)$ to $V(tx_1.begin)$. Events $e_1$, $tx_2.begin$, and $e_2$ are handled in the same way. On processing $e_3$, after $V(t_2)$ has been incremented, the result of $V(e_1) \sqcup V(t_2)$ for capturing the happens-before relation $e_1 \rightarrowtail e_3$ is assigned to $V(t_2)$. It copies $V(e_3)$ from $V(t_2)$. So, we have $V(e_1) \sqsubseteq V(e_3)$ for $e_1 \rightarrowtail e_3$. When executing $e_7$ and capturing $e_6 \rightarrowtail e_7$, because $V(tx_1.begin) \not\sqsubseteq V(e_6)$, $tx_1.begin$ does not happen before $e_6$. According to Theorem 2, such a technique does not report any violation in $\alpha_1$.

We note that tracking all the happens-before relations at the event level for detecting transactional atomicity violations was deemed "infeasible" [16][53]. It is particularly an issue for an execution trace containing large numbers of events and variables. As illustrated in Figure 2, when processing trace $\alpha_1$, such a technique has to perform the timestamp increment operations 19 times and the clock copy operations 17 times.

The happens-before relations captured by traditional VCs ensure that the relation $e' \rightarrowtail e$ holds if and only if $V(e') \sqsubseteq V(e)$ holds. However, this condition is *unnecessarily* stronger than the actual need for detecting atomicity violations. For example, according to our newly formulated Theorem 2, an atomicity checker only needs to determine whether $e' \rightarrowtail e$ holds whenever $e'$ is the beginning event of a transaction.

## 4.2 Detecting Transactional Atomicity Violations

RegionTrack divides each transaction into a sequence of *dynamic subregions* (or *subregions* for short) and makes

**Algorithm 1. Handle Events**

```
procedure begin(t, l)
    V(t) = inc(V(t)) //increment the timestamp of V(t)
    C(t) = new TransactionNode(l)
    C(t).currVC = new VC(V(t)) //create a shadow state of V(t)
    V(C(t).begin) = C(t).currVC //point to the shadow state
end procedure

procedure end(t, l)
    C(t) = null
end procedure

procedure r(t, x)
    if t ≠ T(W(x)) and R(t, x) = null then
        join(W(x), C(t)) //(last write) w(x) ↣ (current read) r(t, x)
        subRegion(C(t))
    end if
    R(t, x) = C(t).currVC //point to the shadow state of V(t)
end procedure

procedure w(t, x)
    if exist lastRead of x then
        for each t' ∈ Tid and t' ≠ t do
            join(R(t', x), C(t)) //(last read) r(t', x) ↣ (current write) w(x)
        end for
    else if t ≠ T(W(x)) then
        join(W(x), C(t)) //(last write) w(x) ↣ (current write) w(x)
    end if
    subRegion(C(t))
    W(x) = C(t).currVC
    for each t' ∈ Tid do //clear last reads
        R(t', x) = null
    end for
end procedure

procedure acq(t, m)
    if t ≠ T(L(m)) then
        join(L(m), C(t)) //(last) rel(m) ↣ (current) acq(m)
        subRegion(C(t))
    end if
end procedure

procedure rel(t, m)
    L(m) = C(t).currVC
end procedure

function subRegion(C(t)) //create a new sub-region if needed
    if C(t).currVC ≠ V(t) then
        C(t).currVC = new VC(V(t)) //create a shadow state of V(t)
    end if
end function
```

all the events in the same subregion sharing the same instance of shadow VC. RegionTrack starts a new subregion for a transaction *tx* whenever a join operation is performed to update the current VC of thread *T*(*tx*).

**Algorithm 2. Transactional Atomicity Violation Detection**

```
45  function join(V(e_x), C(t)) //e_x ↣ the current event e_m of C(t)
46  │ V(t) = V(e_x) ⊔ V(t) //a join operation to propagate V(e_x) to V(t)
47  │ checkHB(C(t), V(e_x)) //check whether or not C(t).begin ↣ e_x
48  end function

49  function checkHB(C(t), V(e_x)) //This operation takes O(1)-time
50  │ if V(C(t).begin)[t] ≤ V(e_x)[t] then //only check at index t
51  │ │ report a transactional atomicity violation on C(t)
52  │ end if
53  end function
```

Algorithm 1 shows how RegionTrack handles each event along an execution trace. Algorithm 2 presents the functions used by Algorithm 1 to detect transactional atomicity violations.

For each *begin*($t$, $l$) event, RegionTrack first increments $V(t)$ (line 2). It creates a new node for the current transaction *tx*, and assigns it to $C(t)$ (line 3). The VC of *tx* is recorded in *currVC* of $C(t)$ (line 4), which is a shadow state of $V(t)$. It refers to the VC of *tx*'s beginning event $V(C(t).begin)$ as *currVC* (line 5). For each *end*($t$, $l$) event, RegionTrack clears the reference to the transaction node $C(t)$ (lines 7-9).

For each read event $r(t, x)$, RegionTrack tracks the cross-thread happens-before relation from the last-write access to the current read on the same variable $x$, providing that no such relation has been captured before (lines 11-12). The relation is captured by the join operation of the corresponding VCs (line 12).

Suppose that $e_m$ is the current event (being executed). Whenever a cross-thread happens-before relation, say $e_x \rightarrowtail e_m$, is captured, RegionTrack checks whether the beginning event of the current transaction happens before $e_x$ (line 47). This checking is based on Lemma 1 in Section 4.5. If this is the case, a *transactional atomicity violation* is reported on the current transaction (line 51). Note that the checking only takes $O(1)$-time (line 50). This is because it only checks whether the timestamp of thread $t$ at the moment of performing $C(t).begin$ is not greater than the corresponding timestamp at the moment of performing $e_x$.

After the join operation, RegionTrack checks whether any new dynamic subregion of the current transaction is needed based on the current $V(t)$ (lines 13 and 41). Specifically, if the join operation updates $V(t)$, then a new dynamic subregion is created to record a new shadow state of $V(t)$ (line 42). RegionTrack lets the underlying garbage collector (GC) collect the shadow states of $V(t)$ if all events no longer reference them. At the end of handling $r(t, x)$, RegionTrack updates the VC of the last-read access $R(t, x)$ by referring it to the VC of the current dynamic subregion $C(t).currVC$ (line 15).

For each write event $w(t, x)$, RegionTrack tracks the cross-thread happens-before relation from the last-read access of each thread on the same variable $x$ to the current write (lines 19-21). It also checks whether $C(t).begin$ happens before each last-read access (lines 49-53). If there is no such last-read access, then RegionTrack tracks the cross-thread happens-before relation from the last-write access on the same variable $x$ to the current write (lines 22-24). If the join operation updates $V(t)$, a new subregion is created (line 25). Then, RegionTrack updates the VC of the last-write access $W(x)$ by referring it to $C(t).currVC$ (line 26). Because all previous last-read accesses on $x$ have happened before the current write, RegionTrack clears all $R(t, x)$ (lines 27-29).

On processing a current read access $r(t, x)$, RegionTrack checks whether there is last-read access, say $e_r$, on $x$ performed by $t$ (i.e., $R(t, x) \neq null$). If this is the case, it knows that the happens-before relation from the last-write access, say $e_w$, on $x$ to the current-read access has already been captured by $e_w \rightarrowtail e_r$ when handling $e_r$.

So, RegionTrack needs not check $r(t, x)$. Similarly, on processing $w(t, x)$, RegionTrack needs not to check the last-write access on $x$ if there exists any last-read access on $x$. This is because if a last-read access happens before the last-write access, then the last-read access should have been cleared when handling the last-write access.

For an event outside any atomic region, RegionTrack adopts the scheme taken by the prior work [16]: create a unary transaction for it, handle the event, and then exit the transaction. The scheme also merges consecutive unary transactions into the same transaction [16]. Because no violation can occur on any unary transactions [5], RegionTrack does not maintain any VC for the beginning event of any unary transaction.

***Example***. Figure 3(a) illustrates how RegionTrack handles the trace $\alpha_1$ in Figure 1(a). On processing $tx_1.begin$, RegionTrack increments $V(t_1)$, creates a shadow state of $V(t_1)$ which is assigned to $tx_1.currVC$, and lets $V(tx_1.begin)$ refer to the current $tx_1.currVC$. Then, on processing $e_1$, $V(t_1)$ needs not to perform any operation, and $V(e_1)$ shares the same VC with $V(tx_1.begin)$ and $tx_1.currVC$. Next, $t_2$ executes $tx_2.begin$, $e_2$, and $e_3$. One happens-before relation $e_1 \rightarrowtail e_3$ is captured through the join operation between $V(t_2)$ and $V(e_1)$ when executing $e_3$, and $V(t_2)$ is updated by this join operation. So, a new shadow state of $V(t_2)$ is created and assigned to $tx_2.currVC$. $V(e_3)$ refers to the current clock $tx_2.currVC$, and $V(e_2)$ remains the same. Therefore, when capturing $e_2 \rightarrowtail e_4$, the happens-before relations indicated by VCs are precise, and the timestamp of $t_1$ (i.e., $V(t_1)[t_1]$) does not propagate to $t_3$. The execution continues, and when capturing $e_6 \rightarrowtail e_7$, no violation is reported on $tx_1$ because the condition $V(tx_1.begin)[t_1] \leq V(e_6)[t_1]$ is not satisfied. RegionTrack increments $V(t)$ for thread $t$ only if $t$ is processing the beginning event of a transaction, which enables the sharing of shadow VCs among events in the same subregion. To handle $\alpha_1$, RegionTrack performs 7 timestamp increment operations and 8 clock copy operations.

On analyzing $\alpha_3$ in Figure 1(c), RegionTrack reports a violation on $tx_1$. Figure 3(b) shows the VCs of thread $t_3$ in $\alpha_3$ (the VCs of $t_1$ and $t_2$ in $\alpha_3$ are the same as those in $\alpha_1$). We can see that when capturing $e_6 \rightarrowtail e_7$ because $V(tx_1.begin)[t_1] \leq V(e_6)[t_1]$ holds, indicating $tx_1.begin \rightarrowtail e_6$. A violation on $tx_1$ is reported.

RegionTrack is built on top of the insight that one needs not to know the cycle (if any) explicitly; instead, we only need to know whether there is any transaction being violated. Moreover, other atomicity checkers, such as the three techniques (i.e., Velodrome, DoubleChecker, and Aerodrome) that reviewed in Section 3, uphold each transaction (expected to be atomic) to be non-divisible. RegionTrack introduces the notion of dynamic subregions of transactions. This notion, in essence, divides transactions in finer granularity.

Once a transaction or a subregion has completed its execution (*aka* finished), RegionTrack needs not to retain the data for this transaction or subregion in its analysis state. We also show by theorems that this strategy does not compromise the soundness and completeness in the detection of transactional atomicity violations.

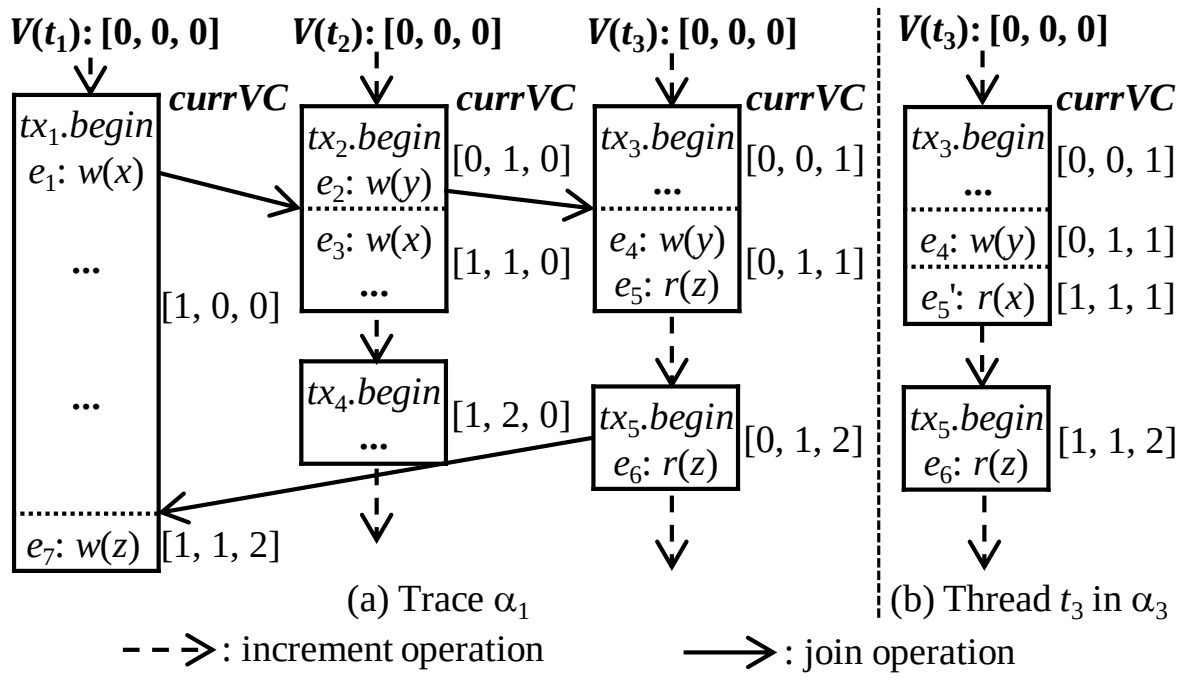

**Figure 3. Illustration of RegionTrack along the trace $\alpha_1$ and $\alpha_3$**

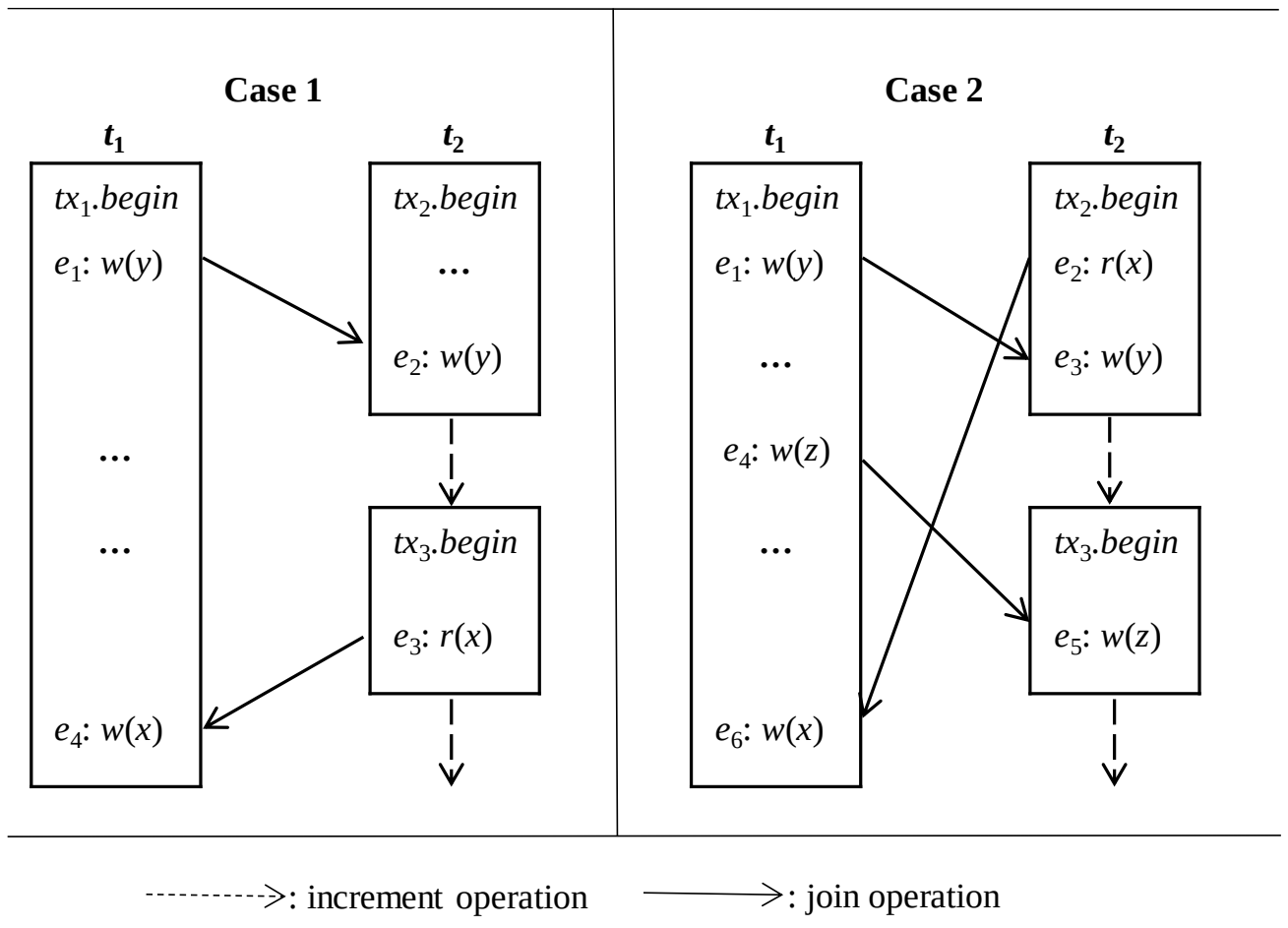

**Figure 4. Two cases of non-serializable traces.**

RegionTrack has a unique design to perform join operations of VCs to maintain cross-thread happens-before relations to ensure that if $e' \rightarrowtail e$, then $V(e') \sqsubseteq V(e)$, which in turn also ensures that if $tx.begin \rightarrowtail e$, then $V(tx.begin) \sqsubseteq V(e)$. Its design on increment operations of VCs (also at the transaction level) can efficiently ensure another condition: If $V(tx.begin) \sqsubseteq V(e)$, then $tx.begin \rightarrowtail e$. In this way, the condition stated in Theorem 1 is fully respected.

## 4.3 Identifying Non-Serializable Traces

### *4.3.1 Motivation*

Each transaction in $\alpha_1$ in Figure 1(a) is serializable in some equivalent trace of $\alpha_1$ (note that different transactions in $\alpha_1$ may be serializable in different equivalent traces of $\alpha_1$). Yet, the whole trace is still non-serializable.

Figure 4 shows two cases of non-serializable traces. The trace is $\langle e_1, e_2, e_3, e_4\rangle$ in Case 1 refers to an increasing cyclic sequence because, along the constituted cyclic sequence, the timestamps of the involved events in the sequence are increasing. Based on the forward propagation property of vector clocks, when Algorithms 1 and 2 detect a transactional atomicity violation, it also indicates a non-serializable trace of Case 1.

In Case 2 of Figure 4, the trace is $\langle e_1, e_2, e_3, e_4, e_5, e_6\rangle$. It refers to as a non-increasing cyclic sequence. Algorithms 1 and 2 do not identify such a non-increasing cyclic sequence because they record happens-before relations at the event level and the subregion level only. Thus, traditional vector clocks may miss recording happens-before relations at the transaction level, which is at a coarser granularity than the two levels above.

Specifically, in Figure 1(a), when executing $e_3$, $V(t_2) = V(e_1) \sqcup V(t_2) = [1, 1, 0]$. Each VC increment operation only increases the timestamp of the current thread by 1. So, one can infer $Trans(e_1) \rightsquigarrow Trans(e_3)$, meaning $tx_1 \rightsquigarrow tx_2$. Similarly, on executing $e_4$, $V(t_3) = V(e_2) \sqcup V(t_3) = [0, 1, 1]$. One can infer $Trans(e_2) \rightsquigarrow Trans(e_4)$, (i.e., $tx_2 \rightsquigarrow tx_3$). Thus, we should have $tx_1 \rightsquigarrow tx_3$ according to the transitivity of THB relation. Nonetheless, we cannot get any information about thread $t_1$ from the vector clock $V(t_3)$. This is because $t_2$ can see the timestamp of $e_1$ of $t_1$ only when $e_3$ executes, which leads to a join operation with $e_2$. However, when $e_4$ executes, $t_3$ can only see the timestamp of $e_2$ (but not that of $e_3$) through a join operation. Events of $t_1$ do not form any HB relations with events of thread $t_3$. Thus, the timestamp of $t_1$ is never propagated to $t_3$. Consequently, the VCs maintained by Algorithms 1 and 2 do not capture such a non-increasing sequence of THB relations. It makes the use of Algorithms 1 and 2 alone to miss detecting non-serializable traces for non-increasing cyclic sequences.

To address this issue, RegionTrack assigns another kind of vector clock, referred to as ***transactional vector clock***, for each *thread* (**not** transaction) to track transactional happens-before relations along a trace.

Consider Case 2 of Figure 4, $tx_1 \rightsquigarrow tx_2$ is captured because of the presence of $e_1 \rightarrowtail e_3$. For ease of our subsequence presentation, transaction $tx_1$ is called ***source* transaction**, and thread $t_1$ is called *source* thread. Similarly, transaction $tx_2$ is called ***sink* transaction**, and thread $t_2$ is called *sink* thread. Moreover, transaction $tx_1$ has an *outgoing* transactional happens-before relation to transaction $tx_2$ while transaction $tx_2$ has an *incoming* transactional happens-before relation from transaction $tx_1$.

Our insight is that an outgoing THB relation of a transaction $tx$ can be added any time along the trace, but an incoming THB relation of $tx$ can only be added when $tx$ is running. (Note that if a cyclic sequence exists, the transaction that starts and ends the sequence must be a currently running transaction of the thread.)

In Case 2 of Figure 4, once $tx_2$ has finished, $tx_2$ cannot have new incoming edges anymore. However, $tx_2$ may be involved in new outgoing THB relations at any later moment. If $tx_2$ is not a currently running transaction, $tx_2$ is impossible to be the last transaction in a cyclic sequence after it has finished.

Based on these observations, we design the *transactional vector clock* for each thread to capture the transactional happens-before relations in a sound and complete manner.

#### *4.3.2 Transactional vector clock*

Transactional vector clock (TVC) of size $n$ is an array of $n$ timestamps. For transactional vector clock $TV(t)$ of thread $t$, RegionTrack lets $TV(t)[t]$ record the *latest* source transaction timestamp of thread $t$ and $TV(t)[u]$ record the *first* sink transaction timestamp of THB relation of any other thread $u$. Consider Case 2 of Figure 4, when executing $e_3$, $e_1 \rightarrowtail e_3$ is formed, implying $tx_1 \rightsquigarrow tx_2$. Therefore, transaction $tx_1$ is the latest source transaction of thread $t_1$, and $TV(t_1)[t_1]$ records the timestamp of transaction $tx_1$. Transaction $tx_2$ is the first sink transaction of thread $t_2$, and $TV(t_1)[t_2]$ records the timestamp of transaction $tx_2$. On executing $e_5$, $e_4 \rightarrowtail e_5$ is formed, implying $tx_1 \rightsquigarrow tx_3$. However, $TV(t_1)[t_2]$ will not be updated to the timestamp of transaction $tx_3$. It is because one can infer $tx_1 \rightsquigarrow tx_3$ based on the program order $tx_2 \rightsquigarrow tx_3$. Hence, $TV(t_1)[t_2]$ still stores the timestamp of the first sink transaction of thread $t_2$, which is $tx_2$. Thus, by checking the TVC of a thread $t$, RegionTrack is aware of the frontier THB relations of the latest source transaction of thread $t$ to other threads in the same trace.

As presented in Section 2.2, a cross-thread THB relation between two transactions can be formed under two conditions. The first condition is met when performing a join operation for the THB relation (e.g., Figure 5 Case 1: $tx_1 \rightsquigarrow tx_2$). The second condition is met by inferring the relation based on a set of other THB relations (e.g., Figure 5 Case 1: $tx_1 \rightsquigarrow tx_3$). For ease of reference, we refer to the latter THB relations in the first condition and the second condition as **direct** and **indirect** THB relations, respectively.

Algorithm 3 deals with the first condition at lines 2-7 and 9-12. During a join operation, it checks whether the current source transaction is the latest source transaction of the source thread. If this is the case, it only stores the timestamp of the first sink transaction of the sink thread.

To propagate the timestamp of a transaction in an indirect THB relation to the transactional vector clock of the latest source transaction, Algorithm 3 performs a mixture of forward- and back-propagations. The purpose of back-propagation for a THB relation $tx_i \rightsquigarrow tx_j$ is to propagate the timestamp of the sink transaction $tx_j$ to each transaction that can see the timestamp of the source transaction $tx_i$. On the other hand, the purpose of forward propagation is to update the timestamp of each transaction that can be seen by the source transaction $tx_i$. As a result, the relation $tx_i \rightsquigarrow tx_j$ is explicitly recorded in the transactional vector clock of $tx_i$, both avoiding repetitive propagation and subsequence inferences to obtain the relation again.

**Algorithm 3. Transactional Vector Clock Operation**

```
function updateTVC(V(e_x), V(t), t)
  source = V(e_x)[T(e_x)]
  sink = V(t)[t]
  if TV(T(e_x))[T(e_x)] = source then //update the TVC of source thread T(e_x)
    if TV(T(e_x))[t] > sink then //only store the first transaction that happens-after source transaction
      TV(T(e_x))[t] = sink
    end if
    forwardPropagate(Tid, T(e_x), t)
  else if TV(T(e_x))[T(e_x)] < source then //only store relations for the latest transaction of source thread
    TV(T(e_x)) = ⊥ //initialize the TVC of source thread T(e_x)
    TV(T(e_x))[T(e_x)] = source
    TV(T(e_x))[t] = sink
    forwardPropagate(Tid, T(e_x), t)
  end if
  backPropagate(Tid, T(e_x), t, source, sink)
end function

function backPropagate(Tid, t1, t2, source, sink)
  Tid1 = Tid \ {t1, t2}
  for each t' ∈ Tid1 do
    if TV(t')[t1] ≤ source then //check whether transaction of t' ⇝ transaction of t1 (source)
      if TV(t')[t2] > sink or TV(t')[t2] = 0 then //only store the first sink transaction
        TV(t')[t2] = sink
      end if
      forwardPropagate(Tid, t', t1)
      Tid1 = backPropagate(Tid1, t', t2, TV(t')[t'], sink) //back propagate to other threads
    end if
  end for
  return Tid1
end function

function forwardPropagate(Tid, t1, t2)
  Tid2 = Tid \ {t1, t2}
  if TV(t1)[t2] ≤ TV(t2)[t2] then //check whether transaction of t1 ⇝ transaction of t2
    for each element t' such that t' ≠ t1 do //forward propagate to source thread
      if TV(t1)[t'] > TV(t2)[t'] or (TV(t1)[t'] = 0 ^ TV(t2)[t'] > 0) then //only store the first sink transaction
        TV(t1)[t'] = TV(t2)[t'] //propagate the transaction timestamp that transaction of t1 cannot see
        if t' ∈ Tid2 then
          Tid2 = forwardPropagate(Tid2, t1, t') //if propagate to a new transaction, update the TVC
        end if
      end if
    end for
  end if
  return Tid2
end function
```

The reason for different strategies for forward- and back-propagations is due to the different creation orders of

**Algorithm 4. Non-serializable Trace and Transaction**

```
function join(V(e_x), C(t))
│ V(t) = V(e_x) ⊔ V(t)
│ updateTVC(V(e_x), V(t), t)
│ checkHB(C(t), V(e_x))
end function

function checkHB(C(t), V(e_x)) //This operation takes O(1)-time
│ if V(C(t).begin)[t] ≤ V(e_x)[t] then  //check for a transactional atomicity violation
│ │ report a violation on C(t)
│ else if TV(t)[t] = V(C(t).begin)[t] ∧ TV(t)[T(e_x)] ≤ V(e_x)[T(e_x)]  then //check for a non-serializable trace
│ │ report a non-serializable trace
│ end if
end function
```

THB relations. For instance, in Case 1 of Figure 5, $tx_2 \rightsquigarrow tx_3$ is constructed before $tx_1 \rightsquigarrow tx_2$. If one only performs a back-propagation, the timestamp of $tx_3$ cannot be propagated to $tx_1$ when $tx_2 \rightsquigarrow tx_3$ is constructed because $tx_1$ can only see the timestamp of $tx_3$ through the forward-propagation when $tx_1 \rightsquigarrow tx_2$ is constructed. On the other hand, in Case 2 of Figure 5 where $tx_1 \rightsquigarrow tx_2$ is constructed before $tx_2 \rightsquigarrow tx_3$. If one only performs a forward-propagation, $tx_1$ cannot see the timestamp of $tx_3$ when $tx_1 \rightsquigarrow tx_2$ is constructed, where the timestamp of $tx_3$ can only be backwardly propagated to $tx_1$ when $tx_2 \rightsquigarrow tx_3$ is constructed.

Thus, after updating the direct transactional happens-before relation, RegionTrack recursively processes forward-propagation (lines 8, 13, 24, and 29-42) and back-propagation (lines 15 and 17-28) to capture the indirect THB relations. Moreover, if the timestamp of a sink transaction can be propagated to another transaction, RegionTrack invokes a round of forward-propagation to update the transaction timestamps that can be seen by that transaction (in Algorithm 3 at line 36).

We use an example to illustrate how the propagation mechanism works. As shown by trace $\alpha_1$ in Figure 6, all TVCs of these three threads are initialized as [0, 0, 0]. On processing $tx_1.begin$, RegionTrack increments $V(t_1)$. Then, on processing $e_1$, no operation is needed. Next, $t_2$ executes $tx_2.begin$, $e_2$, and $e_3$. One HB relation $e_1 \rightarrowtail e_3$ is captured through a join operation. Thus, the (direct) THB relation $tx_1 \rightsquigarrow tx_2$ is captured by updating the TVC of $t_1$ to be [1, 1, 0], where $TV(t_1)[t_1]$ is the timestamp of $tx_1$ and $TV(t_1)[t_2]$ is the timestamp of $tx_2$. On performing a propagation operation, $t_3$ is the only remaining thread to be handled, and $TV(t_3)$ is [0, 0, 0]. Thus, nothing needs to be propagated. Similarly, when capturing $e_2 \rightarrowtail e_4$ and $tx_2 \rightsquigarrow tx_3$, $TV(t_2)$ is updated to be [0, 1, 1]. On invoking *forwardPropagate*, nothing needs to be updated. On invoking *backPropagate*, $t_1$ is the only remaining thread to be handled. Since the condition $TV(t_1)[t_2] \leq TV(t_2)[t_2]$ holds, it indicates the presence of $tx_1 \rightsquigarrow tx_2$. As such, the algorithm backwardly propagates the timestamp of $tx_3$ to $TV(t_1)[t_3]$ as [1, 1, 1], indicating the (indirect) THB relation $tx_1 \rightsquigarrow tx_3$. Thus, through checking the TVC of $t_1$, $tx_1 \rightsquigarrow tx_2$ and $tx_1 \rightsquigarrow tx_3$ have been explicitly recorded.

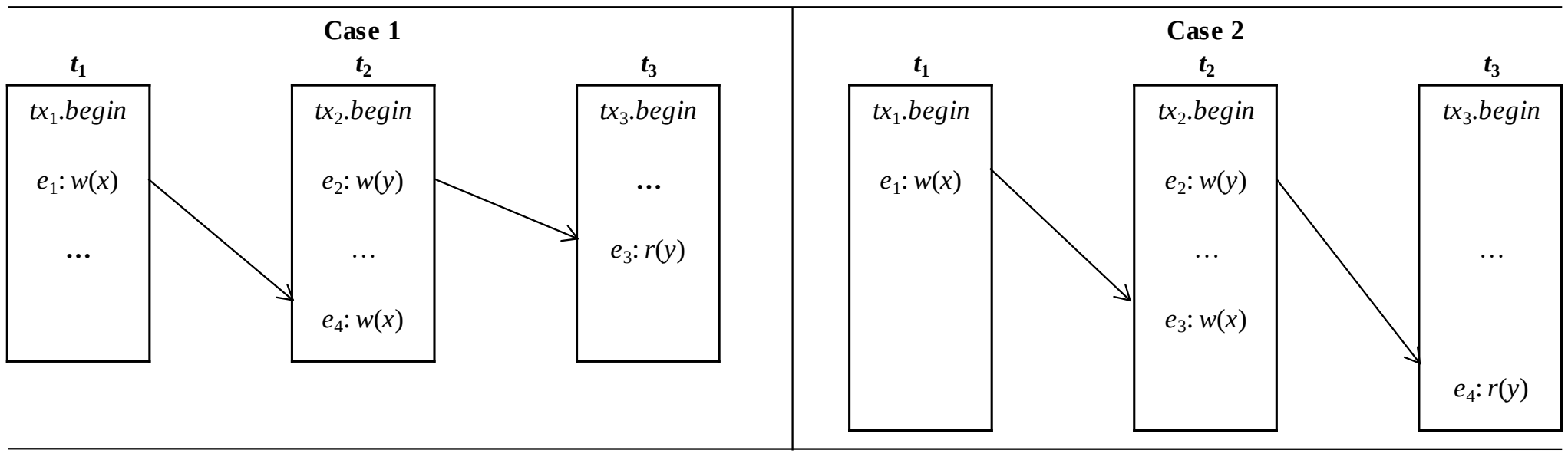

**Figure 5. Two cases of transitive transactional happens-before relations**

Suppose there is another THB relation $tx_4 \rightsquigarrow tx_5$ created before $e_6$ executes. Thus, $TV(t_2)$ is updated to be [0,2,2]. Since $TV(t_1)$ already stores the reversal frontier transactional happens-before relations to $t_2$ and $t_3$, no timestamp will be updated for $TV(t_1)$, eliminating the propagation of this relation.

### *4.3.3 Non-serializable traces*

If Algorithms 1 and 2 report a transactional atomicity violation, the reported transaction must be interleaved by some conflicting events in some transaction(s) performed by some other thread(s). Thus, the trace must be non-serializable. However, if Algorithms 1 and 2 do not report any transactional atomicity violation, the trace may still be non-serializable. In Algorithm 2, RegionTrack checks atomicity violations during the join operation (i.e., when building a happens-before relation). Suppose that RegionTrack is handling the current event $e_m$ in a transaction $tx$. If no transactional atomicity violation is reported on $tx$, $Trans(e_x) \rightsquigarrow tx$ must hold because we must have $e_x \rightarrowtail e_m$ (while $e_x$ is the conflicting event of $e_m$). Thus, to determine whether the trace is serializable, we only need to check whether $tx \rightsquigarrow Trans(e_x)$ holds, according to Theorem 1. Based on the design of the transactional vector clock, RegionTrack can check this condition efficiently.

To enable Algorithm 1 to handle both transactional atomicity violations and non-serializable traces, we replace Algorithm 2 by Algorithms 3 and 4 to form the final RegionTrack technique.

Algorithm 4 works as follows: Whenever a join operation executes, the transactional vector clock will be updated accordingly (line 3). Then, RegionTrack executes the *checkHB* function (line 4) and detects whether there is any transactional atomicity violation (line 7). If there is no such violation reported, Algorithm 4 checks whether the condition $TV(t)[t] = V(C(t).begin)[t] \wedge TV(t)[T(e_x)] \leq V(e_x)[T(e_x)]$ holds (line 9). This condition means that transaction node $C(t)$ has THB relations with transactions of other threads, and the relation $tx \rightsquigarrow Trans(e_x)$ holds, which is what Theorem 1 needs. If the condition holds, it indicates a non-serializable trace.

***Example***. Consider the trace $\alpha_1$ shown in Figure 6. At the beginning, the vector clocks $TV(t_1)$, $TV(t_2)$, and $TV(t_3)$ are all initialized as [0, 0, 0]. On processing $e_3$, $TV(t_1)$ is updated to be [1, 1, 0]. Because the conditions at lines 7 and 9 of Algorithm 4 are not satisfied, no violation is reported, and trace $\alpha_1$ is serializable at this moment. On processing $e_4$, $TV(t_2)$ is updated to [0, 1, 1]. Besides, the conditions at lines 7 and 9 of Algorithm 4 are not satisfied, thus trace $\alpha_1$ is still serializable at this moment. Then, through *backPropagate*, $TV(t_1)$ is updated to [1, 1, 1]. On processing $e_7$, $TV(t_3)$ is updated to [1, 0, 2], indicating $tx_5 \rightsquigarrow tx_1$. RegionTrack checks the condition at line 7, and no transactional atomicity violation is reported. Then, RegionTrack checks the condition at line 9 and $TV(t_1)[t_3] \le V(e_6)[t_3]$ holds, indicating $tx_1 \rightsquigarrow tx_5$. Thus, RegionTrack reports that trace $\alpha_1$ is non-serializable.

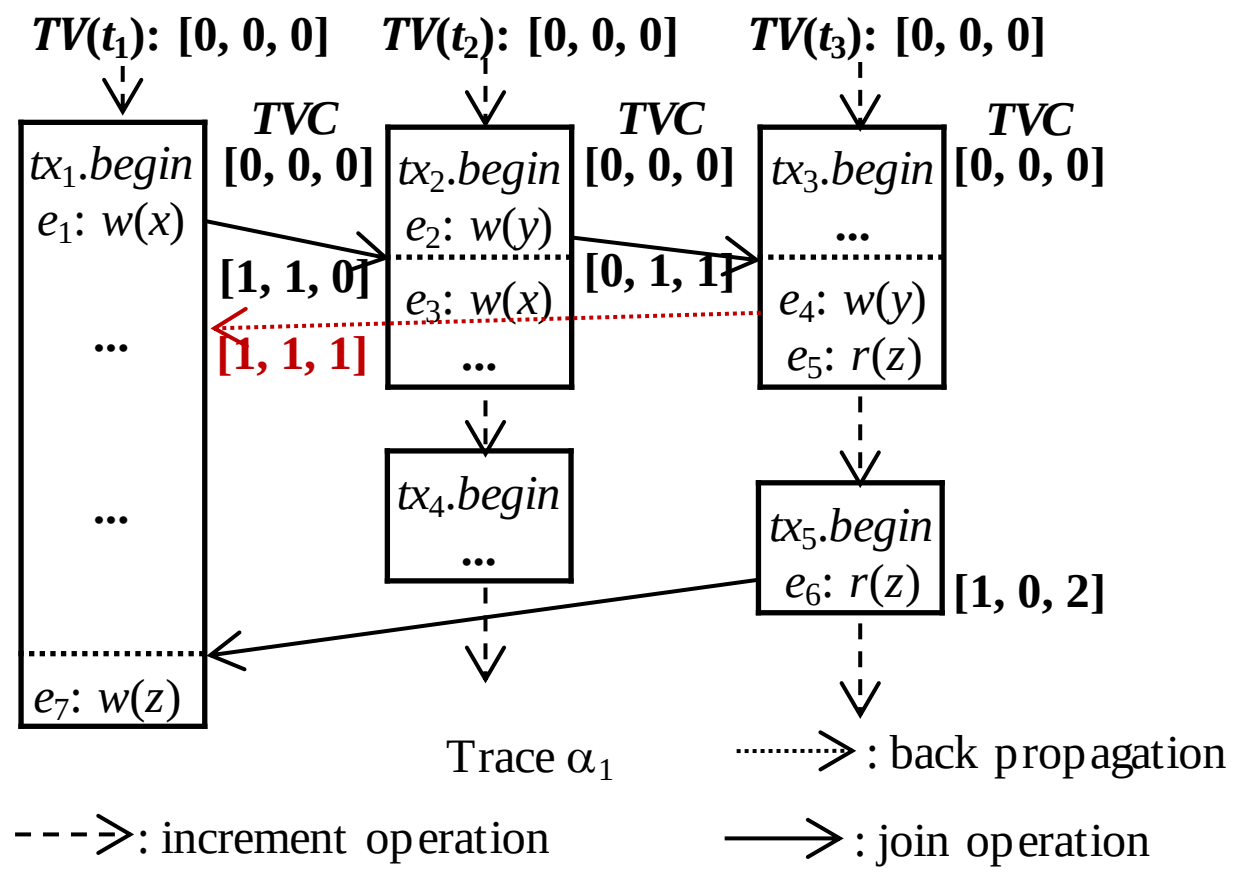

**Figure 6. Illustration of RegionTrack identifying non-serializable traces along the trace $\alpha_1$**

## 4.4 Comparisons with AeroDrome

AeroDrome [52] is an online checker for non-serializable traces, which also utilizes vector clocks to capture the transactional happens-before relations. However, it does not include any tracking procedure to identify transactional atomicity violations. Besides using its traditional vector clock algorithm to capture happens-before relations for events, AeroDrome also updates the vector clocks when each transaction ends. Its idea is that if a transaction *tx* of thread *t* is ending, AeroDrome traverses the vector clocks of all other threads (say thread *u*), all memory locations (say variable *x*), all lock objects (say lock *l*) to check whether *any* of the following conditions hold: $V(tx.begin) \sqsubseteq V(u)$ or $V(tx.begin) \sqsubseteq V(w(x))$ or $V(tx.begin) \sqsubseteq V(r(u,x))$ or $V(tx.begin) \sqsubseteq V(rel(l))$. If the condition holds, the happens-before relation is abstracted to transactional happens-before relation that the vector clock $V(t)$ is joined to the respective thread/variable/lock vector clock. Thus, AeroDrome is able to capture the transactional happens-before relations and detects non-serializable traces according to Theorem 1.

***Example.*** Figure 7(a) illustrates how AeroDrome handles the trace $\alpha_1$ in Figure 1(a). On processing $tx_1.begin$, AeroDrome increments $V(t_1)$, and lets $V(tx_1.begin)$ refer to $V(t_1)$. Then, on processing $e_1$, $V(t_1)$ needs not to perform any operation. Next, $t_2$ executes $tx_2.begin$, $e_2$, and $e_3$. One happens-before relation $e_1 \rightarrowtail e_3$ is captured through the join operation between $V(t_2)$ and $V(e_1)$ when executing $e_3$, and $V(t_2)$ is updated by this join operation. $V(e_3)$ refers to the current clock $V(t_2)$, and $V(e_2)$ remains the same. The happens-before relation $e_2 \rightarrowtail e_4$ is captured through the join operation between $V(t_3)$ and $V(e_2)$ when executing $e_4$. The execution continues, and when $tx_2$ ends, AeroDrome traverses the vector clocks of all threads/locks/variables and determines that $V(tx_2.begin) \sqsubseteq V(t_3)$ and $V(tx_2.begin) \sqsubseteq V(e_4)$ hold. Thus, $V(t_3)$ is updated to be $V(t_3) \sqcup V(t_2)$. Similarly, $V(e_4)$ is updated to be $V(e_4) \sqcup V(t_2)$. When capturing $e_6 \rightarrowtail e_7$, AeroDrome reports a non-serializable trace because $V(tx_1.begin) \sqsubseteq V(e_6)$ holds.

On analyzing $\alpha_3$ in Figure 1(c), AeroDrome also reports a non-serializable trace. Figure 7(b) shows the VCs of thread $t_3$ in $\alpha_3$ (the VCs of $t_1$ and $t_2$ in $\alpha_3$ are the same as those in $\alpha_1$). We can see that when capturing $e_6 \rightarrowtail e_7$ because $V(tx_1.begin) \sqsubseteq V(e_6)$ holds, indicating $tx_1.begin \rightarrowtail e_6$. AeroDrome reports a non-serializable trace.

We recall that AeroDrome and RegionTrack both use vector clocks to capture transactional happens-before relations and identify non-serializable traces. RegionTrack can further point out the atomicity violation on a transaction that can provide useful information to programmers for program debugging.

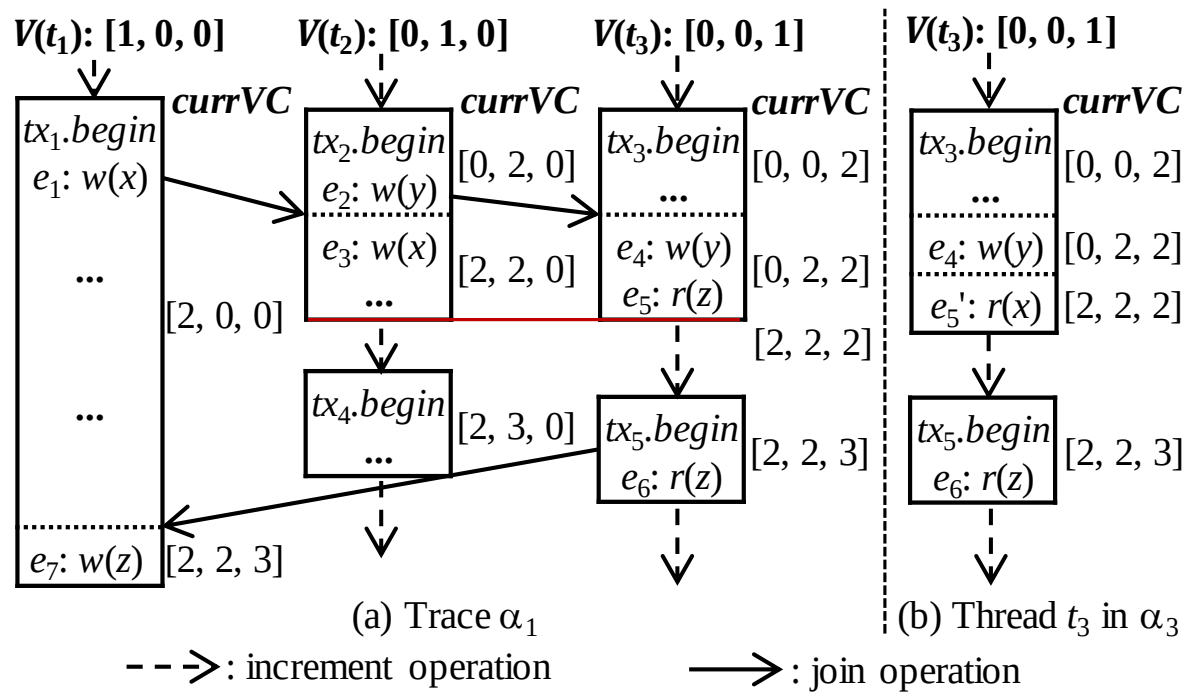

**Figure 7. Illustration of AeroDrome along the trace $\alpha_1$ and $\alpha_3$**

There are many fundamental differences between the two techniques. A key difference between AeroDrome and RegionTrack is that AeroDrome cannot distinguish increasing transactional happens-before relations and non-increasing transactional happens-before relations. Whereas RegionTrack utilizes a novel forward- and back-propagate mechanism to split these two kinds of transactional happens-before relations. It precisely stores them in traditional vector clocks and transactional vector clocks. Another key difference is that RegionTrack need not traverse the vector clocks of other kinds of entities (locks and variables) to determine THB relations in a lazy manner (when a transaction ends). Last but not the least, RegionTrack can both identify non-serializable traces and transactions experiencing transactional atomicity violations.

For time and space complexity, we first denote $n_{non\text{-}end}$, $n_{end}$ and $n_{join}$ as the number of transaction events other than these end events, the number of transaction end events, and the number of join operations, respectively. We also denote the numbers of threads by $t$, memory locations by $V$, and lock objects by $L$.

The time and space complexity of RegionTrack are $O(tn_{join} + t^3n_{join})$ and $O(t(t+tV+L))$, respectively. Whereas the time and space complexity of AeroDrome taken from the original paper [52] are $O(t(n_{non\text{-}end}+(t+L+V)n_{end}))$ and $O(t(t+V+L))$. Note that the time complexity of *checkHB* in Algorithm 4 is O(1) since it only checks two timestamps at lines 7 and 9. All other operations are updated in constant time.

Although the *forwardPropagate* and *backPropagate* functions are recursive, each function behaves like a depth-first search — On every invocation of *forwardPropagate*, the thread set is reduced, and each thread will only be updated once in the entire search. The time complexity of *forwardPropagate* is $O(t)$. Similarly, in *backPropagate*, if the timestamp is backpropagated to a new thread, it will call one round of *forwardPropagate*. Since the thread set is also reduced and the timestamp kept in the transaction vector clock of the thread has been updated, each thread can only be back propagated once. Thus, the time complexity of *backPropagate* is $O(t^2)$. A transactional vector clock keeps $t$ timestamps, which requires each invocation of the function may need to update $t$ timestamps, and each round of forward- and back-propagation is invoked whenever there is a join operation to maintain the transactional happens-before relations.

The key difference in time complexity is the term $t^3n_{join}$ in RegionTrack versus the term $t(t+L+V)n_{end}$ in AeroDrome. Since the time complexity of AeroDrome and RegionTrack consider different variables, it is not quite straightforward to compare these two techniques algebraically.

We are not algorithmic experts. In the sequel, for brevity, we attempt to compare the orders of different variables so that we can compare the time complexity of the two algorithms in a ballpark manner.

The number of threads $t$ in a program is often a small number and bounded by a constant. Moreover, the number of variables is a few orders of magnitude higher than the number of threads in the same program. Thus, we tend to believe that $O(t^2) < O(V)$ holds. In the worst case, each transaction contains at most one non-end event, and each non-end event needs a join operation at the transaction level. (Besides, a transaction may not include events needing join operations.) Thus, we tend to believe that $O(n_{join}) \approx O(n_{non\text{-}end}) \approx O(n_{end})$ is more likely the case. So, the time complexity of RegionTrack = $O(tn_{join} + t^3n_{join}) < O(t(n_{non\text{-}end} + Vn_{join})) \leq O(t(n_{non\text{-}end}+(t+L+V)n_{end}))$ = the time complexity of AeroDrome. In our experiment, RegionTrack did not run slower than AeroDrome (see Section 5).

Moreover, each accessed variable should appear as an access event in at least one transaction, and each such transaction contains one end event. The same is true for locking event. Thus, $O(L+V) \leq O(n_{end})$. Hence, the time complexity of AeroDrome = $O(t(n_{non\text{-}end}+(t+L+V)n_{end})) \approx O(t(n_{end}+(t+L+V)n_{end})) \geq O(t((L + V) +(t+L+V) (L + V)))$ $= O(t(L + V) + t(L + V)^2) = O(t(L + V)^2)$. Hence, we tend to believe that both RegionTrack and AeroDrome are non-linear algorithms.

The space complexity of RegionTrack is higher than that of AeroDrome. In the algorithm of AeroDrome, it includes an optimization on the number of read vector clocks for each memory location. For each memory location, it only stores two read vector clocks rather than $t$'s read vector clocks, which lowers the space complexity order by a factor of $t$. In the algorithm of RegionTrack presented in Section 4, we have not included such optimization. However, we note that the same kind of optimization (proposed by AeroDrome) can be added to RegionTrack to achieve the same space complexity as AeroDrome.

We finally note that the detection of transactional atomicity violations and the detection of non-serializable traces can work independently and together.

## 4.5 Correctness of RegionTrack

This section presents the correctness of RegionTrack. Lemma 1 and Theorem 3 present the correctness of Algorithms 1 and 2. Lemma 2, Lemma 3, and Theorem 4 present the correctness of Algorithms 3 and 4.

**Lemma 1**. Let thread $t_1 = T(tx)$ where $tx$ is a transaction in a trace $\alpha$. Suppose that there is an event $e$ in $\alpha$ such that $e \neq tx.begin$. If $V(tx.begin)[t_1] \leq V(e)[t_1]$, then $tx.begin \rightarrowtail e$.

**Proof**. We first recall that the function $inc(V(t_1))$ increments the timestamp kept at the position $V(t_1)[t_1]$ by 1 when processing $tx.begin$. For an arbitrary event $e_n$ which appears before $tx.begin$ in $\alpha$, $V(tx.begin)[t_1]$ must be greater than $V(e_n)[t_1]$ (i.e., $V(e_n)[t_1] < V(tx.begin)[t_1]$). Thus, given the condition $V(tx.begin)[t_1] \leq V(e)[t_1]$, the event $e$ must appear after $tx.begin$ in $\alpha$. Let $\alpha = \langle ..., tx.begin, ..., e, ...\rangle$. We have the following two cases.

Case 1: Suppose $T(e) = t_1$. According to the program order between $tx.begin$ and $e$, we have $tx.begin \rightarrowtail e$.

Case 2: Suppose $T(e) \neq t_1$. Let $t_2 = T(e)$. Before the execution of $tx.begin$, thread $t_2$ must be in an analysis state where $V(t_2)[t_1] < V(tx.begin)[t_1] \leq V(e)[t_1]$. Because $V(e)$ is the shadow state of $V(t_2)$ at the moment of executing $e$, so the timestamp $V(t_2)[t_1]$ should have been changed from satisfying the condition $V(t_2)[t_1] < V(tx.begin)[t_1]$ to satisfying the condition $V(t_2)[t_1] \geq V(tx.begin)[t_1]$ either before executing $e$ or at the moment of executing $e$. Since the timestamp $V(t_2)[t_1]$ can only be updated through join operations (line 46 of Algorithm 2), in between the execution of $tx.begin$ and $e$, there must exist at least one join operation to propagate the timestamp of thread $t_1$ to thread $t_2$ either directly or indirectly. In Algorithm 2, a join operation is performed whenever there is a cross-thread happens-before relation. So, $tx.begin$ must happen before $e$ (otherwise, the timestamp of thread $t_1$ after having executed $tx.begin$ cannot be propagated to thread $t_2$ either before executing $e$ or at the moment of executing $e$). Therefore, we have proven the lemma. □

Lemma 1 helps to prove the correctness of reporting transactional atomicity violations of RegionTrack. Once the condition at line 8 in Algorithm 4 is satisfied, RegionTrack ensures that the happens-before relation $tx.begin \rightarrowtail e$ exists. Based on Lemma 1, Theorem 3 can be proved as follows.

**Theorem 3**. RegionTrack reports a transactional atomicity violation on a transaction $tx$ in a trace $\alpha$ if and only if $tx$ is not serializable.

**Proof**. Based on Theorem 2, a transaction $tx$ is not serializable if and only if there is an event $e_m \in tx$ and another event $e_x$ in $\alpha$ such that (1) $T(e_x) \neq T(e_m)$, (2) $e_x \rightarrowtail e_m$, and (3) $tx.begin \rightarrowtail e_x$. For each event $e_m \in tx$ in the trace $\alpha$, RegionTrack checks both (1) whether there is a conflicting event $e_x$ where $T(e_x) \neq T(e_m)$ such that $e_x \rightarrowtail e_m$ is formed (lines 11, 18-19, 22, or 32 in Algorithm 1), which are the first two conditions of Theorem 2, and (2) whether the happens-before relation $tx.begin \rightarrowtail e_x$ has been formed (line 50 in Algorithm 2), which is the third condition of Theorem 2. RegionTrack reports a transactional atomicity violation on $tx$ (line 51 of Algorithm 2) if and only if all conditions are satisfied. By Theorem 1, $tx$ is not serializable in $\alpha$ if and only if under the above conditions. So, we have proven Theorem 3. □

After proving the correctness of RegionTrack in reporting transactional atomicity violations, we now prove the correctness of RegionTrack in identifying non-serializable traces. First, we prove Lemma 2 to show the soundness of RegionTrack to capture transactional happens-before relations.

**Lemma 2**. Let thread $t_1 = T(tx)$ and $t_2 = T(tx')$ where $tx$ and $tx'$ are two transactions in trace $\alpha$. Let $C(t_1)$ represent the node of transaction $tx$ and $e$ be an event in transaction $tx'$. Suppose $tx \neq tx'$. If (1) $V(tx.begin)[t_1] \leq V(e)[t_1]$, or (2) $V(tx.begin)[t_1] > V(e)[t_1]$, $TV(t_1)[t_2] \leq V(e)[t_2]$ and $TV(t_1)[t_1] = V(C(t_1).begin)[t_1]$, then $tx \rightsquigarrow tx'$.

**Proof.** First, we recall that the function $inc(V(t_1))$ increments the timestamp kept at the position $V(t_1)[t_1]$ by 1 when processing $tx.begin$. That is, all events in a transaction share the same timestamp of that thread as $V(C(t_1).begin)[t_1]$, and different transactions have different timestamps of its thread. For any event $e'$ belonging to a previous transaction $tx''$ of thread $t_1$, we should have $V(e')[t_1] < V(C(t_1).begin)[t_1]$. Moreover, we recall that in Algorithm 4 at line 3, $TV(t_1)[t_2]$ will only update in the join operation when (1) there is a direct THB relation from transaction $tx$ to transaction of thread $t_2$ at lines 2-7 and 9-12 in Algorithm 3, or (2) it needs to propagate the timestamps kept in $TV$s to $TV(t_1)$ to propagate timestamps among transactions of different threads at lines 8, 13 and 15 in Algorithm 3.

For the first condition, since we know $tx \neq tx'$, we should have $e \neq tx.begin$. According to Lemma 1, if the condition $V(tx.begin)[t_1] \leq V(e)[t_1]$ holds, then $tx.begin \rightarrowtail e$ follows. By the definition of transactional happens-before, we have $Trans(tx.begin) \rightsquigarrow Trans(e)$, which means $tx \rightsquigarrow tx'$.

For the second condition, since we know that the three conditions $V(tx.begin)[t_1] > V(e)[t_1]$, $TV(t_1)[t_2] \leq V(e)[t_2]$ and $TV(t_1)[t_1] = V(C(t_1).begin)[t_1]$ hold, we can further consider the following two subcases:

Subcase 1: Suppose $t_1 = t_2$. Given that $V(tx.begin)[t_1] > V(e)[t_1]$ holds, we should have $V(C(t_1).begin)[t_1] > V(e)[t_1]$. According to program order, $tx'$ must appear before $tx$. Since $t_1 = t_2$, $TV(t_1)[t_2]$ and $TV(t_1)[t_1]$ should keep the same timestamp. If $TV(t_1)[t_1] = V(C(t_1).begin)[t_1]$, we should have $TV(t_1)[t_2] = TV(t_1)[t_1] = V(C(t_1).begin)[t_1]$ $TV(t_1)[t_1] > V(e)[t_1]= V(e)[t_2]$. This is a contradiction to the condition $TV(t_1)[t_2] \leq V(e)[t_2]$, which indicates $t_1 \neq t_2$. Moreover, if $tx'$ must appear before $tx$ in the same thread, we could not have $tx \rightsquigarrow tx'$. Thus, $t_1 \neq t_2$.

Subcase 2: Suppose $t_1 \neq t_2$. Given that $TV(t_1)[t_1] = V(C(t_1).begin)[t_1]$ holds, the transaction $tx$ should be the latest source transaction of thread $t_1$ that has THB relations with transactions of other threads. Since $TV(t_1)[t_2]$ will only be updated when (1) there is a direct THB relation from transaction $tx$ to transaction of thread $t_2$ at lines 2-7, and 9-12 in Algorithm 3; or (2) we need to propagate the timestamps kept in $TV$s to $TV(t_1)$ to propagate the timestamps between transactions of different threads at lines 8, 13 and 15 in Algorithm 3.

We have the following two cases to consider, one for each condition above:

Case 1: Suppose $TV(t_1)[t_2]$ is updated according to the join operation from thread $t_1$ to thread $t_2$ at Algorithm 4 at line 3. Since we have $V(tx.begin)[t_1] > V(e)[t_1]$ and $TV(t_1)[t_2] \leq V(e)[t_2]$, the transactional happens-before relation must be established by some other event $e''$ which appears after $e$ in transaction $tx'$ with some event $e_x$ in transaction $tx$. Moreover, by definition, $TV(t_1)[t_2]$ only stores the timestamp of the first sink transaction of thread $t_2$. When thread $t_2$ executes $e''$ and performs the join operation, $TV(t_1)[t_2]$ will be updated to record the timestamp of transaction $tx'$ and the happens-before relation $e_x \rightarrowtail e''$ establishes, indicating that the transactional happens-before relation $tx \rightsquigarrow tx'$ should hold ($tx'$ is the first sink transaction of thread $t_2$, otherwise $V(tx.begin)[t_1] \leq V(e)[t_1]$ should exist.). Here, $TV(t_1)[t_2] = V(e)[t_2]$.

Case 2: Suppose $TV(t_1)[t_2]$ is updated according to the timestamp propagation procedure over $TV$s at lines 8,

13, and 15 in Algorithm 3 to propagate the required THB relations between transactions of different threads. We first recall that $TV(t_1)[t_2]$ only stores the timestamp of the first sink transaction of thread $t_2$. Let the propagating process be $tx \rightsquigarrow tx_i \rightsquigarrow \ldots \rightsquigarrow tx_j \rightsquigarrow tx'$ where (1) the transaction $tx_i$ is the first transaction of a thread other than $t_1$ and $t_2$ that $tx$ happens-before it, and (2) the transaction $tx_j$ is the first sink transaction of thread $t_2$ that happens-before the transaction $tx'$ where $tx_i$ and $tx_j$ belong to different threads. In Algorithm 3, function *backPropagate* (line 15) and *forwardPropagate* (line 8 and 13) update the timestamp(s) of $TV(t_1)[t_2]$ to propagate the transactional happens-before relations. After the propagation, if $tx_j$ is the first sink transaction of thread $t_2$ that $tx$ happens-before it, then $TV(t_1)[t_2]$ is the timestamp of $tx_j$. Since $TV(t_1)[t_2] \le V(e)[t_2]$ holds, transaction $tx$ must happen-before either transaction $tx'$ or a previous transaction of $tx'$ in thread $t_2$. By the definitions of transactional happens-before and program order, we can infer $tx \rightsquigarrow tx'$. The lemma is proved. □

After proving the soundness of RegionTrack in capturing the transactional happens-before relations, we now prove the completeness of RegionTrack in capturing all the required transactional happens-before relations. Our insight is that if a transaction finishes, it cannot further have any new incoming transactional happens-before relations. This finished transaction cannot involve in a cyclic sequence of transactional happens-before relations that both starts and ends at the transaction.

**Lemma 3**. Let thread $t_1 = T(tx)$ and $t_2 = T(tx')$ where $tx$ and $tx'$ are two transactions in trace $\alpha$. Let $C(t_1)$ represent the node of transaction $tx$ and $e$ be an event in transaction $tx'$. Suppose $tx \neq tx'$ and $tx$ is the currently running transaction of thread $t_1$. If $tx \rightsquigarrow tx'$, then (1) $V(tx.begin)[t_1] \le V(e)[t_1]$, or (2) $V(tx.begin)[t_1] > V(e)[t_1]$, $TV(t_1)[t_2] \le V(e)[t_2]$ and $TV(t_1)[t_1] = V(C(t_1).begin)[t_1]$.

**Proof**. First, we recall that the function $inc(V(t_1))$ increments the timestamp kept at the position $V(t_1)[t_1]$ by 1 when processing *tx.begin*. That is, all events in a transaction share the same timestamp of that thread as $V(C(t_1).begin)[t_1]$, and different transactions have different timestamps of its thread. For any event $e'$ belonging to a previous transaction $tx''$ of thread $t_1$, we should have $V(e')[t_1] < V(C(t_1).begin)[t_1]$. Moreover, we recall that in Algorithm 4 at line 3, $TV(t_1)[t_2]$ will only update in a join operation when (1) there is a direct THB relation from transaction $tx$ to transaction of thread $t_2$ at lines 2-10 in at Algorithm 3, or (2) we need to propagate the timestamps kept in $TV$s to $TV(t_1)$ to propagate THB relations between transactions of different threads at lines 11-12 in Algorithm 3.

Suppose that the algorithm has constructed the relation $tx \rightsquigarrow tx'$. There are two cases to establish this relation according to the definition of transactional happens-before relation.

Case 1: $tx \rightsquigarrow tx'$ establishes because of the direct THB relation. It means that there are two events $e_i \in tx$ and $e_j \in tx'$ (without loss of generality, suppose that $e$ appears after $e_j$) such that $e_i \rightarrowtail e_j$. Because *tx.begin* is the first event in $tx$ and $e_j \rightarrowtail e$ holds (because of program order), we should have $tx.begin \rightarrowtail e$ (transitive happens-before relation). Besides, for any two events $e$ and $e'$ in $\alpha$, the relation $e \rightarrowtail e'$ holds if and only if $V(e)[t] \le V(e')[t]$ holds for each thread $t$. Thus, we should have $V(tx.begin)[t_1] \le V(e)[t_1]$.

If $e$ appears before $e_j$, we cannot infer $tx.begin \rightarrowtail e$, implying $V(tx.begin)[t_1] > V(e)[t_1]$. However, because $tx$ is the currently running transaction of thread $t_1$, when thread $t_2$ executes $e_j$ and the happens-before relation $e_i \rightarrowtail e_j$ establishes, $TV(t_1)[t_1]$ stores the timestamp of transaction $tx$ which is $V(C(t_1).begin)[t_1]$, and $TV(t_1)[t_2]$ stores the timestamp of transaction $tx'$. Because $V(e)[t_2]$ stores the timestamp of transaction $tx'$, we will have $TV(t_1)[t_2] \le V(e)[t_2]$ (indeed, more precisely $TV(t_1)[t_2] = V(e)[t_2]$ in this case).

Case 2, $tx \rightsquigarrow tx'$ holds due to indirect THB relation. We consider two subcases:

Subcase 1: Suppose the sequence to obtain $tx \rightsquigarrow tx'$ is increasing. We prove this case by mathematical induction on the size of the sequence.

Suppose the sequence size is two and the sequence is $tx \rightsquigarrow tx_i \rightsquigarrow tx'$. Since the sequence is increasing, we have four events $e_x \in tx$, $e_i \in tx_i$, $e_{i'} \in tx_i$, and $e_m \in tx'$ satisfying $e_x \rightarrowtail e_i$ and $e_{i'} \rightarrowtail e_m$, where the subscript is the total event order. Moreover, the event order satisfies $x \leq i \leq i' \leq m$ and the event $e$ appears after $e_m$ in $tx'$. In addition, $tx.begin$ is the first event in $tx$. Due to program order, $tx.begin \rightarrowtail e_x$, $e_i \rightarrowtail e_{i'}$ and $e_m \rightarrowtail e$ exist. Thus, according to the transitivity of HB relations, $tx.begin \rightarrowtail e$ holds indicating $V(tx.begin)[t_1] \leq V(e)[t_1]$.

Suppose the sequence size is three and the sequence is $tx \rightsquigarrow tx_i \rightsquigarrow tx_j \rightsquigarrow tx'$. Since the sequence is increasing, we have six events $e_x \in tx$, $e_i \in tx_i$, $e_{i'} \in tx_i$, $e_j \in tx_j$, $e_{j'} \in tx_j$ and $e_m \in tx'$ satisfying $e_x \rightarrowtail e_i$, $e_{i'} \rightarrowtail e_j$ and $e_{j'} \rightarrowtail e_m$. where the subscript is the total event order. Moreover, the event order satisfies $x \leq i \leq i' \leq j \leq j' \leq m$ and the event $e$ appears after $e_m$ in $tx'$. In addition, $tx.begin$ is the first event in $tx$. Due to program order, $tx.begin \rightarrowtail e_x$, $e_i \rightarrowtail e_{i'}$, $e_j \rightarrowtail e_{j'}$, and $e_m \rightarrowtail e$ exist. Thus, according to the transitivity of HB relations, $tx.begin \rightarrowtail e$ holds indicating $V(tx.begin)[t_1] \leq V(e)[t_1]$.

Suppose the sequence size is $m$ and the sequence is $tx \rightsquigarrow tx_i \rightsquigarrow tx_j \rightsquigarrow \ldots \rightsquigarrow tx_k \rightsquigarrow tx'$. Suppose further that the $m - 1$ sequence $tx \rightsquigarrow tx_i \rightsquigarrow tx_j \rightsquigarrow \ldots \rightsquigarrow tx_k$ holds the condition, which means $tx.begin \rightarrowtail e_k$ (where $e_k \in tx_k$). When the $m$th relation $tx_k \rightsquigarrow tx'$ creates, we should have two events $e_{k'} \in tx_k$ and $e_m \in tx'$ satisfying $e_{k'} \rightarrowtail e_m$. Since the whole sequence is increasing, $e_k \rightarrowtail e_{k'}$ exists according to program order. $e$ must be an event appears after $e_m$ in $tx'$. Then, based on the transitivity of HB relation, we will have $tx.begin \rightarrowtail e$ indicating $V(tx.begin)[t_1] \leq V(e)[t_1]$.

Subcase 2: Suppose the sequence to obtain $tx \rightsquigarrow tx'$ is $tx \rightsquigarrow tx_i \rightsquigarrow tx_j \rightsquigarrow \ldots \rightsquigarrow tx_k \rightsquigarrow tx'$ and is non-increasing. Therefore, we cannot find an event sequence implying $tx.begin \rightarrowtail e$, which means $V(tx.begin)[t_1] > V(e)[t_1]$. Moreover, because transaction $tx$ is the currently running transaction of thread $t_1$, we have $TV(t_1)[t_1] = V(C(t_1).begin)[t_1]$ (lines 4 and 11 in Algorithm 3). (We note that only a currently running transaction can be added with more incoming transactional happens-before relations.)

Since having only one/two thread(s) or one THB relation cannot form any non-increasing transitive THB relations, we utilize mathematical induction to prove this subcase on both the size of the transitive process and the number of threads.

For the case of three threads, we need to consider the following cases:

(a) Suppose the size of the transitive process is two and the transitive process is $tx \rightsquigarrow tx_i \rightsquigarrow tx'$ (Note that transaction $tx$, $tx_i$ and $tx'$ belong to different threads).

First, suppose that $tx \rightsquigarrow tx_i$ has been established, and $tx_i \rightsquigarrow tx'$ is the current direct THB relation. When $tx \rightsquigarrow tx_i$ establishes, and without loss of generality, suppose that $tx$ is the currently running transaction of thread $t_1$. So, $TV(t_1)[t_1]$ stores the timestamp of transaction $tx$, and $TV(t_1)[T(tx_i)]$ stores the timestamp of transaction $tx_i$ or the timestamp of the first sink transaction of thread $T(tx_i)$. Currently, nothing needs to be updated during the propagation process (Algorithm 3 at lines 8, 13, and 15).

When $tx_i \rightsquigarrow tx'$ establishes, $tx_i$ can be the currently running transaction or a finished transaction of thread $T(tx_i)$. If $tx_i$ is the current running transaction, then $TV(T(tx_i))[T(tx_i)]$ stores the timestamp of transaction $tx_i$, and $TV(T(tx_i))[t_2]$ stores the timestamp of transaction $tx'$ or the timestamp of the first sink transaction of thread $t_2$ (Algorithm 3 at lines 4, 6, 11, and 12). If $tx_i$ is a finished transaction, then $TV(T(tx_i))$ will not be updated. However, when $tx \rightsquigarrow tx_i$ establishes, $TV(t_1)[T(tx_i)]$ should keep a timestamp not larger than the timestamp of transaction $tx_i$. Thus, $TV(t_1)[t_2]$ will be updated to the timestamp of transaction $tx'$ or remain the same through *backPropagate* (Algorithm 3 line 15). We also recall that $e$ is an event of transaction $tx'$. So, $V(e)[t_2]$ keeps the timestamp of transaction $tx'$. So, we should have $TV(t_1)[t_2] \leq V(e)[t_2]$.

Second, suppose that $tx_i \rightsquigarrow tx'$ is established and $tx \rightsquigarrow tx_i$ is the current direct THB relation in a join operation. Because there is a later incoming transactional happens-before relation added to $tx_i$, $tx_i$ must be the currently running transaction of thread $T(tx_i)$. When $tx_i \rightsquigarrow tx'$ establishes, $TV(T(tx_i))[T(tx_i)]$ stores the timestamp of transaction $tx_i$, and $TV(T(tx_i))[t_2]$ stores the timestamp of transaction $tx'$ or the timestamp of the first sink transaction of thread $t_2$ (Algorithm 3 at lines 4, 6, 11, and 12). At this moment, nothing will be updated during the propagation process (Algorithm 3 at lines 8, 13, and 15). When $tx \rightsquigarrow tx_i$ establishes, $TV(t_1)[t_1]$ stores the timestamp of transaction $tx$, and $TV(t_1)[T(tx_i)]$ stores the timestamp of transaction $tx_i$ or the timestamp of the first sink transaction of thread $T(tx_i)$ (Algorithm 3 at lines 4, 6, 11, and 12). At this moment, during *forwardPropagate* (Algorithm 3 at lines 8 and 13), because $TV(t_1)[T(tx_i)] \leq TV(T(tx_i))[T(tx_i)]$ holds (Algorithm 3 at line 31), $TV(t_1)[t_2]$ should be updated to be $TV(T(tx_i))[t_2]$ (Algorithm 3 at line 34) or keep the first sink transaction's timestamp. Thus, we should have $TV(t_1)[t_2] \leq V(e)[t_2]$.

(b) Suppose the size of the transitive process is three and the transitive process is $tx \rightsquigarrow tx_i \rightsquigarrow tx_j \rightsquigarrow tx'$ (Note that transaction $tx$, $tx_i$ and $tx'$ belong to different threads, but $tx_j$ and $tx'$ both belong to thread $t_2$). Since the incoming edge can only be added when the transaction is currently running and the relation $tx_j \rightsquigarrow tx'$ is an intra-thread edge, relation $tx_i \rightsquigarrow tx_j$ must establish before relation $tx_j \rightsquigarrow tx'$.

Case (b)(i): Suppose relations $tx \rightsquigarrow tx_i$ and $tx_i \rightsquigarrow tx_j$ have been established in order, and $tx_j \rightsquigarrow tx'$ is the current direct THB relation. When $tx \rightsquigarrow tx_i$ establishes, and without loss of generality, suppose that $tx$ is the currently running transaction of thread $t_1$. So, $TV(t_1)[t_1]$ stores the timestamp of transaction $tx$, and $TV(t_1)[T(tx_i)]$ stores the timestamp of transaction $tx_i$ or the timestamp of the first sink transaction of thread $T(tx_i)$. Currently, nothing needs to be updated during the propagation process (Algorithm 3 at lines 8, 13, and 15).

When $tx_i \rightsquigarrow tx_j$ establishes, $tx_i$ can be the currently running transaction or a finished transaction of thread $T(tx_i)$. If $tx_i$ is the current running transaction, then $TV(T(tx_i))[T(tx_i)]$ stores the timestamp of transaction $tx_i$, and $TV(T(tx_i))[t_2]$ stores the timestamp of transaction $tx_j$ or the timestamp of the first sink transaction of thread $t_2$ (Algorithm 3 at lines 4, 6, 11, and 12). If $tx_i$ is a finished transaction, then $TV(T(tx_i))$ will not be updated. However, when $tx \rightsquigarrow tx_i$ establishes, $TV(t_1)[T(tx_i)]$ should keep a timestamp not larger than the timestamp of transaction $tx_i$. Thus, $TV(t_1)[t_2]$ will be updated to the timestamp of transaction $tx_j$ or remain the same through *backPropagate* (Algorithm 3 line 15).

Note that the current direct THB relation $tx_j \rightsquigarrow tx'$ is an intra-thread edge. The implicit intra-thread dependency $tx_j \rightsquigarrow tx'$ establishes when the transaction $tx'$ of thread $t_2$ starts. At this moment, no propagation process will be invoked. However, after the above two relations established, $TV(t_1)[t_2]$ stores a timestamp not larger than the timestamp of transaction $tx_j$. Since the timestamp of $tx'$ is larger than the timestamp of $tx_j$, we will have $TV(t_1)[t_2] \leq V(e)[t_2]$.

Case (b)(ii): Suppose relations $tx_i \rightsquigarrow tx_j$ and $tx_j \rightsquigarrow tx'$ have been established in order, and $tx \rightsquigarrow tx_i$ is the current direct THB relation. Since incoming edge can only be added to the currently running transaction of its thread, transaction $tx_i$ must be the currently running transaction of thread $T(tx_i)$. When $tx_i \rightsquigarrow tx_j$ establishes, $TV(T(tx_i))[T(tx_i)]$ stores the timestamp of transaction $tx_i$, and $TV(T(tx_i))[t_2]$ stores the timestamp of transaction $tx_j$ or the timestamp of the first sink transaction of thread $t_2$ (Algorithm 3 at lines 4, 6, 11, and 12).

The implicit intra-thread dependency $tx_j \rightsquigarrow tx'$ establishes when the transaction $tx'$ of thread $t_2$ starts. At this moment, no propagation process will be invoked. However, during the update of $tx_i \rightsquigarrow tx_j$, $TV(T(tx_i))[t_2]$ stores a timestamp not larger than the timestamp of transaction $tx'$.

When the current THB relation $tx \rightsquigarrow tx_i$ establishes, and without the loss of generality, suppose that $tx$ is the currently running transaction of thread $t_1$. So, $TV(t_1)[t_1]$ stores the timestamp of transaction $tx$, and $TV(t_1)[T(tx_i)]$ stores the timestamp of transaction $tx_i$ or the timestamp of the first sink transaction of thread $T(tx_i)$ (Algorithm 3 at lines 4, 6, 11, and 12). At this moment, during *forwardPropagate* (Algorithm 3 at lines 8 and 13), because $TV(t_1)[T(tx_i)] \leq TV(T(tx_i))[T(tx_i)]$ holds (Algorithm 3 at line 31), $TV(t_1)[t_2]$ should be updated to be $TV(T(tx_i))[t_2]$ (Algorithm 3 at line 34) or keep the first sink transaction's timestamp. Note that the timestamp of transaction $tx'$ is larger than the timestamp of transaction $tx_j$. We also recall that $e$ is an event of transaction $tx'$. So, $V(e)[t_2]$ keeps the timestamp of transaction $tx'$. So, we should have $TV(t_1)[t_2] \leq V(e)[t_2]$.

Case (b)(iii): Suppose relations $tx_i \rightsquigarrow tx_j$ and $tx \rightsquigarrow tx_i$ have been established in order, and $tx_j \rightsquigarrow tx'$ is the current direct THB relation. Since incoming edge can only be added to the currently running transaction of its thread, transaction $tx_i$ must be the currently running transaction of thread $T(tx_i)$. When $tx_i \rightsquigarrow tx_j$ establishes, $TV(T(tx_i))[T(tx_i)]$ stores the timestamp of transaction $tx_i$, and $TV(T(tx_i))[t_2]$ stores the timestamp of transaction $tx_j$ or the timestamp of the first sink transaction of thread $t_2$ (Algorithm 3 at lines 4, 6, 11, and 12).

When the relation $tx \rightsquigarrow tx_i$ establishes, and without the loss of generality, suppose that $tx$ is the currently running transaction of thread $t_1$. So, $TV(t_1)[t_1]$ stores the timestamp of transaction $tx$, and $TV(t_1)[T(tx_i)]$ stores the timestamp of transaction $tx_i$ or the timestamp of the first sink transaction of thread $T(tx_i)$ (Algorithm 3 at lines 4, 6, 11, and 12). At this moment, during *forwardPropagate* (Algorithm 3 at lines 8 and 13), because $TV(t_1)[T(tx_i)] \leq TV(T(tx_i))[T(tx_i)]$ holds (Algorithm 3 at line 31), $TV(t_1)[t_2]$ should be updated to be $TV(T(tx_i))[t_2]$ (Algorithm 3 at line 34) or keep the first sink transaction's timestamp.

The implicit intra-thread dependency $tx_j \rightsquigarrow tx'$ establishes when the transaction $tx'$ of thread $t_2$ starts. At this moment, no propagation process will be invoked. However, during the establishment of the above two relations, $TV(t_1)[t_2]$ stores a timestamp not larger than the timestamp of transaction $tx_j$. Note that the timestamp of transaction $tx'$ is larger than the timestamp of transaction $tx_j$. We also recall that $e$ is an event of transaction $tx'$. So, $V(e)[t_2]$ keeps the timestamp of transaction $tx'$. So, we should have $TV(t_1)[t_2] \leq V(e)[t_2]$.

(c) Suppose the size of the transitive process is $m$ and the sequence of the transitive process is $tx \rightsquigarrow tx_i \rightsquigarrow tx_j \rightsquigarrow \ldots \rightsquigarrow tx_k \rightsquigarrow tx'$ (Note that $tx$, $tx_i$ belong to different threads and not belong to thread $t_2$). Suppose that the transactional vector clock is correctly updated when $m$-1 relations have been established.

Case (c)(i): Suppose that $tx \rightsquigarrow tx_i$ is the $m$-th relation. Since incoming edge can only be added when the transaction is currently running of its thread, transaction $tx_i$ must be the currently running transaction of thread $T(tx_i)$. Thus, when $m$-1 relations have been established, $TV(T(tx_i))[T(tx_i)]$ stores the timestamp of $tx_i$ and $TV(T(tx_i))[t_2]$ stores a timestamp not larger than the timestamp of $tx'$.

When $tx \rightsquigarrow tx_i$ establishes, and without the loss of generality, suppose that $tx$ is the currently running transaction of thread $t_1$. So, $TV(t_1)[t_1]$ stores the timestamp of transaction $tx$, and $TV(t_1)[T(tx_i)]$ stores the timestamp of transaction $tx_i$ or the timestamp of the first sink transaction of thread $T(tx_i)$ (Algorithm 3 at lines 4, 6, 11, and 12). At this moment, during *forwardPropagate* (Algorithm 3 at lines 8 and 13), because $TV(t_1)[T(tx_i)] \leq TV(T(tx_i))[T(tx_i)]$ holds (Algorithm 3 at line 31), $TV(t_1)[t_2]$ should be updated to be $TV(T(tx_i))[t_2]$ (Algorithm 3 at line 34) or keep the first sink transaction's timestamp. We also recall that $e$ is an event of transaction $tx'$. So, $V(e)[t_2]$ keeps the timestamp of transaction $tx'$. So, we should have $TV(t_1)[t_2] \leq V(e)[t_2]$.

Case (c)(ii): suppose that $tx_k \rightsquigarrow tx'$ is the $m$-th relation. If this relation is an intra-thread edge, $TV(t_1)[t_1]$ stores the timestamp of $tx$, $TV(t_1)[T(tx_i)]$ stores a timestamp not larger than the timestamp of $tx_i$ and $TV(t_1)[t_2]$ stores a timestamp not larger than the timestamp of $tx_k$. The implicit intra-thread dependency $tx_k \rightsquigarrow tx'$ establishes when the transaction $tx'$ of thread $t_2$ starts. At this moment, no propagation process will be invoked. However, the timestamp of $tx'$ is larger than the timestamp of $tx_k$. We also recall that $e$ is an event of transaction $tx'$. So, $V(e)[t_2]$ keeps the timestamp of transaction $tx'$. So, we should have $TV(t_1)[t_2] \leq V(e)[t_2]$.

If this relation is an inter-thread edge indicating $tx_k$ also belongs to thread $T(tx_i)$, $TV(t_1)[t_1]$ stores the timestamp of $tx$, $TV(t_1)[T(tx_i)]$ stores a timestamp not larger than the timestamp of $tx_i$. When $tx_k \rightsquigarrow tx'$ establishes, $tx_k$ can be the currently running transaction or a finished transaction of thread $T(tx_i)$. If $tx_k$ is the current running transaction, then $TV(T(tx_k))[T(tx_k)]$ stores the timestamp of transaction $tx_k$, and $TV(T(tx_k))[t_2]$ stores the timestamp of transaction $tx'$ or the timestamp of the first sink transaction of thread $t_2$ (Algorithm 3 at lines 4, 6, 11, and 12). If $tx_k$ is a finished transaction, then $TV(T(tx_k))$ will not be updated. However, when $tx \rightsquigarrow tx_i$ establishes, $TV(t_1)[T(tx_k)]$ should keep a timestamp not larger than the timestamp of transaction $tx_i$. Thus, $TV(t_1)[t_2]$ will be updated to the timestamp of transaction $tx'$ or remain the same through *backPropagate* (Algorithm 3 line 15). We also recall that $e$ is an event of transaction $tx'$. So, $V(e)[t_2]$ keeps the timestamp of transaction $tx'$. So, we should have $TV(t_1)[t_2] \leq V(e)[t_2]$.

We now generalize the results to $n$ threads and need to consider the following cases:

(a) Suppose that the sequence of THB relations to obtain is $tx \rightsquigarrow tx'$ is $tx \rightsquigarrow tx_i \rightsquigarrow tx_j \rightsquigarrow \ldots \rightsquigarrow tx_l \rightsquigarrow tx_n \rightsquigarrow \ldots \rightsquigarrow tx_k \rightsquigarrow tx'$, which involves $n$ threads, $n$-1 transactional happens-before edges, and all transactions in this sequence belongs to different threads. Any relation THB relation in this transitive process is an inter-thread edge.

Suppose that $n$-2 transactional happens-before edges has been established, and the established edges are $tx \rightsquigarrow tx_i \rightsquigarrow tx_j \rightsquigarrow \ldots \rightsquigarrow tx_l$ and $tx_n \rightsquigarrow \ldots \rightsquigarrow tx_k \rightsquigarrow tx'$. Suppose further that these relations have been propagated to respective transactional vector clocks and $tx_l \rightsquigarrow tx_n$ is the current THB relation. Because $tx$ is the currently running transaction of thread $t_1$, $TV(t_1)$ has been updated to store the relation $tx \rightsquigarrow tx_i \rightsquigarrow tx_j \rightsquigarrow \ldots \rightsquigarrow tx_l$. Specifically, $TV(t_1)[t_1]$ stores the timestamp of transaction $tx$, and $TV(t_1)[T(tx_l)]$ stores the timestamp of transaction $tx_l$ or the timestamp of the first sink transaction of thread $T(tx_l)$. Since $tx_n$ has additional incoming THB relations, $tx_n$ must be the currently running transaction of its thread $T(tx_n)$ and $TV(T(tx_n))$ has been updated to store the relation $tx_n \rightsquigarrow \ldots \rightsquigarrow tx_k \rightsquigarrow tx'$. Specifically, $TV(T(tx_n))[T(tx_n)]$ stores the timestamp of $tx_n$, and $TV(T(tx_n))[t_2]$ stores the timestamp of transaction $tx'$ or the timestamp of the first sink transaction of thread $t_2$.

When creating the $n$-1-th transactional happens-before relation $tx_l \rightsquigarrow tx_n$, the timestamp of $tx_n$ should be backwardly propagated to thread $t_1$ (Algorithm 3 line 25) because $TV(t_1)[T(tx_l)]$ stores a timestamp not larger than the timestamp of $tx_l$ (Algorithm 3 line 20). Moreover, there is an invocation of *forwardPropagate*. During *forwardPropagate* (Algorithm 3 line 24), the timestamp stored in transactional vector clock $TV(T(tx_n))$ will be propagated to $TV(t_1)$, since $TV(t_1)[T(tx_n)] \leq TV(T(tx_n))[T(tx_n)]$ holds (Algorithm 3 at line 31). This means that the timestamp of transaction $tx'$ will be propagated to $TV(t_1)[t_2]$ after the propagation process. If transaction $tx'$ is the first sink transaction of thread $t_2$ for transaction $tx$, $TV(t_1)[t_2]$ will store the timestamp of transaction $tx'$. In addition, $V(e)[t_2]$ is the timestamp of $tx'$. Here, we have $TV(t_1)[t_2] = V(e)[t_2]$. On the other hands, if transaction $tx'$ is not the first sink transaction of thread $t_2$ for transaction $tx$, $TV(t_1)[t_2]$ keeps the timestamp of the first sink transaction of thread $t_2$, satisfying $TV(t_1)[t_2] < V(e)[t_2]$. In either case, we should have $TV(t_1)[t_2] \leq V(e)[t_2]$.

(b) Suppose that the sequence of THB relations to obtain is $tx \rightsquigarrow tx'$ is $tx \rightsquigarrow tx_i \rightsquigarrow tx_j \rightsquigarrow \ldots \rightsquigarrow tx_l \rightsquigarrow tx_n \rightsquigarrow \ldots \rightsquigarrow tx_k \rightsquigarrow tx'$, which involves $n$ threads and $m$ transactional happens-before edges ($m \geqslant n$). The transactional happens-before relations refer to both intra-thread and inter-thread dependences. For any intra-thread dependency (say, $tx_l \rightsquigarrow tx_n$), all the incoming transactional happens-before relations to the transaction at the head position (i.e., $tx_l$ for $tx_l \rightsquigarrow tx_n$) should create before the intra-thread THB dependency because no new incoming transactional happens-before relations can be added to an already finished transaction.

Suppose that $m$-1 transactional happens-before edges has been established, and the established edges are $tx \rightsquigarrow tx_i \rightsquigarrow tx_j \rightsquigarrow \ldots \rightsquigarrow tx_l$ and $tx_n \rightsquigarrow \ldots \rightsquigarrow tx_k \rightsquigarrow tx'$. Suppose further that these relations have been propagated to respective transactional vector clocks and $tx_l \rightsquigarrow tx_n$ is the current THB relation. Because $tx$ is the currently running transaction of thread $t_1$, $TV(t_1)$ has been updated to store the relation $tx \rightsquigarrow tx_i \rightsquigarrow tx_j \rightsquigarrow \ldots \rightsquigarrow tx_l$. Specifically, $TV(t_1)[t_1]$ stores the timestamp of transaction $tx$, and $TV(t_1)[T(tx_l)]$ stores the timestamp of transaction $tx_l$ or the timestamp of the first sink transaction of thread $T(tx_l)$. Since $tx_n$ has additional incoming THB relations, $tx_n$ must be the currently running transaction of its thread $T(tx_n)$ and $TV(T(tx_n))$ has been updated to store the relation $tx_n \rightsquigarrow \ldots \rightsquigarrow tx_k \rightsquigarrow tx'$.

Specifically, $TV(T(tx_n))[T(tx_n)]$ stores the timestamp of $tx_n$, and $TV(T(tx_n))[t_2]$ stores the timestamp of transaction $tx'$ or the timestamp of the first sink transaction of thread $t_2$.

We now prove that the case is also true when the $m$-th transactional happens-before relation appears.

When creating the $m$-th transactional happens-before relation $tx_l \rightsquigarrow tx_n$, if this relation is an intra-thread dependency, the establishment of $tx_l \rightsquigarrow tx_n$ should be earlier than the outgoing relation of $tx_n$, which contradicts to the assumption that $tx_l \rightsquigarrow tx_n$, is the current transactional happens-before relation.

When creating the $m$-th transactional happens-before relation $tx_l \rightsquigarrow tx_n$, if this relation is an inter-thread dependency, the timestamp of $tx_n$ should be backwardly propagated to thread $t_1$ (Algorithm 3 line 25) because $TV(t_1)[T(tx_l)]$ stores a timestamp not larger than the timestamp of $tx_l$ (Algorithm 3 line 20). Moreover, there is an invocation of *forwardPropagate*. During *forwardPropagate* (Algorithm 3 line 24), the timestamp stored in transactional vector clock $TV(T(tx_n))$ will be propagated to $TV(t_1)$, since $TV(t_1)[T(tx_n)] \leq TV(T(tx_n))[T(tx_n)]$ holds (Algorithm 3 at line 31). This means that the timestamp of transaction $tx'$ will be propagated to $TV(t_1)[t_2]$ after the propagation process. If transaction $tx'$ is the first sink transaction of thread $t_2$ for transaction $tx$, $TV(t_1)[t_2]$ will store the timestamp of transaction $tx'$. In addition, $V(e)[t_2]$ is the timestamp of $tx'$. Here, we have $TV(t_1)[t_2] = V(e)[t_2]$. On the other hands, if transaction $tx'$ is not the first sink transaction of thread $t_2$ for transaction $tx$, $TV(t_1)[t_2]$ keeps the timestamp of the first sink transaction of thread $t_2$, satisfying $TV(t_1)[t_2] < V(e)[t_2]$. In either case, we should have $TV(t_1)[t_2] \leq V(e)[t_2]$.

By mathematical induction, RegionTrack is able to capture the transactional happens-before relation for transaction $tx$ through the recursive invocation of *backPropagate* and *forwardPropagate* completely. □

The purposes of Lemma 2 and Lemma 3 are to show that RegionTrack is able to capture required transactional happens-before relations in a sound and complete manner. Then, we prove Theorem 4.

**Theorem 4**. RegionTrack reports a non-serializable trace on trace $\alpha$ if and only if the trace $\alpha$ is not serializable.

**Proof**. Based on Theorem 1, RegionTrack checks for the contradiction conditions for non-serializable traces. For each event $e_m \in tx$ in trace $\alpha$, RegionTrack checks both (1) whether there is a conflicting event $e_x \in tx'$ where $T(e_x) \neq T(e_m)$ such that $e_x \rightarrowtail e_m$ is formed, implying $tx' \rightsquigarrow tx$, and (2) whether the transactional happens-before relation $tx \rightsquigarrow tx'$ has been formed. Specifically, Algorithm 1 at line 11, line 19, line 22, and line 32 ensures that $T(e_x) \neq T(e_m)$ holds before the join operation. Algorithm 4 at line 2 performs a join operation between $e_x$ and $e_m$, and so we have $e_x \rightarrowtail e_m$ which implies that $tx' \rightsquigarrow tx$ holds. Then, in the join operation procedure, Algorithm 4 first invokes *updateTVC* to capture $tx' \rightsquigarrow tx$ at line 3 and then invokes function *checkHB* at line 4. Function *checkHB* checks whether (1) *tx.begin* $\rightarrowtail e_m$ holds at line 8, which implies the transactional happens-before relation $tx \rightsquigarrow tx'$ is increasing, or (2) $tx \rightsquigarrow tx'$ is non-increasing at line 10. As proved in Lemma 2, if RegionTrack identifies two transactions that form a transactional happens-before relation, these two transactions must form such a THB relation. In addition, as proved in Lemma 3, when transaction $tx$ is the currently running transaction of the thread, RegionTrack captures all transactional happens-before relations starting from $tx$ to transactions of other threads. According to Theorem 1 [16], trace $\alpha$ is not serializable if and only if these two conditions are met (i.e., $tx' \rightsquigarrow tx$ and $tx \rightsquigarrow tx'$). So, we have proven Theorem 4. □

Based on the above theorems, we prove the soundness and completeness of RegionTrack in reporting transactional atomicity violations and identifying non-serializable traces.

As mentioned in Section 2.4, a transactional atomicity violation indicates a non-serializable trace, whereas a non-serializable trace cannot imply a transactional atomicity violation. From the theory of the four techniques:

Velodrome, RegionTrack, DoubleChecker and Aerodrome, all these four techniques are able to identify the non-serializable traces. However, DoubleChecker and AeroDrome cannot distinguish transactional atomicity violations. Velodrome can identify some but not all the transactional atomicity violations among the non-serializable traces. Besides identifying all non-serializable traces, RegionTrack also detects all transactional atomicity violations based on its theoretical guarantees.

# 5 Evaluation

## 5.1 Implementation

Our implementation contained two parts: online and offline analyses. To evaluate performance, we have implemented Algorithms 1, 3 and 4 of RegionTrack (RT) based on the existing implementation (referred to as Velo) of Velodrome [16], which was written by the authors of [5] and built on top of Jikes RVM 3.1.3 [2][4]. Jikes RVM is an open-source Java virtual machine written in Java. The current implementation of AeroDrome (Aero) was in a different framework Rapid [54], and this implementation is an offline version that cannot be used to evaluate performance. Thus, we also implemented Aero according to the algorithm presented in [52] based on Velo to conduct a fair comparison. RT and Aero adopted the same instrumentation of Velo to identify the regular and unary transaction nodes, insert read and write barriers at field accesses, and monitor program synchronizations. Following [5], we denoted this instrumentation framework as Empty. Besides, we also denote RT-Trace as the framework that only enables the detection of non-serializable traces in Algorithms 1, 3 and 4. We downloaded the implementation of DoubleChecker's single-run mode (DC) from the website [6]. DC was built on top of Octet [8] (a tool capturing cross-thread dependences with high efficiency). Note that the authors of DC did not implement Velo on top of Octet.

Velo maintained the last access references to transaction nodes as weak references. The Java garbage collector (GC) collected those transaction nodes that were transitively unreachable from each thread's current transaction reference. Once a transaction finished, RT cleared the reference to the transaction node immediately. For the last access references to shadow VCs, RT treated them as strong references because RT needed these shadow VCs when there were events still referring to them.

The implementation of RT [1] straightforwardly followed Algorithms 1, 3 and 4, and we implemented a VC or a TVC as an array of integers. Each VC operation was implemented as an update of integers. Each *RVMThread* object maintained a current VC and a TVC. The size of VC grew if needed when performing join operations of VCs. For each subregion, RT allocated new memory space for *currVC* of that subregion, and GC could collect the obsolete *currVC*. The differences between the online version of RT and Aero were that Aero did not maintain a TVC for each *RVMThread* object but added two maps to track the live memory locations and lock objects for the following traverse at each transaction end. We discussed in Section 4.4 that AeroDrome includes an optimization on the number of read vector clocks for each memory location, which was implemented in Aero, and RT did not include this optimization.

The experiments in [5] did not control which execution traces generated by executing the benchmark subject over the same input (which the same execution traces are needed for using the iterative refinement methodology used in DC). To facilitate analysis on the same trace, we have additionally implemented an offline analysis version for these four techniques based on Java to evaluate the correctness of four detection algorithms. We first utilized Empty to collect the program execution traces, passed each collected trace to each of RT, Velo, DC and Aero, and compared their detection results. The offline analysis consumed much memory, as the offline analysis needed to record the information of the whole trace into memory, while online analysis could collect unreferenced objects

**Table 1. Statistics of benchmarks**

| Benchmark Subject | Jar file size (KB) | # of Threads* | # of Transaction Nodes* | # of Read* | # of Write* | # of Locks* |
|---|---|---|---|---|---|---|
| $eclipse_6$ | 41821.5 | 18 | 2,030,000 | 143,000,000 | 12,600,000 | 654,000 |
| $hsqldb_6$ | 824.8 | 43 | 175,000 | 11,600,000 | 1,640,000 | 211,000 |
| $lusearch_6$ | 182.4 | 6 | 193,000 | 113,000,000 | 32,500,000 | 119,000 |
| $xalan_6$ | 1027.1 | 10 | 7,020,000 | 77,900,000 | 11,700,000 | 529,000 |
| $avrora_9$ | 2085.8 | 5 | 42,700,000 | 437,000,000 | 192,000,000 | 1,150,000 |
| $jython_9$ | 4068.3 | 4 | 92 | 52,600,000 | 14,100,000 | 4,130,000 |
| $luindex_9$ | 355.4 | 2 | 95 | 6,480,000 | 2,010,000 | 482 |
| $lusearch_9$ | 271.6 | 4 | 1,510,000 | 114,000,000 | 28,800,000 | 207,000 |
| $pmd_9$ | 1091.8 | 3 | 89 | 2,370,000 | 397,000 | 714 |
| $sunflow_9$ | 1016.9 | 6 | 70,200 | 233,000,000 | 29,100,000 | 685 |
| $xalan_9$ | 3143.7 | 4 | 4,210,000 | 72,500,000 | 10,200,000 | 449,000 |
| crypt | 28.1 | 3 | 65 | 23,500,000 | 57 | 11 |
| lufact | 40.0 | 3 | 372,000 | 809,000 | 1,060 | 9 |
| series | 17.6 | 3 | 1,910,000 | 20,100 | 26 | 9 |
| sor | 15.2 | 3 | 36 | 7,710,000 | 997,000 | 9 |
| sparsematmult | 15.6 | 3 | 37 | 360,000,000 | 251,000 | 8 |
| moldyn | 38.9 | 3 | 1,070,000 | 1,310,000,000 | 8,810,000 | 12 |
| montecarlo | 158.2 | 3 | 597,000 | 232,000,000 | 41,100,000 | 14 |
| raytracer | 71.0 | 3 | 33,300,000 | 2,910,000,000 | 365,000,000 | 24 |
| | | **Total:** | **95,300,000** | **6,120,000,000** | **753,000,000** | **7,450,000** |

* Dynamic data was collected on using the Empty instrumentation with small input. Each value of dynamic data is a mean of 10 trials and rounded to three significant digits.

as garbage during program execution. However, the offline comparisons were used to show that the four techniques analyzed the same execution trace. Thus, we only evaluated performance for online analysis.

Each of these four techniques (Velo, DC, Aero, and RT) can identify non-serializable traces, but only three (Velo, DC and RT) of them have the procedures to detect transactional atomicity violations. Hence, we conducted experiments of non-serializable trace detection for all techniques and excluded Aero in the experiments of transactional atomicity violation detection. Moreover, we utilized RT-Trace to evaluate the performance of Aero.

We had tested our implementations by a few small programs.

## 5.2 Benchmarks

We performed the experiment on the DaCapo benchmark suite [7] and Java Grande Forum benchmark suite [48]. In total, our experiment used the following 19 programs that Jikes RVM 3.1.3 currently execute successfully in our environment. $eclipse_6$, $hsqldb_6$, $lusearch_6$, and $xalan_6$ from DaCapo 2006-10-MR2; and $avrora_9$, $jython_9$, $luindex_9$, $lusearch_9$, $pmd_9$, $sunflow_9$, and $xalan_9$ from DaCapo 9.12-bach; and crypt, lufact, series, sor, sparsematmult, moldyn, montecarlo and raytracer from Java Grande Forum.

Table 1 shows the descriptive statistics and runtime characteristics of the benchmarks. We followed prior experiments [5] to use the *small* input size of the DaCapo benchmarks since both DoubleChecker and Velodrome run out of memory with a larger input size for some benchmarks. We also used the input size A of the Java Grande Forum benchmarks. All runtime data of the subjects were collected on the Empty framework. We followed [5][17] to set the number of trials to 10 and reported the mean result of these trials. The first two columns show the name and jar file size (excluding the size of dependent libraries) of the benchmarks. The next two columns show the number of threads in the executions and the number of (regular and unary) transaction nodes. The last three column shows the numbers of read, write, and lock accesses encountered in the executions.

Currently, we only conducted experiments on 19 subjects from two different versions of DaCapo benchmark [7]: DaCapo 2006-10-MR2 and DaCapo 9.12-bach, and Java Grande Forum Benchmark [48]. All subjects ran successfully using JVM. However, some subjects failed when running with Empty on our virtual machines: chart from DaCapo 2006-10-MR2, and tomcat, tradebeans, tradesoap from DaCapo 9.12-bach. Hence, following [5], we selected 11 subjects from DaCapo benchmark suites and 8 subjects from Java Grande Forum Benchmark suite that ran successfully with Empty on our virtual machines. In addition, we also attempted to run the test suite microbenchmark suite [51] and SPECjvm2008 [49], they were not available online right now and had internal errors when running on Jikes RVM, respectively. We have also attempted to find modern Java subjects that can be run on Jikes RVM (without any instrumentation) that we used for the experiment. Despite having tried many subjects, none of them can be run successfully. We summarized the experiment results of the current subjects presented in Table 1.

## 5.3 Experimental Setup

For performance evaluation and trace collection, we ran the experiment on an Ubuntu Linux 12.04 x86_64 virtual machine built on a server with two 2.20GHz Intel Xeon E7-4850 v3 processors. The virtual machine was configured with two logical processors (2 cores), 16GB memory, and OpenJDK 1.6. For detection correctness evaluation, we ran the experiment on an Ubuntu Linux 18.04 x86_64 virtual machine built on a server with two 2.20GHz Intel Xeon E7-4850 v3 processors. The virtual machine was configured with two logical processors (2 cores), 128GB memory, and OpenJDK 11.

For online analysis part of the experiment, following [5], we compiled all four detection tools (Velo, RT, DC, Aero) using the production configuration (i.e., FastAdaptiveGenImmix) in Jikes RVM. This configuration includes the optimizing compiler. Using this configuration is closer to the production environment (i.e., more efficiency) and was also adopted in the experiment reported in [5]. Similar to [5], the Jikes RVM was configured with 2600MB memory.

For offline analysis part of the experiment, we adopted the iterative refinement methodology [5][14][15][16] to execute each technique on each subject to detect atomicity violations so that all three techniques can analyze each same trace. Specifically, the iterative refinement methodology produced an atomicity specification for each subject as follows: First, except methods that were intended to be non-atomic (e.g., **main()** and **Thread.run()**), all other methods were put into an atomicity specification, and we used the initial specification provided by Biswas et al. [5]. Then, a method was excluded from this specification for the next trial of analysis on the same trace if any of its transactions in the current trial are reported by the technique as non-serializable. If no new atomicity violation was reported after 2 successive trials, the iterative refinement process ended [5]. Note that we set the number of successive trials as 2 since the offline iterative refinement methodology analyzes the same trace in each trial, which is different from dynamically analyzing different traces [5]. We repeated the experiment over 100 collected traces of each benchmark.

For online analysis part of experiment, to assess the performance of each technique, for every subject, the intersection of the distinct methods detected online by all three techniques was used as the final atomicity specification (to avoid bias). Following [5][16], we ran each technique on each subject 10 times and computed the mean results. Note that, we were able to execute DC on $avrora_9$ using this final specification for $avrora_9$, while for a small number of specifications, DC ran out of memory. However, DC still ran out of memory and Aero caused exception in GC for some benchmarks even with the final specification during the online analysis. For offline analysis, we were able to collect and analyze all benchmarks.

**Table 2. Average number of distinct transactional atomicity violations reported by all four techniques, and the comparisons between pair of techniques that can detect transactional atomicity violations**

| Benchmark Subject | Average # of non-serializable traces | | | | Average # of distinct violations | | | Comparisons | | | | | |
|---|---|---|---|---|---|---|---|---|---|---|---|---|---|
| | | | | | | | | Velo | | RT | | DC | |
| | Velo | RT | DC | Aero | Velo | RT | DC | missed | error | missed | error | missed | error |
| eclipse$_6$ | 1 | 1 | 1 | 1 | 180 | 181 | 183 | 1 | 0 | 0 | 0 | 0 | 2 |
| hsqldb$_6$ | 1 | 1 | 1 | 1 | 5 | 5 | 5 | 0 | 0 | 0 | 0 | 0 | 0 |
| lusearch$_6$ | 1 | 1 | 1 | 1 | 1 | 2 | 2 | 1 | 0 | 0 | 0 | 0 | 0 |
| xalan$_6$ | 1 | 1 | 1 | 1 | 61 | 61 | 61 | 0 | 0 | 0 | 0 | 0 | 0 |
| avrora$_9$ | 1 | 1 | 1 | 1 | 21 | 21 | 23 | 0 | 0 | 0 | 0 | 0 | 2 |
| jython$_9$ | 0 | 0 | 0 | 0 | 0 | 0 | 0 | 0 | 0 | 0 | 0 | 0 | 0 |
| luindex$_9$ | 0 | 0 | 0 | 0 | 0 | 0 | 0 | 0 | 0 | 0 | 0 | 0 | 0 |
| lusearch$_9$ | 1 | 1 | 1 | 1 | 18 | 18 | 18 | 0 | 0 | 0 | 0 | 0 | 0 |
| pmd$_9$ | 0 | 0 | 0 | 0 | 0 | 0 | 0 | 0 | 0 | 0 | 0 | 0 | 0 |
| sunflow$_9$ | 1 | 1 | 1 | 1 | 11 | 11 | 11 | 0 | 0 | 0 | 0 | 0 | 0 |
| xalan$_9$ | 1 | 1 | 1 | 1 | 47 | 47 | 48 | 0 | 0 | 0 | 0 | 0 | 1 |
| crypt | 1 | 1 | 1 | 1 | 1 | 1 | 1 | 0 | 0 | 0 | 0 | 0 | 0 |
| lufact | 0 | 0 | 0 | 0 | 0 | 0 | 0 | 0 | 0 | 0 | 0 | 0 | 0 |
| series | 1 | 1 | 1 | 1 | 1 | 1 | 1 | 0 | 0 | 0 | 0 | 0 | 0 |
| sor | 1 | 1 | 1 | 1 | 1 | 1 | 1 | 0 | 0 | 0 | 0 | 0 | 0 |
| sparsematmult | 1 | 1 | 1 | 1 | 1 | 1 | 1 | 0 | 0 | 0 | 0 | 0 | 0 |
| moldyn | 0 | 0 | 0 | 0 | 0 | 0 | 0 | 0 | 0 | 0 | 0 | 0 | 0 |
| montecarlo | 1 | 1 | 1 | 1 | 2 | 2 | 2 | 0 | 0 | 0 | 0 | 0 | 0 |
| raytracer | 0 | 0 | 0 | 0 | 0 | 0 | 0 | 0 | 0 | 0 | 0 | 0 | 0 |
| **Total:** | **13** | **13** | **13** | **13** | **350** | **352** | **357** | **2** | **0** | **0** | **0** | **0** | **5** |

## 5.4 Experimental Results

### *5.4.1 Effectiveness*

Table 2 shows the mean numbers of non-serializable traces reported by the four tools over 100 traces of each benchmark: Velo, RT, DC, Aero. Each reported trace represented that the corresponding program execution was non-serializable. Overall speaking, Velo, RT, DC and Aero were able to detect all non-serializable traces.

Table 2 also summaries the average number of distinct atomicity violations (i.e., distinct methods in source code) reported during the iterative refinement of atomicity specifications by the three atomicity checkers over 100 repeated trials, respectively: Velo, RT, DC. Each reported violation represented a method that at least one execution of that method (i.e., a transaction) was not serializable (or a reported violation is a false positive yielded by DC). We note that a distinct violation reported in one trace may not always appear in other traces at run-time due to the internal non-determinstic behavior of each subject.

On average, RT detected more violations than Velo, and DC detected more violations than RT, which were in line with the theory. Overall speaking, RT detected all violations that Velo detected. For DC, the number of violations reported was more than those reported by RT. (Note that DC is complete but unsound.)

Table 2 shows the comparisons of reported violations between the three techniques as well. For a collected trace, RT detected all the violations reported by Velo. RT also detected violations that were missed by Velo on eclipse$_6$ and lusearch$_6$. On eclipse$_6$, DC reported false positives. DC also had differences in reported violations on avrora$_9$ from the other two techniques. Since DC was unsound in detecting transactional atomicity violations, the large number of violations reported by DC would require subsequent confirmations of true positives by some follow-up techniques, where we are unaware of automated tools (except RT) that can precisely separate them from the true positives. Both Velo and RT do not require follow-up confirmation.

**Table 3. Average number of dynamic transactional atomicity violations reported by techniques, and the average number of iterative refinements conducted by techniques for each trace.**

| Benchmark | # of dynamic transactional atomicity violations | | | | | # of iterative refinement | | |
|---|---|---|---|---|---|---|---|---|
| Subject | Velo (A) | A/B (in %) | RT (B) | DC (C) | C/B (in %) | Velo | RT | DC |
| eclipse$_6$ | 19,677 | 7.83 | 251,500 | 427,322 | 170 | 27 | 21 | 21 |
| hsqldb$_6$ | 762 | 9.22 | 8,267 | 9,827 | 119 | 4 | 4 | 4 |
| lusearch$_6$ | 3 | 0.0004 | 643,325 | 658,907 | 103 | 2 | 3 | 3 |
| xalan$_6$ | 13,250 | 1.30 | 1,024,043 | 1,275,667 | 125 | 13 | 13 | 13 |
| avrora$_9$ | 814,156 | 9.91 | 8,377,052 | 9,219,170 | 113 | 8 | 8 | 8 |
| jython$_9$ | 0 | - | 0 | 0 | - | 1 | 1 | 1 |
| luindex$_9$ | 0 | - | 0 | 0 | - | 1 | 1 | 1 |
| lusearch$_9$ | 18 | 16.0 | 109 | 139 | 128 | 11 | 11 | 11 |
| pmd$_9$ | 0 | - | 0 | 0 | - | 1 | 1 | 1 |
| sunflow$_9$ | 32 | 50.9 | 63 | 7,823 | 125 | 7 | 7 | 7 |
| xalan$_9$ | 1,964 | 0.69 | 288,625 | 348,995 | 121 | 14 | 14 | 14 |
| crypt | 1 | 100 | 1 | 1 | 100 | 2 | 2 | 2 |
| lufact | 0 | - | 0 | 0 | - | 1 | 1 | 1 |
| series | 1 | 100 | 1 | 1 | 100 | 2 | 2 | 2 |
| sor | 1 | 100 | 1 | 1 | 100 | 2 | 2 | 2 |
| sparsematmult | 1 | 100 | 1 | 1 | 100 | 2 | 2 | 2 |
| moldyn | 0 | - | 0 | 0 | - | 1 | 1 | 1 |
| montecarlo | 1,866 | 86.6 | 2,155 | 2,155 | 100 | 2 | 2 | 2 |
| raytracer | 0 | - | 0 | 0 | - | 1 | 1 | 1 |
| **Total:** | **851,732** | - | **10,595,143** | **11,950,019** | - | **102** | **107** | **107** |

Table 3 shows the average number of dynamic atomicity violations reported during the iterative refinement of atomicity specifications by the three atomicity checkers: Velo, RT, DC. Each such violation represented an instance of a method that was not serializable (or a reported violation is a false positive yielded by DC). Overall speaking, there were significant discrepancies between Velo and the other two techniques. Table 3 also shows the percentage that the numbers of dynamic violations detected by Velo and DC over the number of dynamic violations detected by RT, respectively. From Table 3, Velo detected no more than 50% of dynamic violations that can be detected by RT over all benchmarks of DaCapo benchmark suite while missed reporting 13.4% of dynamic violations that can be detected by RT on montecarlo of Java Grande Forum benchmark suite. This result indicates that the detection probability of Velo was lower than RT by about a half on average. DC detected more dynamic violations than RT, ranging from 3% to 70% more for DaCapo benchmark suite.

Table 3 also presents the average number of iterative analysis performed by Velo, DC and RT on a trace. On hsqldb$_6$, xalan$_6$, avrora$_9$, jython$_9$, luindex$_9$, lusearch$_9$, pmd$_9$, sunflow$_9$, xalan$_9$, crypt, lufact, series, sor, sparsematmult, moldyn, montecarlo, and raytracer, these three tools iterated the same number of refinements to analyze the same trace. On eclipse$_6$, Velo needed more iterations than RT and DC to detect transactional atomicity violations. The reason is that Velo kept at most one edge between two transactions. If one transaction was long-lived, then it was relatively easy for Velo to miss reporting transactional atomicity violations for this transaction in the first iteration. Velo had to wait for some other nested transactions to be exposed first before it can detect the transactional atomicity violations in a target transaction in a later iteration of iterative analysis. Hence, Velo required more rounds of iteration than RT and DC to analyze the same trace. On lusearch$_6$, RT and DC required more iterations than Velo for analysis. The reason was that Velo missed the violations in later iterations and thus terminated the iterative refinements, while RT and DC detected these violations.

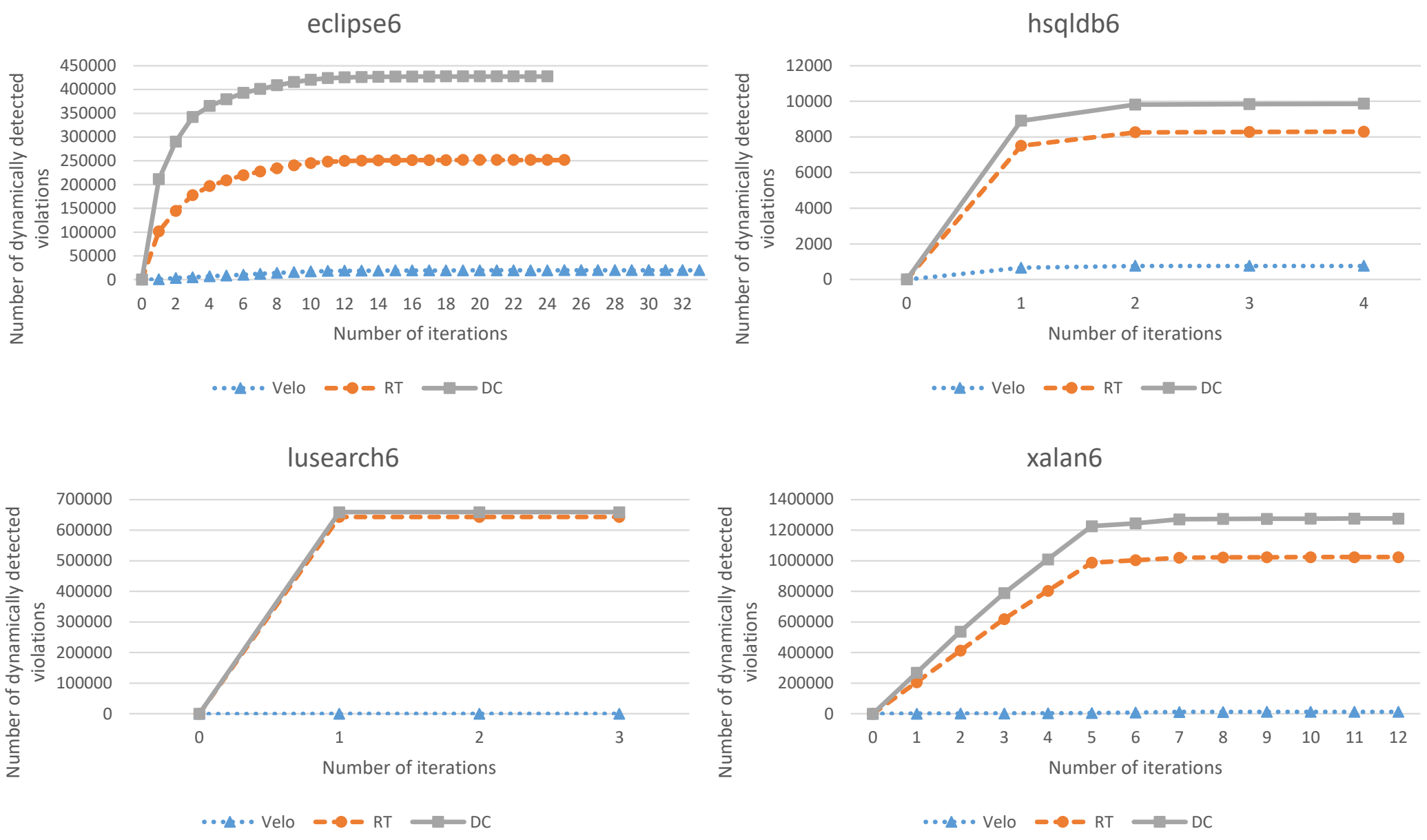

**Figure 8. Accumulated number of detected dynamic violations over iterations on eclipse$_6$, hsqldb$_6$, lusearch$_6$ and xalan$_6$.**

Figure 8 and Figure 9 summarize the accumulated number of dynamic transactional atomicity violations over the number of iterations detected by these three techniques: Velo, DC and RT, which is the average over 100 traces. They show the results on 9 subjects because on the remaining 7 subjects, there is either 0 or 1 instance detected by each technique (see Table 3). These results were presented in two figures to fit the plots within individual pages. Overall speaking, there are large discrepancies between Velo and the other two techniques. Although rarely missing to report static violations, Velo only detected a tiny fraction of all dynamic violations compared to DC and Velo. On benchmarks eclipse$_6$, hsqldb$_6$, lusearch$_6$, xalan$_6$, avrora$_9$, lusearch$_9$ and xalan$_9$, Velo missed a large number of dynamic transactional atomicity violations. This implies a very low probability of Velo to report transactional atomicity violations precisely. On benchmark montecarlo, Velo detected a large number of dynamic transactional atomicity violations. Except on lusearch$_6$, DC reported significantly a greater number of dynamic violations than RT.

Recall that RT is sound and complete in detecting transactional atomicity violations. We thus used the set of transactional atomicity violation instances reported by RT as a reference to check whether an instance reported by other tools is included in that set. It seems to us that DC appears to report many transactional atomicity violation instances that are suspicious to be false positive with respect to the traces they are located in. Especially on

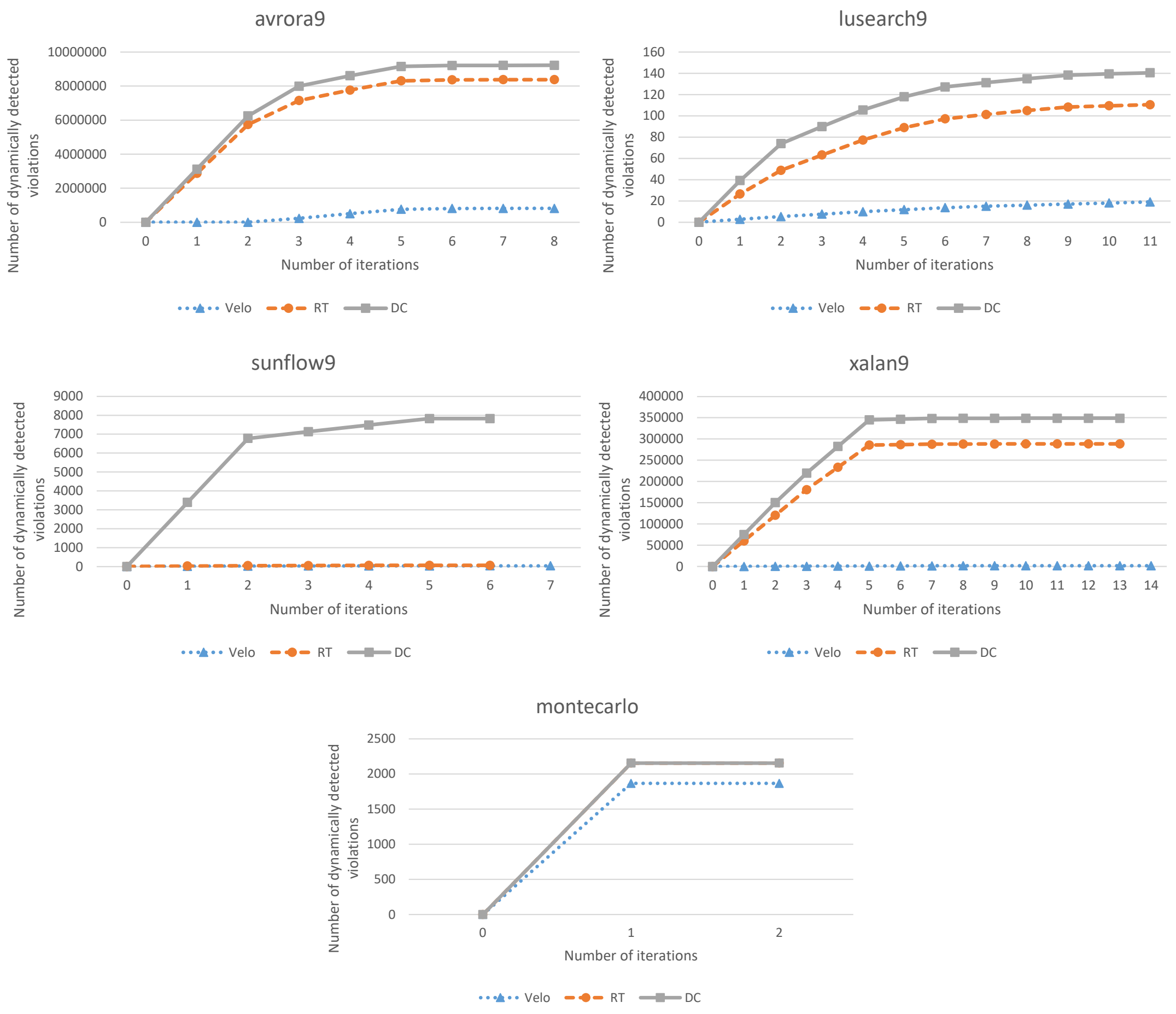

**Figure 9. Accumulated number of detected dynamic violations over iterations on avrora$_9$, lusearch$_9$, sunflow$_9$, xalan$_9$ and montecarlo.**

sunflow$_9$, DC reported around 7,800 dynamic violations, while RT only reported 63 dynamic violations.

Figure 10 (a) and Figure 10 (b) summarize the average imprecision rates of Velo and DC against RT for each function over 100 collected traces of each benchmark, respectively. The average imprecision rate of Velo against RT for each function is computed as follows: (the number of dynamic transactional atomicity violation detected by RT – the number of dynamic transactional atomicity violation detected by Velo) ÷ the number of dynamic transactional atomicity violation detected by RT. Similarly, the average imprecision rate of DC against RT for each function is computed as follows: (the number of dynamic transactional atomicity violation detected by DC – the number of dynamic transactional atomicity violation detected by RT) ÷ the number of dynamic transactional atomicity violation detected by DC.

The imprecision rate of each function is sorted from smallest to largest. In Figure 10 (a), Velo missed reporting dynamic atomicity violations less than 30% for around 50 functions. As for a majority of functions, Velo missed reporting dynamic atomicity violations more than 30%. There are more than 50 functions that Velo missed more than 90% of dynamic atomicity violations. This indicates a high probability of Velo to miss reporting transactional atomicity violations for these functions. In Figure 10 (b), DC reported 15% more dynamic atomicity violations for about 100 functions. In addition, the amount of dynamic atomicity violation instances reported by DC for 50 functions almost double that reported by RT. We also found that for 8 functions, the differences between RT and DC in terms of the number of reported violation instances are more than 20 folds.

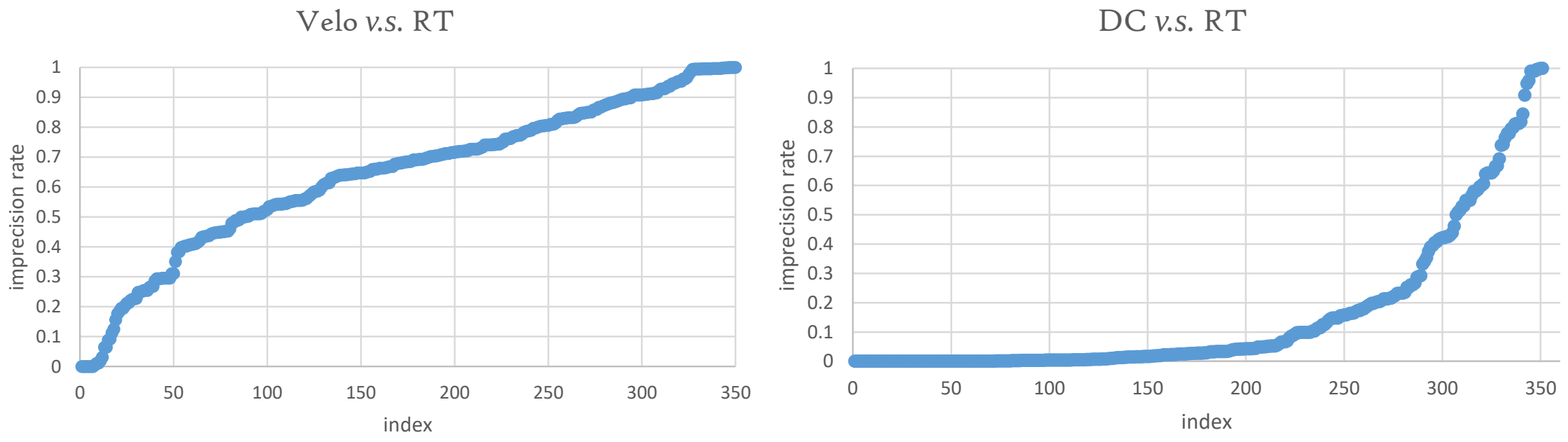

**Figure 10. (a) Average imprecision rate of Velo against RT for each function; (b) Average imprecision rate of DC against RT for each function**

In summary, although Velo only missed reporting two distinct violations on average (as shown in Table 2) in the experiment, the results of Figure 8, Figure 9 and Figure 10 indicate that Velo has a much lower probability of detecting dynamic transactional atomicity violations than RT. Moreover, an obvious difference in performance between DC and RT was that DC may report far more violation instances than RT in some functions.

***Discussion***. To conduct a fair comparison and let these three techniques analyze the same trace, we implemented an offline version of these three techniques. Following [5], we implemented the three techniques in an iterative refinement way. We note that following [16], both Velo and RT can be implemented to deal with nested atomic blocks, avoiding iterative refinements. The key idea is only to start a transaction for the outermost atomic block. The labels and timestamps of nested blocks executed within that transaction are recorded in a stack. When the nested block finishes, respective labels and timestamps are pushed out of the stack. With such modifications, both Velo and RT can process nested blocks. Nonetheless, DC cannot avoid iterative refinements based on the above modification because DC utilizes a different instrumentation scheme. DC contains two phases of cycle detection: imprecise cycle detection and precise cycle detection. During the imprecise cycle detection, DC identifies a complete set of transactions possibly involved in cycles. It logs respective read and write accesses for precise analysis. Then, during the precise cycle detection, DC strictly replays the event sequences based on the logged accesses for the involved transactions and identifies precise cycles according to the replayed trace. To deal with nested blocks, DC may require modification on instrumentation framework, analysis algorithm, and additional logs. Therefore, we implemented all three techniques using the same iterative refinement [5].

### *5.4.2 Memory Overheads*

Table 4 summarizes the memory overheads (collected via the *Linux time* command, which is the maximum resident set size of an execution) of Velo, DC, RT and Aero. In Table 4, Base means the results of the benchmark subject executing directly on the un-instrumented virtual machine. The memory consumed on each subject by Base is shown in the first column of Table 4. The memory overhead is the ratio of the memory used by a technique to the memory used by Base (i.e., memory overhead = technique's memory consumption ÷ Base's memory consumption [17]). The second to fourth columns present the memory overhead incurred by each technique.

**Table 4. Memory overhead and slowdown in online analysis**

| Benchmark Subject | Base (MB) | Memory Overhead | | | Base (sec.) | Slowdown | | |
|---|---|---|---|---|---|---|---|---|
| | | Velo | RT | DC | | Velo | RT | DC |
| $eclipse_6$ | 856 | 1.39 | 1.40 | 3.99 | 17.4 | 1.71 | 1.36 | 3.49 |
| $hsqldb_6$ | 469 | 1.18 | 1.14 | 2.14 | 2.40 | 1.54 | 1.43 | 1.33 |
| $lusearch_6$ | 396 | 1.54 | 1.51 | 4.97 | 2.09 | 3.91 | 3.30 | 2.20 |
| $xalan_6$ | 562 | 1.83 | 1.36 | 8.25 | 2.63 | 3.97 | 2.82 | 6.12 |
| $avrora_9$ | 531 | 3.20 | 1.49 | 5.85 | 8.78 | 10.6 | 6.06 | 6.39 |
| $jython_9$ | 635 | 1.11 | 1.11 | 2.18 | 7.43 | 1.49 | 1.40 | 1.16 |
| $luindex_9$ | 401 | 1.15 | 1.15 | 1.35 | 1.39 | 1.55 | 1.52 | 1.17 |
| $lusearch_9$ | 419 | 1.56 | 1.35 | 8.39 | 2.42 | 3.77 | 3.00 | 2.54 |
| $pmd_9$ | 493 | 1.13 | 1.13 | 1.11 | 1.74 | 1.11 | 1.00 | 1.00 |
| $sunflow_9$ | 408 | 1.39 | 1.46 | 4.40 | 2.40 | 7.71 | 7.18 | 3.00 |
| $xalan_9$ | 502 | 1.36 | 1.36 | 6.00 | 2.88 | 3.05 | 2.36 | 2.81 |
| crypt | 326 | 1.00 | 1.00 | 1.10 | 0.54 | 3.75 | 3.08 | 1.83 |
| lufact | 241 | 1.02 | 1.03 | 1.22 | 0.35 | 1.08 | 1.15 | 1.09 |
| series | 225 | 1.00 | 1.00 | 1.16 | 2.04 | 1.05 | 1.12 | 1.00 |
| sor | 320 | 1.00 | 1.02 | 1.14 | 1.35 | 1.05 | 1.22 | 1.14 |
| sparsematmult | 296 | 1.45 | 1.46 | 1.18 | 0.62 | 23.95 | 26.38 | 2.07 |
| moldyn | 244 | 2.78 | 2.59 | >2GB | 1.75 | 23.81 | 21.41 | - |
| montecarlo | 674 | 1.03 | 1.10 | 4.05 | 2.27 | 4.09 | 3.93 | 2.95 |
| raytracer | 314 | 1.80 | 1.67 | >2GB | 1.73 | 46.38 | 46.04 | - |
| **Mean:** | **-** | **1.47** | **1.34** | **3.87** | **-** | **7.67** | **7.15** | **2.43** |

From Table 4, RT incurred the least memory overhead on 7 out of 19 subjects than Velo and DC. On average, RT incurred 1.34x memory overhead and did not incur more than 2.59x memory overhead on each subject. The memory overheads of DC were much higher, where DC incurred 3.87x memory overhead on average, which was larger than RT and Velo by 2.88x and 2.64x. On eight subjects (i.e., $eclipse_6$, $lusearch_6$, $xalan_6$, $avrora_9$, $lusearch_9$, $sunflow_9$, $xalan_9$, and montecarlo), DC consumed significantly more memory than Velo and RT. On $pmd_9$, crypt, series and sor, the memory consumptions among these three techniques are almost the same. The memory overhead incurred by DC also exhibited significant variations ranging from 1.1x to 8.25x except for moldyn and raytracer. On moldyn and raytracer, DC ran out of memory (>2GB). On eight subjects (i.e., $eclipse_6$, $lusearch_6$, $xalan_6$, $avrora_9$, $lusearch_9$, $sunflow_9$, $xalan_9$ and montecarlo), DC incurred more than 3.5x memory overhead. On 17 subjects, DC never incurred less memory overhead than RT. Velo incurred 1.47x memory overhead on average which is slightly larger than RT. On $avrora_9$, Velo consumed significantly more memory than RT.

Table 5 shows the memory overhead on each subject by using RT-trace and Aero. From Table 5, the memory overhead of subjects $eclipse_6$, $hsqldb_6$, $lusearch_6$, $xalan_6$, $avrora_9$, $lusearch_9$, $xalan_9$, moldyn, montecarlo and raytracer could not be collected, since Aero caused exceptions on these subjects and could not complete successfully. For the rest subjects $jython_9$, $luindex_9$, $pmd_9$, $sunflow_9$, crypt, lufact, series, sor, and sparsematmult, RT-trace incurred 1.15x memory overhead on average and did not incur more than 2.44x memory overhead on each subject, while Aero incurred 2.41x memory overhead on average. On subject $sunflow_9$ and $jython_9$, Aero incurred significantly larger memory overhead than RT-trace. On subjects $luindex_9$, $pmd_9$, crypt,

**Table 5. Memory overhead and slowdown in online analysis**

| Benchmark Subject | Base (MB) | Memory Overhead | | Base (sec.) | Slowdown | |
|---|---|---|---|---|---|---|
| | | RT-trace | Aero | | RT-trace | Aero |
| $eclipse_6$ | 856 | 1.36 | - | 17.4 | 1.31 | EG |
| $hsqldb_6$ | 469 | 1.15 | - | 2.40 | 1.36 | EG |
| $lusearch_6$ | 396 | 1.54 | - | 2.09 | 3.19 | EG |
| $xalan_6$ | 562 | 1.35 | - | 2.63 | 2.71 | EG |
| $avrora_9$ | 531 | 1.48 | - | 8.78 | 5.69 | EG |
| $jython_9$ | 635 | 1.11 | 3.77 | 7.43 | 1.39 | 3.94 |
| $luindex_9$ | 401 | 1.15 | 1.19 | 1.39 | 1.39 | 1.52 |
| $lusearch_9$ | 419 | 1.35 | - | 2.42 | 2.85 | EG |
| $pmd_9$ | 493 | 1.13 | 1.14 | 1.74 | 1.00 | 1.35 |
| $sunflow_9$ | 408 | 1.44 | 9.97 | 2.40 | 6.90 | 12.72 |
| $xalan_9$ | 502 | 1.34 | - | 2.88 | 2.27 | EG |
| crypt | 326 | 1.00 | 1.00 | 0.54 | 2.98 | 3.92 |
| lufact | 241 | 1.04 | 1.04 | 0.35 | 1.28 | 1.73 |
| series | 225 | 1.00 | 1.00 | 2.04 | 1.10 | 1.25 |
| sor | 320 | 1.01 | 1.08 | 1.35 | 1.04 | 1.42 |
| sparsematmult | 296 | 1.45 | 1.43 | 0.62 | 27.39 | 29.46 |
| moldyn | 244 | 2.44 | - | 1.75 | 22.93 | EG |
| montecarlo | 674 | 1.10 | - | 2.27 | 4.18 | EG |
| raytracer | 314 | 1.70 | - | 1.73 | 49.87 | EG |
| **Mean:** | - | **1.33** | - | - | **7.41** | - |

* “EG” represents triggering exceptions in garbage collection (GC).

lufact, series, sor, and sparsematmult, the memory consumptions among Aero and RT-trace were similar.

### 5.4.3 *Slowdown Overheads*

Table 4 also illustrates the slowdown on each subject by using each technique. We collected the CPU time via the *Linux time* command. The slowdown incurred by each technique is reported as technique’s time spent ÷ Base’s time spent [17], and the results incurred by each technique are shown in the last three columns of Table 4. The fifth column of Table 4 presents the time spent on each subject by Base.

From Table 4, Velo incurred the heaviest slowdown among all the subjects except $eclipse_6$, $xalan_6$, lufact, series, sor, and sparsematmult. On subjects $eclipse_6$ and $xalan_6$, DC used the longest execution time, while RT ran fastest on each. Velo incurred a 10.6x slowdown on $avrora_9$ and DC incurred at least 6x slowdown on 2 subjects (i.e., $xalan_6$ and $avrora_9$). However, on subjects sparsematmult, moldyn and raytracer, Velo and RT incurred more than 20x slowdown. RT incurred more than 6x slowdown on $avrora_9$ and $pmd_9$. The slowdown incurred by Velo varied from 1.05x to 46.38x. On two subjects ($xalan_6$ and $avrora_9$), RT was faster than Velo by at least 1.4x. On $eclipse_6$, $lusearch_6$, $lusearch_9$, $pmd_9$, $xalan_9$, crypt and moldyn, Velo incurred a larger slowdown than RT by at least 1.11x. On subjects $hsqldb_6$, $jython_9$, $luindex_9$, $sunflow_9$, montecarlo and raytracer, RT was slightly faster than Velo by 1.01x to 1.08x, respectively while RT incurred a larger slowdown than Velo by 1.07x to 1.17x on subjects lufact, series, sor and sparsematmult. Overall speaking, RT was faster than Velo by 1.08x.

For DC, the time overhead was 2.43x on average, which was faster than RT. The slowdown of subjects moldyn and raytracer could not be collected for DC, as DC ran out of memory on these two subjects. Moreover, RT was

**Table 6. Run-time characteristics of each technique**

| Benchmark Subject | # of SubRegions | # of Joins | Max # of Transaction Nodes at Run-time | # of End Event | # of Memory Locations | # of Lock Objects | # of Edges* | | # of Logs |
|---|---|---|---|---|---|---|---|---|---|
| | RT | RT | RT | Aero | Aero | Aero | Velo | DC | DC |
| $eclipse_6$ | 1,310 | 306,000 | 78 | 218,000 | 816,000 | 945 | 270,000 | 167,000 | 25,100,000 |
| $hsqldb_6$ | 596 | 12,400 | 12 | 50,400 | 358,000 | 495 | 15,500 | 15,600 | 2,500,000 |
| $lusearch_6$ | 8 | 18 | 6 | 599 | 30700 | 37 | 7 | 17 | 34,400,000 |
| $xalan_6$ | 15,000 | 292,000 | 9 | 20,900 | 56,300 | 169 | 139,000 | 134,000 | 29,100,000 |
| $avrora_9$ | 441,000 | 2,600,000 | 10 | 52,500 | 6,450 | 17 | 875,000 | 2,360,000 | 457,000,000 |
| $jython_9$† | 0 | 0 | 2 | 44 | 2,880,000 | 212 | 0 | 0 | 16,900,000 |
| $luindex_9$† | 0 | 0 | 2 | 44 | 59,500 | 41 | 0 | 0 | 2,000,000 |
| $lusearch_9$ | 33 | 76 | 4 | 55 | 1,910,000 | 110 | 1,910 | 77 | 35,000,000 |
| $pmd_9$† | 0 | 0 | 2 | 43 | 59,700 | 14 | 0 | 0 | 378,000 |
| $sunflow_9$† | 209 | 42,900 | 4 | 35,300 | 13,200,000 | 31 | 15,600 | 358 | 32,500,000 |
| $xalan_9$ | 7,190 | 138,000 | 3 | 96,300 | 74,700 | 154 | 52,000 | 33,200 | 26,000,000 |
| crypt† | 7 | 23 | 9 | 15 | 55 | 4 | 11 | 13 | 166 |
| lufact† | 4 | 16 | 3 | 20 | 47 | 3 | 6 | 9 | 1,190 |
| series† | 4 | 11 | 3 | 13 | 20 | 2 | 8 | 11 | 83 |
| sor† | 4 | 14 | 9 | 3 | 36 | 2 | 8 | 8 | 997,000 |
| sparsematmult† | 3 | 19 | 9 | 15 | 54 | 3 | 7 | 6 | 251,000 |
| moldyn | 5 | 45 | 3 | 19,000 | 61,700 | 1 | 47,400 | - | - |
| montecarlo | 18,800 | 52,200 | 3 | 132,000 | 424,000 | 3 | 42,100 | 28,600 | 41,600,000 |
| raytracer | 5 | 2,560 | 3 | 7,040 | 73,300 | 1 | 10,900 | - | - |
| Total: | 485,000 | 3,450,000 | 174 | 633,000 | 20,100,000 | 2,250 | 1,480,000 | 2,740,000 | 704,000,000 |

* Only show the number of cross-thread edges. Intra-thread edges can be lightweight, and the number of nodes can imply the number of intra-thread edges.

† This symbol represents that Aero can complete the analysis of the whole program execution without producing GC exception. For the remaining subjects, columns 5-7 show the numers of end events, memory locations and lock objects when the subject raised exceptions.

able to run faster than DC on four subjects (i.e., $eclipse_6$, $xalan_6$, $avrora_9$ and $xalan_9$). Furthermore, in order to run successfully on $avrora_9$ in the experimental environment (and in the experiment in [5] alike), DC required our manual tuning on the subject to final specifications. We note that DC used a more efficient event profiling framework Octet [8] than the Empty framework used by both Velo and RT, and DC did not do further analysis on localizing the error but reports the transaction which completes the cycle as the transaction to blamed with.

Table 5 also illustrates the slowdown on each subject by using RT-trace and Aero. From Table 5, on subjects $eclipse_6$, $hsqldb_6$, $lusearch_6$, $xalan_6$, $avrora_9$, $lusearch_9$, $xalan_9$, moldyn, montecarlo and raytracer, Aero raised exceptions in garbage collection caused by recursive errors and thus could not complete. For the remaining subjects $jython_9$, $luindex_9$, $pmd_9$, $sunflow_9$, crypt, lufact, series, sor, and sparsematmult, RT-trace incurred 4.95x slowdown on average while Aero incurred 6.37x slowdown on average. On subject $jython_9$ and $sunflow_9$, Aero ran significantly slower than RT-trace by 2.84x and 1.85x, respectively. On subjects $pmd_9$, crypt, lufact, series and sor, Aero incurred a larger slowdown than RT-trace by more than 1.32x. RT-trace ran slightly faster than Aero on subjects $luindex_9$ and sparsematmult by 1.10x and 1.08x, respectively.

In summary, compared to RT and Velo, DC made a heavy tradeoff between memory overhead and slowdown. Compared to Velo, RT improved both memory overhead and slowdown. RT did not compromise its effectiveness

**Table 7. Results on avrora$_9$ with large input**

| | Base | Empty | RT | Velo | DC | Aero |
|---|---|---|---|---|---|---|
| Memory Usage | 611MB | 682MB | 1532MB | >2GB | >2GB | - |
| Time Spent | 67s | 162s | 413s | - | - | EG |
| # of Threads | N.A. | 28 | 28 | - | - | - |

- Both Velo and DC ran out of memory on avrora$_9$ with large input size.
- Aero caused exception in GC on avrora$_9$ with large input size.

in precisely detecting transactional atomicity violations and identifying non-serializable traces. Although both RT-trace and Aero were able to detect all non-serializable traces, RT-trace incurred lower memory and slowdown overheads than Aero.

*5.4.4 Further Analysis*

Table 6 shows the statistics of each technique at runtime. The second column presents the number of subregions in RT which is the number of additional vector clocks other than thread vector clock and TVCs. The third column shows the number of cross-thread join operations performed by RT, which is also the number of cross-thread happen-before relations on conflicting events. The fourth column presents the maximum number of transaction nodes at run-time of RT which is the resident number of transaction nodes during execution. The fifth to seventh columns show the number of transaction end events, memory locations and lock objects of Aero at run-time, respectively. For failed subjects, these number are collected until Aero caused exceptions. The eighth and nineth columns illustrate the number of cross-thread edges of Velo and DC at run-time, respectively. The last column presents the number of read/write logs maintained by DC.

Velo, RT and Aero were implemented on the same instrumentation framework. In terms of memory overhead, the main difference between Velo and RT was that Velo needed to maintain a transactional HB graph in the memory explicitly. So even though a transaction finished, it might still be kept in an HB graph due to the presence of references to it, and the Garbage Collector (GC) process still could not collect it. For RT and Aero, once a transaction finished, the GC process can collect it immediately because RT and Aero did not need this node for further analysis. On the other hand, RT divided transactions into dynamic subregions and kept many vector clocks to track the update of the transaction state, while Aero needed to store two maps to track all live memory locations and lock objects for traversing at the end events.

DC incurred larger memory overheads than Velo and RT because DC additionally maintained read/write logs [5]. From Table 6, on 8 out of 19 subjects (i.e., eclipse$_6$, lusearch$_6$, xalan$_6$, avrora$_9$, lusearch$_9$, sunflow$_9$, xalan$_9$ and montecarlo), DC maintained 20+ million read/write logs when analyzing these subjects. The number of read/write logs was larger than the number of transaction nodes shown in Table 1 by at least one order of magnitude. As shown in Table 4, on these subjects, DC at least doubled the memory overheads incurred by both Velo and RT; and on these subjects, DC incurred at least 3.5x memory overhead.

As for slowdown, the main difference between Velo and RT was in detecting transactional atomicity violations. Whenever a transaction had a new incoming cross-thread HB edge, Velo verified the atomicity of that transaction by checking whether there was a cycle both starting and ending at this transaction in the HB graph. So, as the number of edges increased, Velo would spend more time on graph traversal, whereas RT would take a constant-time operation to verify the atomicity of a transaction when a join operation was performed on that transaction. In addition, the non-serializable trace analysis of RT will only be invoked if the detected transaction is serializable. Comparing the 4$^{th}$ column of Table 6 and the 4$^{th}$ column of Table 1, at run time, RT kept a small portion of all the transaction nodes in memory at any one time, which also speeded up its analysis.

From Table 1 and Table 6, $eclipse_6$, $xalan_6$ and $avrora_9$ had the largest numbers of transaction nodes and cross-thread transactional happens-before dependencies among all subjects, but RT was able to run faster than Velo on these three subjects. The performance difference between Velo and RT on each of $jython_9$, $luindex_9$ and $pmd_9$ was small. We observed that these three subjects have the least number of transaction nodes and cross-thread edges among all subjects.

DC used the instrumentation framework Octet[8] which was significantly different from Velo and RT. Octet efficiently captures the cross-thread dependencies by only inserting synchronized instrumentation when the access potentially involves in cross-thread conflicting dependencies. However, the instrumentation framework of Velo and RT insert synchronized instrumentation for every memory access, except when the access is well-synchronized. Comparing the numbers of accesses in the last three columns of Table 1 with the numbers of cross-thread edges in the $5^{th}$ and $6^{th}$ columns of Table 6, Octet of DC apparently incurs much less slowdown than the corresponding instrumentation of Velo and RT. Moreover, DC contains an imprecise cycle detection and a precise cycle detection. For imprecise cycle detection, DC invokes it when a transaction ends rather than a cross-thread edge creates. The imprecise cycle detection detects strongly connected components and filters out non-promising transactions for precise cycle detection. After imprecise cycle detection, DC invokes the precise cycle detection whenever a cross-thread edge creates. DC contains a two-phase detection that slows its overall analysis. Based on these designs, DC incurred almost similar slowdowns as RT, while these two techniques both ran faster than Velo.

Aero incurred a larger slowdown than RT-trace, and as presented above, Aero can complete its analysis on 9 out of 19 subjects, which are marked with the symbol † in the $1^{st}$ column of Table 6. Although Aero is stated to be a linear algorithm to identify non-serializable traces, we observed from the experiment that the performance obstacle of Aero is in the traverse of each thread, memory location and locks at each transaction end. In the offline experiment, all vector clocks of memory locations and lock objects were stored in central maps, which was convenient for traversal. However, in the online experiment environment, there was no such scheme to directly access the vector clocks of all live memory locations and lock objects. Thus, at each write event or lock release event, Aero needed to track the live memory locations and lock objects for the following traversal process. During the traverse, to ensure all vector clocks were updated atomically, the vector clock of each thread, each memory location or each lock object should be checked and updated in a critical section, which slows down Aero. Moreover, with the growth in the number of live memory locations and lock objects, the traversal analysis triggered at each transaction end occurrence appears to spend a longer time, influencing the program execution itself and caused exceptions. From Table 6, most subjects that successfully ran by Aero (i.e., $jython_9$, $luindex_9$, $pmd_9$, crypt, lufact, series, sor and sparsematmult) had a small number of transaction end events and lock objects. As a whole, Aero ran slower than RT-trace on these subjects. This observation is in line with the time complexity of AeroDrome and RegionTrack, where the difference mainly lies in $t^2 n_{join}$ versus $(t+L+V)n_{end}$. Note that the number of threads is normally a small value while the number of memory locations is a large value (e.g., on $sunflow_9$, there were 6 threads, 43k join operations for RT, 35k transaction end events, 31 lock objects, and 13-million memory locations), which makes Aero ran slower than RT-trace in the experiment. In addition, we found that subjects $jython_9$, $luindex_9$ and $pmd_9$ did not have any join operations but have transaction end events. Hence, the non-serializable trace analysis of RT was not invoked at all, whereas Aero needed to traverse and update its vector clocks accordingly at each transaction end event.

Overall speaking, compared to Velo, DC and Aero, RT efficiently detected all transactional atomicity violations and non-serializable traces for a given trace without needing follow-up analysis.

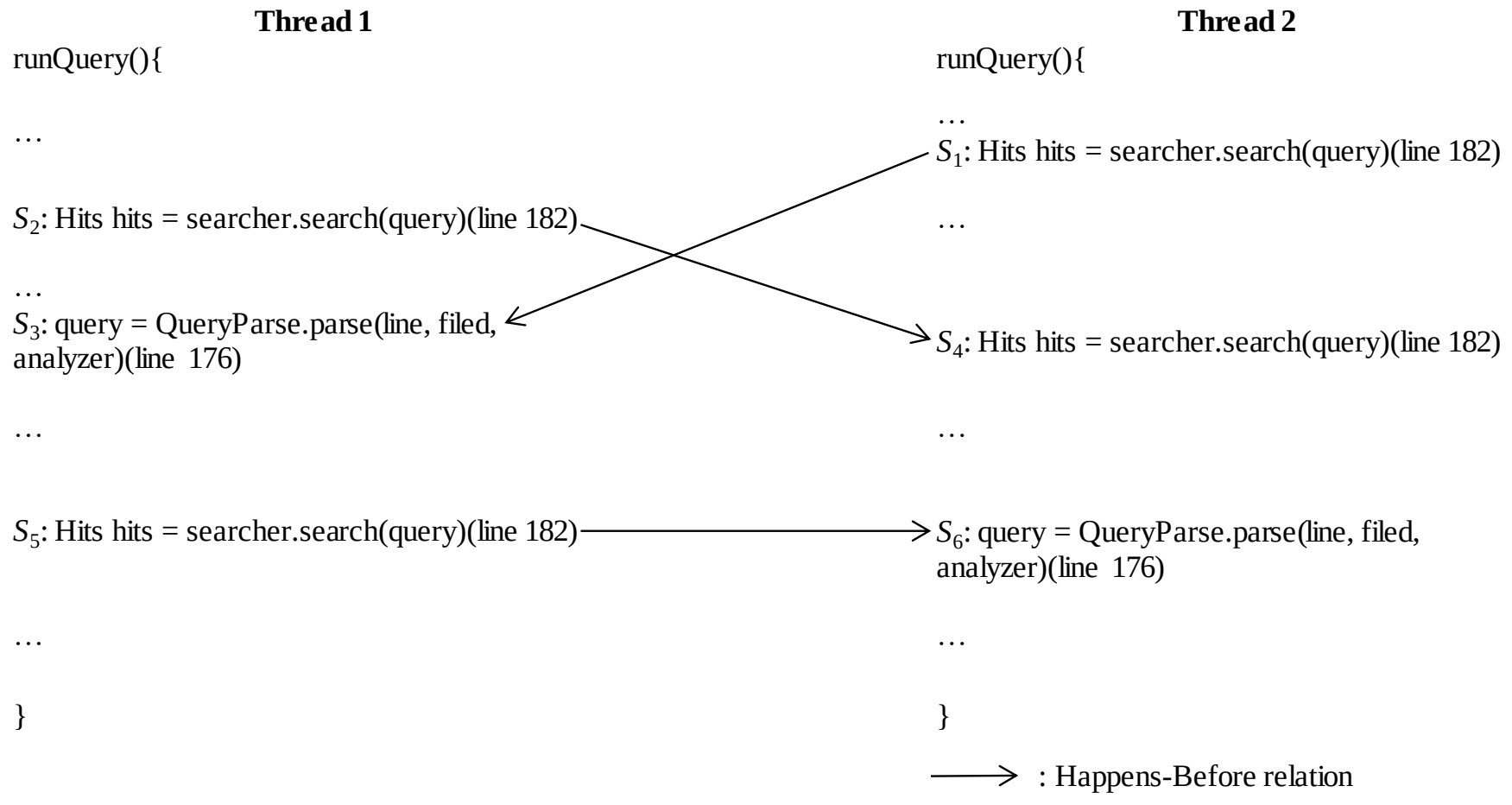

**Figure 11. Transactional Atomicity Violation in Lusearch of DaCapo-2006-10-MR2 missed reporting by Velodrome but detected by RegionTrack.**

### *5.4.5 Handling Long Execution Trace*

We chose avrora9 to study the memory consumption of each technique on analyzing long execution traces because avrora9 generated one of the longest execution traces, as shown in Table 1. We ran avrora9 with the large input size [7] under each technique using the procedure described in Section 5.3. Table 7 shows the results. The trace generated by using the large input size was longer than the trace generated using small input by one order of magnitude in terms of the number of transaction nodes and the number of accesses.

From Table 7, both Velo and DC ran out of memory (i.e., >2GB, Jikes RVM is limited to a heap of 2GB). Aero caused exceptions in GC and thus could not complete at run time. However, RT completed its analysis by consuming only 1532MB memory. It demonstrates that RT provides a feasible way to analyze program executions that previous techniques experienced difficulties in analyzing them.

### *5.4.6 Found Bugs*

From the transactional atomicity violations detected in the experiments, RegionTrack precisely reported all transactional atomicity violations. Figure 11 illustrates a transactional atomicity violation in Lusearch of DaCapo-2006-10-MR2. The arrow line represents a happens-before relation between the head and the tail statements. The violation was missed by Velodrome but detected by RegionTrack. The buggy interleaving is shown in Figure 11. After the statement $S_3$ has executed, the instance of *runQuery* in thread 1 establishes a happens-before relation with the instance of *runQuery* in thread 2. Similarly, after the statement $S_4$ has executed, the instance of *runQuery* in thread 2 establishes another happens-before relation with the instance of *runQuery* in thread 1. Both Velodrome and RegionTrack captured these two relations. However, after the execution of the statement $S_6$, there is an increasing cyclic sequence of THB relations (i.e., increasing cycles) according to the relation sequence: $S_1 \rightarrowtail S_3 \rightarrowtail S_5 \rightarrowtail S_6$. Thus, RegionTrack reported a transactional atomicity violation on the instance of *runQuery* in thread 2. Velodrome missed this violation because it can only record the relation $S_2 \rightarrowtail S_4$ between these two executions and misses the relation $S_5 \rightarrowtail S_6$.

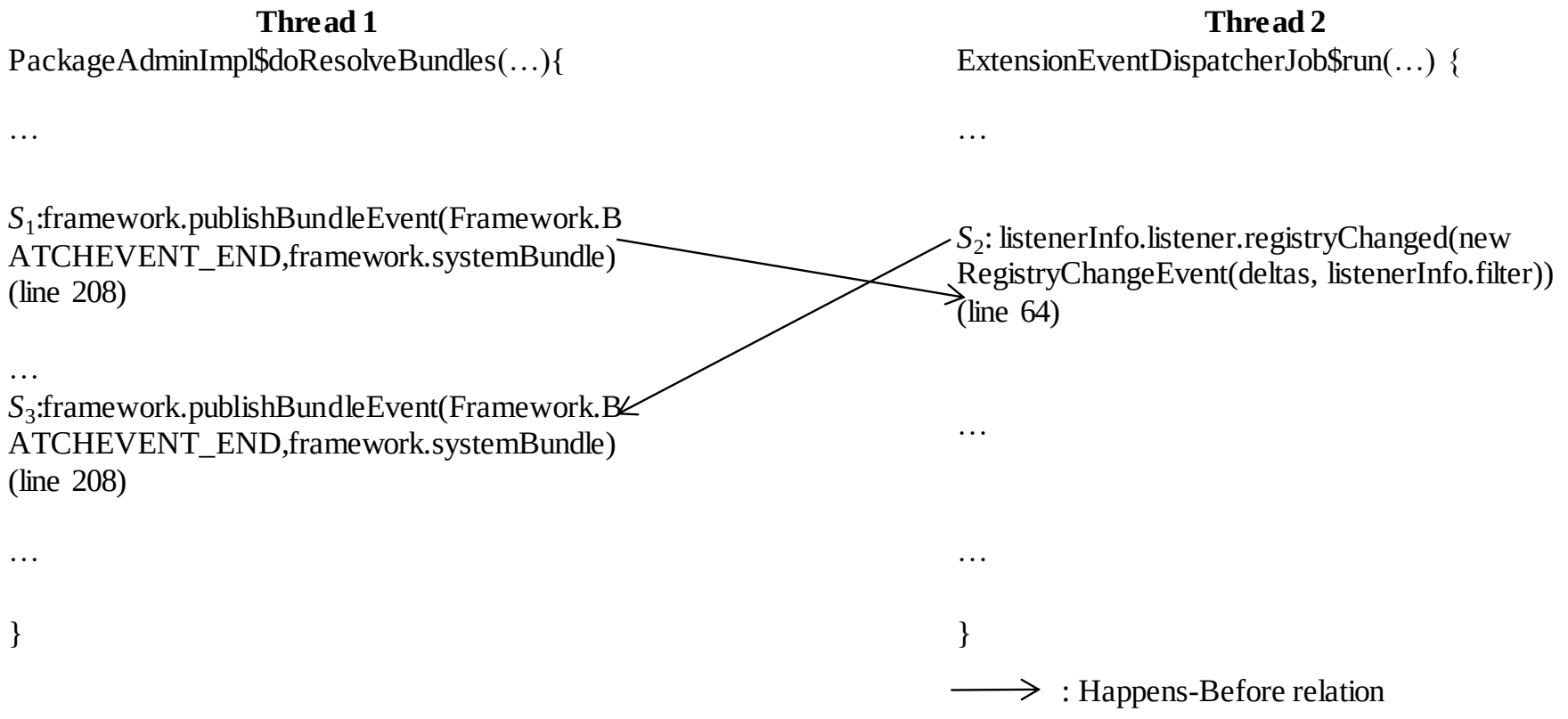

**Figure 12. Transactional Atomicity Violation in Eclipse of DaCapo-2006-10-MR2 falsely reported by DoubleChecker.**

Figure 12 shows a false positive of transactional atomicity violation reported by DoubleChecker in Eclipse of DaCapo-2006-10-MR2. When thread 2 executes the function *ExtensionEventDispatcherJob$run*, the duration of statement $S_2$ is long. Thus, a happens-before firstly establishes between statement $S_2$ and statement $S_3$ in the execution of function *PackageAdminImpl$doResolveBundles* of thread 1 (i.e., $S_2 \rightarrowtail S_3$), and then another happens-before relation between statement $S_1$ and statement $S_2$ is built (i.e., $S_1 \rightarrowtail S_2$). Therefore, DoubleChecker detects a cycle and reports a transactional atomicity violation on *ExtensionEventDispatcherJob$run,* which completes the cycle. But, the detected cycle is not increasing, indicating the execution of *ExtensionEventDispatcherJob$run* is still serializable. Therefore, DoubleChecker reports a false positive of transactional atomicity violation, while RegionTrack correctly identifies this case, and reports a non-serializable trace.

## 5.5 Threats to Validity

Although the experiments can be successfully conducted on 19 subjects from two different benchmark suite: DaCapo benchmark suite [7] and Java Grande Forum benchmark suite [48], there all still some subjects that cannot be executed, such as chart from DaCapo 2006-10-MR2, and tomcat, tradebeans, tradesoap from DaCapo 9.12-bach. The errors of these subjects fall into two categories: toolkit load exception (chart) and server initialization exception (tomcat, tradebeans and tradesoap). The microbenchmark suite [51] is not available online and JikeRVM has some internal errors to execute SPECjvm2008 benchmark [49]. The use of other benchmark suite and subjects may produce different results. Using more subjects can improve the generalization of the results. Some errors in executing these benchmarks are due to the limitations of JikesRVM [59]. Conducting online experiments on a different framework may obtain more empirical results.

For offline experiments, we only collected 100 different traces for each subject. Consider the interleaving space of program execution. These collected traces only occupied a tiny fraction of the whole interleaving space. We observed from the statistics among the profiled traces of each subject that their statistics do not differ significantly. Having said that, larger experiments may obtain more generalized results. In addition, the instrumentation itself will also influence the execution of a program. The traces we collected may be different from the one the program originally executed. However, the reason for conducting offline analysis is that we cannot control the executing trace of online analysis. Thus, during the offline analysis, each technique processed the same execution trace for a fair comparison. In the experiment, we were not able to collect runtime statistics when configuring RVM to have all four tools enabled at the same time. It was mainly due to the memory constraint imposed on RVM. Using a different JVM runtime may allow the collection of different statistics, which, if it can be done, it will help cross-validate the results of this experiment.

We did not re-implement DC using the Velo framework as the main advantage of DC is to use its underlying framework (Octet [7]) to profile events and transactions. Porting all tools to be run on Octet or porting DC to be run on Velo's underlying framework may produce results. Moreover, Octet was designed for fast profiling (faster than the framework provided with the Velo implementation), the interpretation of the results on DC running faster than RT should be careful.

The original implementation of Aero was an offline version that could not be used to evaluate the memory overhead and slowdown overheads of its online performance. Thus, we implemented RT and Aero ourselves. We referenced the implementation of Velo and followed its style and data structures whenever possible to lower the chance of results due to the choices of data structures and low-level algorithms and implementation code. We have validated the implementations to ensure their correctness. However, since the frameworks of Velo, DC and Octet were not written by us and the source codes were small. There may be latent bugs in them. However, we did not observe abnormality from the experimental results, which shows that the results were consistent with what we expected according to their theory on producing false negatives and false positives among these four techniques. Currently, Velo, RT and Aero share the same instrumentation framework, which added two words for each object and static filed: one references the vector clock write to the filed, and the other references the vector clock read the field. Therefore, if we implemented an online version of Aero, the new tool needed to maintain additional maps to track the memory locations and lock objects for the following traverse at the transaction end event. Instrumenting the object and static filed by storing the vector clocks in a central map may help to successfully execute the failed subjects of Aero online.

We have compared the four tools on detection precision, runtime overhead and space overhead using multiple metrics. Measuring them using other metrics may produce different insights on these four techniques.

We were not able to run Aero on all 19 benchmarks in the experiment. The comparison to Aero in this experiment was thus limited to these nine relatively small-scale programs (traces). However, from the algebraic term of their time complexity, we can see that the main difference is a tradeoff between $t^2$ and $t+L+V$ and a tradeoff between $n_{join}$ and $n_{end}$. From Table 6, we can see that $n_{join}$ and $n_{end}$ are similar in scale on 9 subjects successfully completed by Aero. From Table 1, $t$ ranges from 2 to 43, and from Table 6, the total number of memory locations and lock objects ranges from 22 to more than 13 million. We tend to believe that overall speaking, RT could be more efficient than Aero on practical benchmarks.

We chose to use offline analyses when needing to compare techniques on the same trace. An alternative approach is to employ a deterministic replay technique (e.g., [40]) so that the same trace can be replayed for different techniques to analyze. We originally attempted to use TPLAY [40], the latest state-of-the-art deterministic replay technique for this purpose. However, we found that TPLAY was still insufficient to enable an experiment on iterative refinement methodology to run smoothly. More specifically, in our experiments presented above, benchmarks like $\text{eclipse}_6$ required continuously reproducing the same trace more than 21 times to analyze the involved transactions iteratively. This means that TPLAY needs to deterministically replay the same trace for more than 21 times to conduct an online iterative refinement analysis and to run the whole experiment smoothly, such as in our cases, we need to repeat the trial attempts by 100 times, it demands a high probability of successful reproduction of the recorded traces. Moreover, the overhead of deterministic replay was still very high [40]. Hence, if we used such a replayer, we would end up with many failed attempts, which would lengthen our experiment and may pose a threat in our data analysis if we would not clean up the garbage correctly. Therefore, we leave it as future work.

## 6 Related Work

Velodrome [16] and DoubleChecker [5] have extensively been reviewed and compared in the above sections. In the sequel, we review other closely related work.

Farzan and Madhusudan [13] also build a transactional HB graph in checking atomicity. They summarize the effect of completed transactions to minimize the graph by throwing in transitive edges and absorbing their event content into active transactions. In addition, they leverage this method and propose a solution to the model checking problem for checking atomicity in concurrent Boolean programs. Their analysis is offline and targets to check atomicity for concurrent Boolean programs. Similar to the online technique RegionTrack, which can offload the burden to a garbage collector to collect terminated transactions once completed, their technique summarizes the HB graph when transactions terminate to reduce memory overhead. Their technique also aims to locate cycles in the summarized transactional HB graph. The slowdown overhead incurred by this technique should be similar to that of Velodrome, which we believe that it would be slightly larger than that of RegionTrack. Moreover, their analysis detects non-serializable traces but cannot identify non-serializable transactions. The present paper has shown that efficiently identifying such transactions in a sound and complete manner is challenging.

In checking atomicity, Atomizer [14][15] synthesizes Lipton's theory of reduction [22] and the lockset algorithm [30] to reason about the standard synchronization idiom (i.e., mutual-exclusion locks). Atomizer actively checks whether each step of each step follows a guaranteed serializable pattern and also searches atomicity violations that may appear in other interleavings. Thus, Atomizer may produce false positives since it depends on the lockset algorithm, whereas RegionTrack is both sound and complete in detecting non-serializable traces. Moreover, a lockset algorithm is more efficient than VC-based tracking, Atomizer should incur smaller slowdown overheads than RegionTrack.

Wang and Stoller detected non-serializable traces based on unserializable patterns. They aimed to detect predictive violations in other possible executions. However, their technique is not precise and produces false positives. They further proposed a commit-node algorithm [42] and a block-based algorithm [43] to eliminate false positives, but these algorithms may still report false positives. Moreover, these algorithms incurred much larger slowdown overheads than RegionTrack at run-time. Some other techniques [9][10][11][12][19][33][35] also detect non-serializable traces using predictive analysis. Since predicting the whole trace can be prohibitively expensive, these techniques limit to sliced causality, hybrid analysis, or single-variable/multi-variable atomicity violation detection. Based on an observed trace, these predictive techniques report violations not only for currently observed trace but also for other possible interleavings of the trace. However, these techniques may lead to false positives and also false negatives in reporting non-serializable traces, and cannot detect transactional atomicity violations. In addition, these techniques are offline and need to log the execution trace first for their prediction purposes. Some of them (e.g., [19]) rely on constraint solvers to find a solution which may limit the scalability of these techniques. These predictive techniques apparently incurred larger slowdown overheads than RegionTrack due to its complicated analysis.

Techniques such as CTrigger [24][28] and PENELOPE [36] employ a two-phase strategy to expose atomicity violations [18][20][27]. They predict suspicious instances of violations in a given trace in the first phase and examine these instances by scheduling confirmation runs in another phase, whereas RegionTrack precisely and completely detect transactional atomicity violation in the given trace. In particular, CTrigger examines a few runs of a program and observes a large set of atomicity violation interleavings. Then, CTrigger prunes away infeasible interleavings based on order synchronization and mutual exclusion synchronizations. For the remaining interleavings, CTrigger ranks them according to its occurrence probability and gives preferences to those interleavings with lower probability. PENELOPE predicts the possible atomicity violations by finding the cut-points in the first phase. Then, it generates alternate schedules that reach the cut-points concurrently to expose the atomicity violations which heuristically has the maximum consistency to the original execution. Thus, CTrigger and PENELOPE can detect atomicity violations that may not appear in the current observed execution run, but they can only detect violations due to two threads and one variable. In contrast, RegionTrack only monitors one observed execution and detects violations involving any number of threads and variables that occurred in that execution trace. The monitoring and prediction phases of PENELOPE are efficient and do not incur large slowdown overheads. However, when it comes to the rescheduling phase, PENELOPE incurred significant slowdown overheads to reschedule the feasible interleavings. ASP [45] and ASR [46] insert a prioritization phase and a reduction phase, respectively, between the prediction and confirmation to make this class of techniques cost-effective. Swan [34] makes use of given buggy execution traces to generate thread schedules to expose bugs. These techniques generate different execution traces to expose potential atomicity violations. RegionTrack targets observed traces and can benefit from the traces generated by these techniques since the generated traces can be sent to RegionTrack to detect violations soundly and completely in the traces. Compared to RegionTrack which runs along with the program execution, the main slowdown overheads of these two-phase techniques are caused by rescheduling all generated feasible interleavings, which in essence run much longer than RegionTrack.

Intruder [29] synthesizes test cases to detect atomicity violations. EnfoRSer [32] enforces the atomicity of code regions at runtime. SOFRITAS [36] proposes a new Ordering-Free Region serializability consistency model to prevent atomicity violations. G. Agha and K. Palmskog [31] introduces a novel algorithm to infer the concurrency structure of programs from their traces. NodeAV [47] proposes a method to detect atomicity violations for Node.js applications. Deterministic replay techniques [39][40][41] can also help to debug atomicity violations. TSVD is a thread-safety violation detector through active testing, which does not track any synchronization operations and happens-before relations [50]. Their research directions are different from RegionTrack.

Dynamic race detectors use VCs to precisely track happens-before relations which are expensive in time and space overhead. To provide a fast path analysis for the common cases, FastTrack [17] proposes a lightweight *epoch* to only record the timestamp on the thread of read/write events, which requires only constant space and supports constant-time operations compared to the traditional VC algorithms. Similar to FastTrack [17], RegionTrack uses VCs to track happens-before relation. However, when verifying the atomicity of a transaction, the timestamp on the transaction's thread (which is different from the thread of compared event) is used in the comparison, so the epoch approach cannot be applied to the detection of transactional atomicity violations. SlimState [44] improves FastTrack by compressing the shadow states on array-type data structures. It relies on the notion of *epoch*, which cannot be applied to track happens-before relations for checking of transactional atomicity. Therefore, these techniques incurred fewer memory and slowdown overheads than RegionTrack according to the utilization of epoch.

AeroDrome [52] proposes a VC-based algorithm to identify non-serializable traces for multithreaded programs, which shares a similar idea of RegionTrack. We have discussed it in Section 4.4 and compared it to RegionTrack in the experiment. AeroDrome encodes all THB relations in the same vector clock of each thread and thus can only identify whether the execution trace is serializable, while RegionTrack can further locate the transaction violating atomicity which the algorithm (Algorithms 1 and 2 presented in this paper) has been presented in the PhD thesis [53] of an author of this paper. Due to the optimization of the number of read vector clocks of each memory location, AeroDrome incurs lower memory overheads than RegionTrack. Although RegionTrack is a non-linear algorithm, it does not run slower than AeroDrome in practice.

Maiya and Kanade [26] identify a major bottleneck in computing the HB relations and introduce a new and efficient data structure, called *event graph*. Through an event graph, one only keeps a subset of the HB relations to efficiently infer the relation between any pair of events, which improves the overhead of VCs' algorithm. It is interesting to explore to what extent RegionTrack can be efficient if the integration of their event graph to the design of RegionTrack is feasible.

## 7 Conclusion

In this paper, we have presented a novel online atomicity checker named RegionTrack. RegionTrack dynamically localizes transactional atomicity violations along observed execution traces and identifies non-serializable traces of multithreaded programs. We prove its soundness and completeness by theorems. RegionTrack is designed on top of the transitivity of happens-before relations for both conflicting events and transactions. It can discard each finished transaction and its associating happens-before relations and vector clocks right after its novel mechanism of timestamp propagation. We have evaluated RegionTrack to be both memory- and time-efficient. Future work includes a study on further optimization of the timestamp propagation mechanism and integration with other concurrency bug detection techniques such as data race checkers, atomicity violation detection at the memory location level, and linearizability checkers. We will further develop the concept of timestamp propagation.

## ACKNOWLEDGEMENTS

This research is supported in part by the GRF of HKSAR Research Grants Council (project nos. 11214116, 11200015 and 11201114), the HKSAR ITF (project no. ITS/378/18), the CityU MF_EXT (project no. 9678180), the CityU SRG (project nos. 7004882, 7005122, and 7005216), and NSFC of China (project nos. 61772056 and 61690202).